\documentclass[amsmath,amssymb,pra,aps,showpacs,superscriptaddress,twocolumn]{revtex4-2}
\usepackage{amsmath,amsfonts,amssymb,amsthm,graphics,graphicx,epsfig,bbm}
\usepackage[colorlinks=true,citecolor=blue,linkcolor=red,urlcolor=blue]{hyperref}

\usepackage{color}
\usepackage{graphicx}
\usepackage{dcolumn}
\usepackage{float} 
\usepackage{bbm}

\usepackage{setspace}

\def \markColorOne {Plum}
\def \markColorTwo {Red}
\def \markColorThree{blue}

\def \markColorOne {Black}
\def \markColorTwo {Black}
\def \markColorThree{Black}

\usepackage[dvipsnames]{xcolor}

\newcommand \braket[2]{\left|#1 \right>\left<#2 \right|}

\newcommand \dket[1]{\left|\left.#1 \right>\right>}
\newcommand \dbra[1]{\left<\left<#1 \right.\right|}
\newcommand \dketbra[2]{\left|\left.#1 \right>\right>\left<\left< #2\right. \right|}
\newcommand \dbraket[2]{\left<\left< #1 \mid  #2\right> \right>}

\begin{document}

\title{Photon-resolved Floquet theory II:  Open quantum systems }

\author{Georg Engelhardt}
\email{engelhardt@sustech.edu.cn}

\affiliation{Shenzhen Institute for Quantum Science and Engineering, Southern University of Science and Technology, Shenzhen 518055, China}
\affiliation{International Quantum Academy, Shenzhen 518048, China}
\affiliation{Guangdong Provincial Key Laboratory of Quantum Science and Engineering, Southern University of Science and Technology, Shenzhen, 518055, China}

\author{JunYan Luo}

\affiliation{Department of Physics, Zhejiang University of Science and Technology, Hangzhou 310023, China}

\author{Victor M. Bastidas}

\affiliation{Physics and Informatics Laboratory, NTT Research, Inc.,
940 Stewart Dr., Sunnyvale, California, 94085, USA}

\affiliation{Department of Chemistry, Massachusetts Institute of Technology, Cambridge, Massachusetts 02139, USA}

\author{Gloria Platero}

\affiliation{Instituto de Ciencia de Materiales de Madrid ICMM-CSIC, 28049 Madrid, Spain}

\date{\today}

\pacs{
  }

\begin{abstract}
 	Photon-resolved Floquet theory keeps track of the photon exchange of a quantum system with a coherent driving field. It thus complements the standard full-counting statistics  that counts the  number of photons exchanged  with incoherent photon modes giving rise to dissipation. In this paper, we introduce a unifying framework  describing both situations. We develop methods suitable for an analytical evaluation of low-order cumulants of photonic probability distributions. Within this framework we analyze the two-mode Jaynes-Cummings model to demonstrate that the Photon-resolved Floquet theory and the standard full-counting statistics make consistent statistical predictions. Interestingly, we find that the photon-flux fluctuations diverge for vanishing dissipation, which can be related to an  entanglement effect between the driven matter system and the driving field. 
  To substantiate our results, we use our framework to describe efficient photon up-conversion in an ac-driven lambda system, that is characterized by a high signal-to-noise ratio. 
 	 As the framework is non-perturbative and predicts  fluctuations, it paves the way towards  non-perturbative spectroscopy, which will assist to improve metrological methods. 
\end{abstract}

\maketitle

\allowdisplaybreaks


\section{Introduction}

Full-counting statistics (FCS) is a successful theoretical framework finding broad applications in many fields of quantum physics~\cite{Sanchez2007,Sanchez2008,Flindt2009,Bastianello2018,Ridley2019,Pollock2022,Gerry2023,Mcculloch2023}. FCS was originally developed by Mandel in quantum optics to describe spontaneous photon emission~\cite{mandel1979sub}, and has been later adapted to electron transport through mesoscopic systems by Levitov and Lesovik~\cite{levitov1996electron}. The formalism also describes thermal transport in heat engines and other thermal devices~\cite{Restrepo2018}. More recently, a Floquet version of the FCS has been developed to describe electron transport through  periodically-driven quantum dots~\cite{kleinherbers2018revealing,benito2016fullCounting,Honeychurch2020,PicoCortes2019}. Moreover, within the framework of FCS, feedback control protocols have been developed which can suppress transport noise~\cite{Xu2023}.

However, the standard FCS fails to correctly predict the photon exchange with coherent photonic driving fields which is important in stimulated absorption and emission processes. The FCS formalism is fundamentally based on two-point projective measurements that project the state of photons at the beginning and the end of the time evolution to extract the photon statistics. Formally, the initial projection destroys coherences  of the photonic field~\cite{Schoenhammer2007}. Thus, the resulting state after the projection is not a coherent state. To circumvent this problem,  the Photon-resolved Floquet theory (PRFT) has been recently developed, whose formalism avoids the initial projective measurement~\cite{Engelhardt2024}.

By modifying concepts and methods of the standard FCS,  the PRFT is capable to  predict the photonic probability distribution by integration of the semiclassical equations of motion, allowing for a simple analytical and numerical treatment. Intriguingly, the PRFT makes correct predictions in a deep-quantum regime beyond the capability of common semiclassical methods. It predicts  light-matter entanglement in the Floquet-state basis, which has been suggested as a resource  for  highly-efficient quantum communication protocols using coherent states instead of single photons~\cite{Engelhardt2024}. These predictions are beyond typical semiclassical light-matter frameworks such as methods from standard non-linear spectroscopy~\cite{Mukamel1995} or perturbation theory in the Floquet basis~\cite{Yan2019,Engelhardt2021,Kumar2020,Tiwari2023}.

\begin{figure*}
	\includegraphics[width=\linewidth]{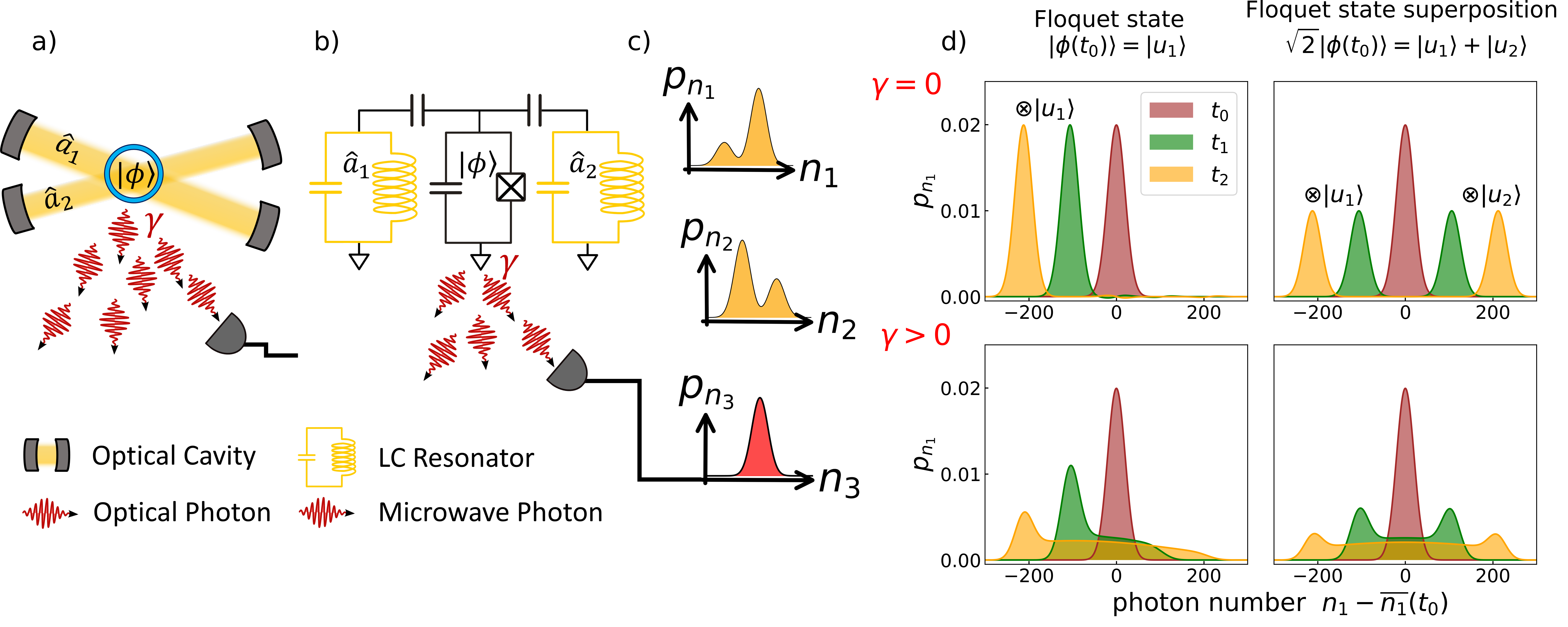}
	\caption{(a) and (b) Sketch of a  quantum system in  state $\left| \phi\right>$ which is driven by two coherent light fields and coupled to a continuum of  optical and microwave photon modes, respectively, with coupling strength $\gamma$. In the case of optical photons, the atom is coupled to two optical cavities and a continuum environment of photonic modes, while in the case of microwave photons, a nonlinear LC resonator (transmon) is coupled to two LC resonators acting as cavities.  Due to the light-matter interaction, photons become redistributed between the driving modes and the reservoir. c) The resulting photonic probability distributions can be measured and predicted within the framework introduced in Sec.~\ref{sec:unifyingFrameworkOfCountingStatistics}.  d) Photonic probability distribution  of the coherent photon mode $k=1$ in closed ($\gamma =0$) and open ($\gamma>0$) quantum systems for three different light-matter interaction durations $t_0$, $t_1$, $t_2$. For $\gamma =0$, the photonic dynamics is determined by the Floquet states: (l.h.s.) photonic probability distribution for an initial Floquet state; (r.h.s.) photonic probability distribution for an initial superposition of two Floquet states. For long times, the initial superposition state leads to a light-matter entangled state  in the  Floquet-state basis. For $\gamma >0$, the dissipation induces jumps between the Floquet states. As a consequence, the light-matter entanglement  depicted in
		the r.h.s. is gradually destroyed.  The dynamics has been calculated for the Jaynes-Cummings model in Sec.~\ref{sec:jcModel}, for $\epsilon_{\Delta} =0$, $\Omega_1=\Omega_2$, $\varphi_1=0$, $\varphi_2 =\pi/2$, and $\gamma =0.01 \Omega_1$. }
	\label{figOverview}
\end{figure*}

	Unification of the standard FCS and the PRFT is needed to describe systems where both spontaneous and stimulated emissions are present, which is the aim of the present work. Sketches of  typical setups consisting of a quantum systems, two driving modes, and a dissipative photonic environment (i.e., a bath) in a cavity and circuit quantum electrodynamics experiments are  depicted in Fig.~\ref{figOverview}(a) and Fig.~\ref{figOverview}(b), respectively.
	The first challenge is to modify the PRFT, that has been originally developed for closed quantum systems, to  open quantum systems, in which the FCS is valid. The second challenge is to ensure that all photons are consistently counted 
	under the constraint of photon conservation:  Energy lost in the coherent photon modes must eventually re-appear in the incoherent ones. It is interesting to note that, while the FCS has been originally developed in quantum optics, the theoretical framework has revealed its full potential in mesoscopic electron transport, where charges can tunnel in a multi-terminal setup. The PRFT takes advantage of the rich methodology of mesoscopic transport, and establishes a quantum-optical framework that describes photon flux between various photonic modes, which can be likewise viewed as a multi-terminal setup.

In a series of two papers, we investigate in detail basic questions of the PRFT framework. In the first paper, we analyze the scaling properties of the PRFT in closed quantum systems when approaching the semiclassical limit of the light-matter interaction~\cite{Engelhardt2024c}. In this paper,  we investigate the PRFT in open quantum systems.
In doing so, we unify the standard FCS and the PRFT for quantum systems which are both coherently driven and coupled to a continuum of incoherent photon modes. Moreover, we develop three analytical methods to characterize the photonic probability distributions: (i)  expansion of the characteristic polynomial of the Liouvillian in a power series of the counting field; (ii)  generalizing the stationary perturbation theory known  for Hermitian system to non-Hermitian  ones; and (iii)  modifying  well-established adiabatic elimination procedures.

Taking a two-mode generalization of the Jaynes-Cummings model as an example, we demonstrate that our unified formalism consistently describes the photon statistics of coherent and incoherent photon modes. \textcolor{\markColorThree} {The findings are illustrated in Fig.~\ref{figOverview}(d). In an isolated quantum system, light and matter becomes entangled in the Floquet basis. When dissipation is present, the entanglement decreases during the time evolution.}
 Intriguingly,  this effect  leads to diverging photon flux fluctuations for a small dissipation as we will show in this paper. As a more sophisticated application, we apply the formalism to an ac-driven lambda system, and show how to deploy it as a frequency-conversion device of coherent light characterized by a high signal-to-noise ratio.

 Importantly,  the theory proposed in the present work  paves  the way to a non-perturbative spectroscopic framework that allows for statistical predictions beyond   standard methods from non-linear spectroscopy~\cite{Mukamel1995} and quantum optics~\cite{Scully1997}, which is of utmost importance in chemistry~\cite{Swofford1978} and biology~\cite{Krafft2017}. For instance, we derive exact expressions for the photon flux and its fluctuations  in the paradigmatic two-mode Jaynes-Cummings model, that are non-perturbative in all parameters. These findings will constitute a basis for improved metrological methods with impact on other areas of physics such as dark-matter detection~\cite{Smith1990}. In fact, increasing effort is committed to develop highly-sensitive  dark-matter detectors based on co-magnetometers, that use lasers  to measure the magnetic state of atoms via the Faraday effect~\cite{Budker2014,Bloch2020}.  Floquet systems are deployed to improve the sensitivity of   so-called Floquet co-magnetometers~\cite{Bloch2022}. Similar approaches are used to improve the standard of atomic clocks~\cite{Yin2021,Hutson2023}.    The theoretical framework developed here will pave the way  to develop stochastic   models that predict  measurement statistics, and can thus help to  further enhance the sensitivity in such experiments. 

Besides  optics, the framework developed here can also find application in fast read-out protocols required in  quantum computation devices based on superconducting flux qubits~\cite{Blais2021} and silicon quantum dots~\cite{Vigneau2023}. In particular, recent developments in the field of arrays of superconducting qubits~\cite{Kjaergaard2020,Blais2021} exploit the coupling of the qubit to resonators to perform dispersive readout of the qubit state~\cite{Walter2017}. In addition, the qubit is coupled to external control lines and tunable couplers that enable to apply microwave pulses to drive the dynamics of the system~\cite{Chen2014,Geller2015}. As the qubit is coupled to control lines and resonators for read out, it is also under the effect of the external environment~\cite{Gu2017}, understanding the statistics of the photons is of extreme interest in the field~\cite{Hofer2016,Brange2019,Nesterov2020}.
Dispersive read-out schemes typically 
employ off-resonance configurations, which limits the read-out speed. The exact photon statistics predicted by the   PRFT would allow the development of close-to-resonance read-out protocols, which could  significantly reduce the read-out time.

This paper is organized as follows. In Sec.~\ref{sec:unifyingFrameworkOfCountingStatistics}, we  introduce a general framework that  unifies the standard FCS and the PRFT. In Sec.~\ref{sec:jcModel}, we apply the framework to a  multimode Jaynes-Cummings model and investigate the fate of the light-matter entanglement in  open quantum systems. In Sec.~\ref{sec:acDrivenLambdaSystem}, we investigate an ac-driven lambda system for its application as an efficient frequency conversion device. In Sec.~\ref{sec:conclusions}, we conclude and discuss our results. We provide detailed derivations and introductions to various analytical methods in the Appendices.

\section{Unifying framework of FCS and PRFT }

\label{sec:unifyingFrameworkOfCountingStatistics}

Here, we establish a unifying framework allowing for simultaneously counting coherent and incoherent photon processes. Moreover, we explain how to evaluate cumulants in the long-time limit using Floquet theory in terms of the eigenvalues of the stroboscopic Liouvillian operator.

\subsection{System}

\label{sec:system}

 Figure~\ref{figOverview}(a) depicts a quantum system driven by $N_{ \text{D}}$ coherent light fields (e.g., in optical cavities), that is coupled to $N_{ \text{B}}$ modes, to which it incoherently emits photons. To describe both effects on an equal footing, we start from  a fully-quantized light-matter Hamiltonian
\begin{equation}
\hat H  = \hat H_{\text{M}} + \sum_{k=1}^{N_{ \text{D} }+N_{\text{B} } }\omega_k \hat a_k^\dagger  \hat a_k  +  \sum_{k=1}^{N_{ \text{D} }+N_{\text{B} } }  g  \hat V_k \left(  \hat a_k^\dagger   + \hat a_k \right),
\label{eq:hamiltonian:quantum}
\end{equation}
where $\hat H_{\text{M}}$ denotes the  matter Hamiltonian, and the photonic creation $\hat a_k^\dagger$ and annihilation $\hat a_k$ operators  quantize the electromagnetic field. The light-matter interaction is described by  $\hat V_k$ acting on the matter system, where $g$ parameterizes the interaction strength. At this point it is important to emphasize the generality of our approach. The Hamiltonian in Eq.~\eqref{eq:hamiltonian:quantum} is a general description of a system coupled to a bosonic enviromnent. Besides in quantum optics, one encounters this configuration in circuit quantum electrodynamics,  where LC resonators analogous to cavities can be coupled to transmons acting as qubits~\cite{Zhang2019}, which is sketched in Fig.~\ref{figOverview}(b). In fact, the Hamiltonian we discuss in this work is also relevant for the coupling of a superconducting qubit to a collection of acoustic modes that can be exploited as hardware-efficient implementation of a quantum random access memory (QRAM)~\cite{Hann2019}.

 The photonic modes are distributed into two categories: 
The first $N_{\text{D}}$ photonic modes coherently drive the matter system. Their joint initial state shall be given by a product of coherent states
\begin{eqnarray}
\left| A_0  \right> &=&  \prod_{k=1}^{N_\text{D}} \left|\alpha_k e^{i\overline \varphi_k }\right>  \nonumber \\
 &=& \sum_{\boldsymbol n }  a_{\boldsymbol n} \left|\boldsymbol n \right>,
\label{eq:coherentIntialState}
\end{eqnarray}
that are defined as the eigenstates of the annihilation operators  $\hat a_k \left|\alpha_k e^{i\overline  \varphi_k }\right>  =\alpha_k e^{i\overline \varphi_k }  \left|\alpha_k e^{i\overline \varphi_k }\right> $ with real-valued amplitudes $\alpha_k$ and phases $\overline  \varphi_k$. In the second equality, we have expressed the photonic initial state in the Fock basis $ \left|\boldsymbol n \right>  = \bigotimes_{k=1}^{N_{\text{D}}}  \left| n_k \right>   $, where $\boldsymbol n =(n_1,\dots n_{\text{D}})$ denotes a vector of photon numbers in each coherent photonic mode $k$.

 In the semiclassical limit $\alpha_k \gg1$, it is justified to replace the photonic operators by their expectation values in the Heisenberg picture $\hat a_k(t) \rightarrow \alpha_k e^{-i\omega_k t + i \overline  \varphi_k}$. This establishes a semiclassical Hamiltonian that  describes the dynamics of the matter system~\cite{Engelhardt2024}, which we consider in the present work. Conceptually, we focus here on coherent light in a cavity setup, while application of the PRFT to propagating light fields (e.g., laser light) will be considered in relation to spectroscopy in a future work. However, we note  that these two situations are almost identical on a formal level.

The photonic modes  $k= N_{\text{D} } +1, \dots, N_{\text{D}} + N_{\text{B}} $ form a continuum (i.e., a bath with $N_{\text{B} }\gg 1$) and are assumed to be initially in a thermal state
\begin{eqnarray}
\rho_{\text{B}}  &=& \frac{1}{Z} \exp\left(-\beta \sum_{k=N_{\text{D}}+ 1}^{N_{\text{D}} +N_{\text{B} }} \omega_k \hat a_k^\dagger  \hat a_k \right) \nonumber \\
&=& \sum_{\boldsymbol \nu}  	p_{\boldsymbol  \nu}(0)  \left|\boldsymbol \nu \right> \left< \boldsymbol \nu \right|,
\label{eq:thermalInitalState}
\end{eqnarray}
where $\beta$ denotes the inverse temperature and $Z$ is the partition function. In contrast to a coherent state, the thermal state is diagonal in the photon-number basis. Thus, these modes act incoherently on the matter system and give rise to dissipation.  In the second line, we have introduced a Fock-state representation of the initial state, $ \left|\boldsymbol \nu \right>  = \bigotimes_{k=N_{\text{D}}+1}^{N_{\text{D}}+N_{\text{B}}}  \left| \nu_k \right>   $, where $\boldsymbol \nu =(\nu_{N_{\text{D}}+1},\dots \nu_{N_{\text{D}}+N_{\text{B}}})$ denotes a vector of photon numbers in each incoherent photonic mode $k$ with $\hat a_k^\dagger  \hat a_k  \left| \nu_k \right>  = \nu_k  \left| \nu_k \right> $. The probability that the bath is  initially in  photonic state labeled by $\boldsymbol \nu$ is denoted by $	p_{\boldsymbol  \nu}(0)  $.

Throughout the paper, the initial state of the joint light-matter system is assumed to be
\begin{eqnarray}
\rho_{\text{tot}}(0) &=& \rho_{\text{M} }(0) \otimes \rho_B \otimes \left| A_0 \right> \left< A_0  \right| \nonumber \\
&=& \sum_{\boldsymbol \nu} 	p_{\boldsymbol  \nu}(0)    \rho_{\text{tot},\boldsymbol \nu} (0),
\label{eq:initialState}
\end{eqnarray}
where $\rho_{\text{M} }(0)$ is the initial state of the matter system. For later purpose, we have expanded the initial state in terms of the photon numbers $\boldsymbol  \nu$ of the incoherent modes, where $p_{\boldsymbol  \nu}(0)$ is the photonic probability distribution and 
\begin{equation}
 \rho_{\text{tot},\boldsymbol \nu} (0) =  \rho_{\text{M} }(0) \otimes \left| A_0 \right> \left< A_0  \right|  \otimes \left|\boldsymbol \nu \right> \left< \boldsymbol \nu \right| 
 \label{eq:conditionalDensityMatrix}
 \end{equation}
denotes the corresponding conditioned density matrix.

\subsection{Coherent and incoherent counting of photons }
\label{sec:countingOfPhotons}

{\color{\markColorThree}
The aim of both the standard FCS and the PRFT is to  predict the statistics of the photon number operators $\hat n_k = \hat a_k^\dagger \hat a_k $  of the incoherent and coherent photonic modes, respectively, as a function of time. To enable the numerical and analytical treatment of the  large number of photonic modes in the system,  both methods consider only the dynamics of the  (generalized) reduced density matrix of the matter system~\cite{Engelhardt2024,Schoenhammer2007,levitov1996electron}.  

However, because of the different form of the  initial states of the coherent modes in Eq.~\eqref{eq:coherentIntialState} and the incoherent modes in Eq.~\eqref{eq:thermalInitalState},  the photons in coherent and incoherent modes are counted in a distinct fashion: The joint probability distribution unifying the  standard FCS and the PRFT is then defined by
\begin{equation}
\overline p_{\boldsymbol  n  , \Delta \boldsymbol  \nu   }(t) \equiv \sum_{\boldsymbol  \nu} p_{\boldsymbol  n , \boldsymbol  \nu +\Delta \boldsymbol  \nu \mid \boldsymbol  \nu  }(t) p_{\boldsymbol  \nu}(0)
\label{eq:def:jointProbabilityDistribution}
\end{equation}
with
\begin{equation}
	p_{\boldsymbol  n , \boldsymbol  \nu +\Delta \boldsymbol  \nu \mid \boldsymbol  \nu  }(t) = \text{tr} \left[ \hat P_{\boldsymbol n }   \hat P_{\boldsymbol \nu + \Delta\boldsymbol \nu  }  \rho_{\text{tot},\boldsymbol \nu}(t) \right] ,
	\label{eq:conditionalProbabilities}
\end{equation}
 which quantifies the conditional  probability that we measure $\boldsymbol  n$ photons in the coherent modes and   $\boldsymbol  \nu + \Delta\boldsymbol  \nu $ photons  in the incoherent modes at time $t$ given that that there  have been initially $\boldsymbol  \nu$ photons in the incoherent modes. Thereby, $\rho_{\text{tot},\boldsymbol \nu}(t) =\hat U(t)\rho_{\text{tot},\boldsymbol \nu}(0)\hat U^\dagger(t) $ denotes the conditioned density matrix at time $t$, which has been evolved by the time evolution operator $\hat U(t)=\exp \left(-i\hat H t\right)$ corresponding to the Hamiltonian in Eq.~\eqref{eq:hamiltonian:quantum}. The Fock state projection operators are defined by $\hat P_{\boldsymbol n }  = \left|\boldsymbol n \right> \left< \boldsymbol n\right|$ and $\hat P_{\boldsymbol \nu }  = \left|\boldsymbol \nu \right> \left< \boldsymbol \nu \right|$, respectively.

 As we see from Eq.~\eqref{eq:def:jointProbabilityDistribution}, the counting statistics of the coherent and incoherent photons is defined in a distinct way. This reflects two conceptually different measurement protocols. In the standard FCS, the state of the incoherent modes is measured at the begin and the end of the measurement process, which is usually dubbed as a two-point projective measurement~\cite{Schoenhammer2007,levitov1996electron}. The standard FCS is thereby interested in the statistics of the difference of the photon numbers $\Delta \boldsymbol \nu$ at the begin and the end  of the time evolution. In contrast, in the PRFT, the state of the coherent modes is measured only at the end of the measurement process. A projective measurement at the begin of the time-evolution would destroy the coherences in the Fock basis, which would be inconsistent with the initial coherent state assumed in Eq.~\eqref{eq:coherentIntialState}.

In general, the  joint photonic probability distribution of the standard FCS and the PRFT can be obtained by a  Fourier transformation of the  moment-generating function
	\begin{equation}
	\overline p_{\boldsymbol  n, \Delta \boldsymbol  \nu }(t) = \int_{-\pi}^{\pi} \frac{ d \boldsymbol\chi }{(2\pi)^{N_{\text{D}}} } \int_{-\pi}^{\pi} \frac{d\boldsymbol\xi}{(2\pi)^{N_{\text{B}}} } M_{\boldsymbol  \chi , \boldsymbol \xi   } (t)e^{i( \boldsymbol n \cdot \boldsymbol \chi + \Delta \boldsymbol \nu \cdot \boldsymbol \xi  )},
	\label{eq:relation_momGenFkt_probabilites}
	\end{equation}
where the notation $\int_{-\pi}^{\pi} d \boldsymbol\chi =\int_{-\pi}^{\pi} d\chi_1\dots \int_{-\pi}^{\pi}d\chi_{N_{\text{D}}}$ denotes  a multidimensional integration (likewise for $\boldsymbol\xi $).
As shown in Appendix~\ref{app:defMomentGenFct}, the moment-generating function corresponding to the probability distribution in Eq.~\eqref{eq:def:jointProbabilityDistribution} can be expressed as
\begin{eqnarray}
		M_{\boldsymbol  \chi,\boldsymbol \xi } (t) 
		&\equiv&\text{tr} \left[  e^{-i\left( \boldsymbol  \chi\cdot \hat {\boldsymbol n}_{\text{D}} + \boldsymbol  \xi \cdot \hat {\boldsymbol n}_{\text{B}}   \right)  }   \hat   U (t)  e^{i\boldsymbol  \xi\cdot \hat {\boldsymbol n}_{\text{B}}   }    \rho_{\text{tot}} (0)   \hat U^{\dagger} (t)  \right].
		\label{eq:def:momentGeneratingFkt_exact}
		\end{eqnarray}
Thereby, we have introduced the notation $\boldsymbol  \chi \cdot \hat {\boldsymbol n}_{\text{D}}  =   \sum_{k=1}^{N_{\text{D}}}  \chi_k \hat n_{k}   $ and $\boldsymbol  \xi\cdot \hat {\boldsymbol n}_{\text{B}}  =   \sum_{k= N_{\text{D}}+1}^{N_{\text{D}}+ N_{\text{B}}}  \xi_k \hat n_{k}   $, where the so-called counting fields $\chi_k$ and $ \xi_k$ are tracking the photons exchanged with the coherent driving fields and the incoherent bath modes, respectively. For convenience, the counting fields have been arranged  in vectors  $ \boldsymbol  \chi = \left(\chi_{1} ,\dots, \chi_{N_{\text{D}}}  \right) $ and $ \boldsymbol  \xi = \left(\xi_{N_{\text{D}}+1} ,\dots, \xi_{N_{\text{D}} +N_{\text{B}} }  \right) $.  The operation $\text{tr}\left[\bullet \right]$ in Eq.~\eqref{eq:def:momentGeneratingFkt_exact} denotes the trace over all degrees of freedom, i.e.,  the matter system, driving modes, and bath.

 }

Alternatively, one can describe the probability distribution in terms of its cumulants, which can be calculated via the cumulant-generating function $K_{\boldsymbol  \chi,\boldsymbol \xi   } (t)  \equiv \log M_{\boldsymbol  \chi,\boldsymbol \xi   } (t) $.
The cumulants are  defined via
\begin{equation}
\kappa_l^{(k)}  (t) = \left.\frac{d^{l}}{d(-i\zeta_k)^{l}}   K_{\boldsymbol \chi,\boldsymbol \xi}( t )\right|_{\boldsymbol \chi=0, \boldsymbol \xi=0}  ,
\label{eq:def:cumulantsGeneral}
\end{equation}
where $\zeta_k$ refers to $\chi_k$ for $k\leq N_{\text{D}}$ and or  $\xi_k$ for $k> N_{\text{D}}$, respectively.  The first two cumulants 
\begin{eqnarray}
\kappa_1^{(k)}  (t) &=& \left< \hat n_k(t)\right>  = \overline n_{k}(t),\nonumber\\
\kappa_2^{(k)}  (t) &=&  \left< \hat n_k^2(t)\right> -\left< \hat n_k(t)\right>^2 =\sigma_k^2(t)
\end{eqnarray}
are the mean $ \overline n_{k}$ and the variance $\sigma_k^2$ of the photonic distribution, respectively.  Likewise, one can calculate correlations between different photon modes by evaluating  mixed derivatives.

\subsection{Semiclassical dynamics in an open quantum systems}

\label{sec:openQuantumSystems}

Because of the large number of  photon modes in the fully-quantized Hamiltonian and the infinite-sized Hilbert space of each mode, the dynamics of the moment-generating function in Eq.~\eqref{eq:def:momentGeneratingFkt_exact} cannot be exactly calculated in general. Instead, effective approaches must be applied to determine its time evolution.

%
%

{\color{\markColorTwo}
As we explain  in Appendices~\ref{sec:exact_momGenFct} and \ref{sec:semiclassicalLimit}, within a semiclassical representation of the coherent photonic modes in Sambe space,  the moment-generating function in Eq.~\eqref{eq:def:momentGeneratingFkt_exact} can be exactly expressed as
\begin{eqnarray}
M_{\boldsymbol  \chi, \boldsymbol \xi   } (t)  = M_{\text{dy}  ,\boldsymbol  \chi  ,\boldsymbol \xi} (t) M_{ \boldsymbol  \chi,\boldsymbol \xi  } (0),
\label{eq:unifyingMomentGeneratingFkt}
\end{eqnarray}
where the moment-generating function at initial time $t=0 $ is given by
\begin{equation}
M_{\boldsymbol  \chi,\boldsymbol \xi  } (0) =\left< A_0  \right|   e^{-i\boldsymbol  \chi\cdot \hat {\boldsymbol n}_{\text{D}   } }\left| A_0  \right>.
\label{eq:initial_momentGeneratingFunction}
\end{equation}
The initial moment-generating function contains the information about the photon statistics of the  coherent modes at time $t=0$. In contrast, $M_{\boldsymbol  \chi, \boldsymbol \xi   } (0)$  does not depend on the  state of the incoherent modes. This reflects the projective counting protocol of the incoherent photons which starts counting at $t=0$.

The change of the photon statistics due to the light-matter interaction is described by the dynamical moment-generating function, which is given by
\begin{eqnarray}
M_{\text{dy}  ,\boldsymbol  \chi ,\boldsymbol \xi } (t) =  \text{tr}\left[\rho_{\boldsymbol  \chi ,\boldsymbol \xi } (t) \right]
\label{eq:dynamcalMomentGeneratingFkt}.
\end{eqnarray}
Thereby,  $\rho_{\boldsymbol  \chi , \boldsymbol \xi  } (t)$  denotes a generalized reduced density matrix of the matter subsystem only, and not the coherent and incoherent photonic modes.  Its time-evolution can thus be described as an open quantum system, which makes a numerical simulation feasible.

Several methods have been developed to describe the time-evolution of reduced density matrices in  open quantum systems, such as the quantum master equation~\cite{Breuer2002,Schaller2014a}, the Redfield equation~\cite{Redfield1965}, the polaron-transformed master equation~\cite{Silbey1984,Xu2016}, the hierarchy equations of motion~\cite{Ishizaki2005,Hou2014}, the quantum Langevin approach~\cite{Gardiner2004,Wiseman2010,Scully1997},  quasi adiabatic propagator path integral methods~\cite{Makri1995,Richter2017}, and other more advanced approaches~\cite{Becker2022}. The detailed equation of motion for the reduced density matrix depends thereby on the deployed method and the system under investigation. As we illustrate in Sec.~\ref{sec:blochEquations} for the Jaynes-Cummings model, a perturbative approach in the coupling to the incoherent photonic modes in the Born-Markov approximation typically gives rise to a generalized quantum master equation of the following form
\begin{eqnarray}
 \frac{d}{dt}   \rho_{ \boldsymbol  \chi,\boldsymbol \xi  }  &=&  -i \left[ \hat{\mathcal H}_{\overline {\boldsymbol \varphi}   + \frac{\boldsymbol \chi}{2}  } (t) \rho_{ \boldsymbol  \chi,\boldsymbol \xi  }  -   \rho_{ \boldsymbol  \chi,\boldsymbol \xi  } \hat{\mathcal  H}_{\overline {\boldsymbol \varphi}   - \frac{\boldsymbol \chi}{2} } (t)   \right] \nonumber  \\
 &&+ \mathcal \sum_j  \gamma_j D_{\boldsymbol \xi } \left[ S_j \right]\rho_{ \boldsymbol  \chi,\boldsymbol \xi  } ,
 \label{sec:quantumMasterEquation}
\end{eqnarray}
where the semiclassical Hamiltonian 
\begin{equation}
\hat {\mathcal H}_{\boldsymbol \varphi}(t)   =\hat  H_{\text{M}} +  \sum_{k=1}^{N_{\text{D} } } 2 g_k \hat V_k \alpha_k \cos\left(  \omega_k t   - \varphi_k  \right)
\label{eq:hamiltonian:semiclassical}
\end{equation}
describes the action of  the coherent modes in Eq.~\eqref{eq:hamiltonian:quantum} on the matter system. The impact of the incoherent modes is encoded into the dissipator $D_{\boldsymbol \xi } \left[\hat S_j \right]\rho$ with system operators $\hat S_j$, which are determined by  the  $\hat V_k$ and the exact form of the Hamiltonian in Eq.~\eqref{eq:hamiltonian:quantum}. The inital condition of the time evolution is given by $\rho_{\boldsymbol  \chi, \boldsymbol \xi  } (0) = \rho_{\text{M}}(0)$.
}

{\color{\markColorThree}
  As we can see in Eq.~\eqref{sec:quantumMasterEquation},  the density matrix  $\rho_{\boldsymbol  \chi,\boldsymbol \xi } $ does not follow a Hermitian time evolution: (i) the coherent dynamics for the forward time evolution is determined  by $\mathcal H_{\overline {\boldsymbol \varphi}   + \frac{\boldsymbol \chi}{2} }  $, while the backward time evolution is determined by $\mathcal H_{\overline {\boldsymbol \varphi}   - \frac{\boldsymbol \chi}{2} }  $; (ii) the dissipator deviates from the Lindblad form for a non-vanishing counting field $\boldsymbol \xi$. Since  consequently  $\left| \text{tr}\left[\rho_{ \boldsymbol  \chi,\boldsymbol \xi  } \right]\right| \leq 1$  as shown in Appendix~\ref{sec:upperBoundOfTrace}, these objects are strictly speaking no density matrices. Yet, these generalized density matrices can be considered as mathematical tools that trace the photons exchanged with the photonic modes. For $(\boldsymbol  \chi, \boldsymbol \xi  )=(\boldsymbol  0 ,\boldsymbol  0)$, they reduce to the common density matrix of the matter system, i.e., $\rho_{\boldsymbol  0 ,\boldsymbol  0 }(t)=\rho_{\text{M}}(t)$.

}

 Importantly,  we can define a dynamical cumulant-generating function by $K_{\text{dy} ,\boldsymbol  \chi,\boldsymbol \xi  } (t)  \equiv \log M_{\text{dy} ,\boldsymbol  \chi,\boldsymbol \xi } (t) $. Using Eqs.~\eqref{eq:def:cumulantsGeneral} and \eqref{eq:unifyingMomentGeneratingFkt}, we find that  its derivatives
 \begin{eqnarray}
 \kappa_{\text{dy} ,l}^{(k)}  (t) &=& \left.\frac{d^{l}}{d(-i\zeta_k)^{l}}   K_{\text{dy} ,\boldsymbol  \chi,\boldsymbol \xi }( t )\right|_{\boldsymbol \chi=0,\boldsymbol \xi=0}   \nonumber \\
   &=& \kappa_{l}^{(k)}  (t) - \kappa_{l}^{(k)}  (0)
 \end{eqnarray}
  describe the difference of the cumulants at the beginning and the end of the time evolution. From now on, these derivatives will be referred to as dynamical cumulants.

Noteworthy, the time derivative of the first dynamical cumulant of the coherent modes can be also calculated in a semiclassical fashion without resorting to the PRFT via
\begin{equation}
	\frac{d}{dt} \kappa_{\text{dy} ,k}^{(1)}  =  \frac{d}{dt}n_{k}(t)  = \left< \frac{d}{d\varphi_k} \hat {\mathcal H}_{\overline {\boldsymbol \varphi}}(t) \right>_t,
	\label{eq:semiclassicalCurrent}
\end{equation}
where the expectation value is taken in the  state of the matter system at time $t$, and $\frac{d}{d\varphi_k}  \hat {\mathcal H}_{\boldsymbol \varphi=\overline {\boldsymbol \varphi}}(t)$ denotes photon-flux operator~\cite{Long2021,Crowley2019,Crowley2020}. We note that the calculation of higher-order cumulants via the photon-flux operator  is more delicate as explained in detail in the companion paper~\cite{Engelhardt2024c}.

\subsection{Long-time limit}

\label{sec:longTimeLimit}

By interpreting the reduced density matrix as a vector, the quantum master equation in Eq.~\eqref{sec:quantumMasterEquation} can be written in matrix form as
\begin{equation}
\frac{d}{dt}\rho_{\boldsymbol  \chi,\boldsymbol \xi}(t) =  \mathcal L_{\boldsymbol  \chi,\boldsymbol \xi} (t ) \rho_{\boldsymbol  \chi,\boldsymbol \xi}(t)
\end{equation}
with a time-dependent Liouvillian $\mathcal L_{\boldsymbol  \chi,\boldsymbol \xi} (t )$. We note that many computational methods for open quantum systems can be brought in this convolution-less form. 

We now focus on the important case of time-periodic systems with $\mathcal L_{\boldsymbol  \chi,\boldsymbol \xi} (t ) =\mathcal L_{\boldsymbol  \chi,\boldsymbol \xi} (t +\tau )  $ for some period $\tau$. According to  Floquet theory, the general solution  can be  written as
\begin{equation}
\rho_{\boldsymbol  \chi,\boldsymbol \xi}(t) =  \mathcal B_{\boldsymbol  \chi,\boldsymbol \xi} (t )\exp \left( \mathcal A_{\boldsymbol  \chi,\boldsymbol \xi}  t\right) \rho_{\text{M} }(0),
\label{eq:generalFloquetSolution}
\end{equation}
where $\mathcal B_{\boldsymbol  \chi,\boldsymbol \xi} (t ) =\mathcal B_{\boldsymbol  \chi,\boldsymbol \xi} (t+\tau )$ inherits the time-periodicity of the underlying equations of motion. In particular, $\mathcal B_{\boldsymbol  \chi,\boldsymbol \xi} (n\tau ) =\mathbbm 1$ as required by the initial condition $\rho_{\boldsymbol  \chi,\boldsymbol \xi}(0) =\rho_{\text{M} }(0) $.
The long-time dynamics is determined by the time-independent stroboscopic  Liouvillian $\mathcal A_{\boldsymbol  \chi,\boldsymbol \xi}$. This is a  non-Hermitian matrix, whose right and left eigenstates are defined by, respectively,
\begin{eqnarray}
\mathcal A_{\boldsymbol  \chi,\boldsymbol \xi} \dket{ u_{\mu,\boldsymbol  \chi,\boldsymbol \xi} }  & =& \lambda_{\mu;\boldsymbol  \chi,\boldsymbol \xi} \dket{ u_{\mu,\boldsymbol  \chi,\boldsymbol \xi}  } , \nonumber \\
\dbra{  u_{\mu,\boldsymbol  \chi,\boldsymbol \xi}  } \mathcal A_{\boldsymbol  \chi,\boldsymbol \xi}   &=& \dbra{  u_{\mu,\boldsymbol  \chi,\boldsymbol \xi}  }   \lambda_{\mu; \boldsymbol  \chi,\boldsymbol \xi} ,
\label{eq:effectiveLiouvillianEigenvalues}
\end{eqnarray}
 where $\mu$ labels the eigenstates. 
The left and right eigenvalues $\lambda_{\mu; \boldsymbol  \chi,\boldsymbol \xi  }  $ are identical and the eigenvectors fulfill the orthogonality relation $\left<\left< u_{\mu_1} \mid u_{\mu_2}\right> \right>=\delta_{\mu_1,\mu_2} $.  In contrast to the Hermitian case, the left and right eigenvectors are not related by Hermitian conjugation.
The effective Liouvillian can thus  be expanded as
\begin{equation}
\mathcal A_{\boldsymbol  \chi,\boldsymbol \xi} =   \sum_{\mu =0}^{d^2-1} \lambda_{\mu;\boldsymbol  \chi,\boldsymbol \xi} \dketbra{  u_{\mu,\boldsymbol  \chi,\boldsymbol \xi}} {  u_{\mu,\boldsymbol  \chi,\boldsymbol \xi} } ,
\label{eq:effectiePropagatorExpansion}
\end{equation}
where $d$ is the dimension of the matter  system. The eigenvalues are sorted according to their real parts for $(\boldsymbol  \chi,\boldsymbol \xi) = (\boldsymbol 0,\boldsymbol 0) $ , i.e., $ \text{Re} \,\lambda_{\mu;0,0}  > \text{Re} \,\lambda_{\mu';0,0}   $ for $\mu<\mu'$. As the dissipative system will approach a periodic stationary state in the long-time limit, the first eigenvalue fulfills $ \lambda_{0;0,0}  = 0 $. The real parts of the other eigenvalues $\mu>0$ are negative and describe the transient dynamics. \textcolor{\markColorTwo}{In this work, we assume that the eigenvalue with $ \text{Re} \,\lambda_{0;0,0} =0$ is unique, while noting that counterexamples related to dissipative phase transitions in many-body systems exist~\cite{Kessler2012,Minganti2018}.  }

According to Eqs.~\eqref{eq:generalFloquetSolution}  and ~\eqref{eq:effectiePropagatorExpansion},  the long-time behavior is determined by the $\mu=0$ eigenvalue, such that the dynamical moment-generating function at stroboscopic times $t_n=n\tau$ (integer $n$) asymptotically becomes 
\textcolor{\markColorOne}{
\begin{eqnarray}
 M_{\text{dy},\boldsymbol  \chi,\boldsymbol \xi}( t_n ) &\rightarrow & e^ {\lambda_{0;\boldsymbol  \chi,\boldsymbol \xi} t_n}  .
 \label{eq:cumulantGenertingFunctionLongTimes}
\end{eqnarray}
}
 We note that for times $t\neq n\tau $, the time-periodic matrix $ \mathcal B_{\boldsymbol  \chi,\boldsymbol \xi}(t)$ describes only  bounded  photonic flux oscillations, and does not contribute to a large-scale photon flux.

From  relation~\eqref{eq:cumulantGenertingFunctionLongTimes}, we can determine the  cumulants of the photonic modes in the long-time limit
\textcolor{\markColorOne}{
\begin{eqnarray}
\kappa_{\text{dy},l}^{(k)}(t_n)  &=&  \left.\frac{d^l}{d(-i\zeta_k)^l } \lambda_{0;\boldsymbol  \chi,\boldsymbol \xi} \right|_{\boldsymbol \chi=0,\boldsymbol \xi=0} t_n  .
 \label{eq:def:cumulantsLongTime}
\end{eqnarray} }
 The cumulants increase linearly in time because of the exponential dependence of the moment-generating function in Eq.~\eqref{eq:cumulantGenertingFunctionLongTimes}. 
 
 Using the linear time-dependence  of the first and second cumulants in Eq.~\eqref{eq:def:cumulantsLongTime}, we define the asymptotic  mean photon flux and its variance by
\begin{subequations}
	\label{eq:def:photonFluxAndVar}
\begin{eqnarray}
	I_k &=& \lim_{t_n\rightarrow \infty }\frac{1}{t_n}\kappa_{\text{dy},1}^{(k)}(t_n) ,\label{eq:def:photonFlux} \\
	\sigma_{I,k}^2   &=& \lim_{t_n\rightarrow \infty }\frac{1}{t_n}\kappa_{\text{dy},2}^{(k)}(t_n),
	\label{eq:def:photonVar}
\end{eqnarray}
\end{subequations}
which are time-independent observables.
We note that this limit is well-defined not only for stroboscopic times $t_n=n\tau$, but holds also when considering the periodic time dependence described by $\mathcal B_{\boldsymbol  \chi,\boldsymbol \xi} (t )$ in Eq.~\eqref{eq:generalFloquetSolution}, as it does  describe bounded periodic  oscillations of the first two cumulants that do not  grow linearly in time.

 In general, it is very challenging to determine the eigenvalues and eigenvectors of the effective Liouvillian exactly. In Appendix~\ref{app:cumulantEvaluationLongTimeLimit}, we introduce three methods which can be used to perturbatively determine the eigenvalue $ \lambda_{0;\boldsymbol \chi, \boldsymbol \xi}  = 0 $ required to construct the cumulants in the long-time limit in Eq.~\eqref{eq:cumulantGenertingFunctionLongTimes}, which have their specific advantages and disadvantages depending on the investigated  system.

\subsection{PRFT in closed quantum systems}
\label{sec:photonResolvedFloquetTheory}

While the implications of the standard FCS have been frequently investigated~\cite{Flindt2009,Bastianello2018,Ridley2019,Pollock2022,Gerry2023,Mcculloch2023,mandel1979sub,levitov1996electron,Schoenhammer2007}, the PRFT has been only recently developed~\cite{Engelhardt2024,Engelhardt2024c}. Here, we review  shortly the light-matter entanglement predicted by the PRFT in closed quantum systems including $N_{\text{D}} $ coherent modes and  neglecting the incoherent ones, as this is the basis to understand the photonic dynamics in open quantum systems.

The time evolution operator for  the periodically-driven Hamiltonian in Eq.~\eqref{eq:hamiltonian:semiclassical} (i.e., with commensurate frequencies $\omega_k$ for the modes $k=1,\dots,N_\text{D}$)  can be written as
\begin{equation}
 \hat { \mathcal U}_{\boldsymbol  \varphi}(t) = \hat { \mathcal U}_{\text{kick},\boldsymbol \varphi}(t)  \exp\left(-i \hat {\mathcal H}_{\text{Fl},\boldsymbol  \varphi }t\right).
\end{equation}
The kick operator  $  \hat { \mathcal U}_{\text{kick},\boldsymbol\varphi }(t)   =   \hat { \mathcal U}_{\text{kick},\boldsymbol \varphi }(t+\tau)  $ is time periodic with a period $\tau$ and describes periodic changes of the system state. 
The Hermitian Floquet Hamiltonian can be expanded as  $ \hat  {\mathcal H}_{\text{Fl},\boldsymbol  \varphi} = \sum_\mu E_{\mu, \boldsymbol \varphi} \left| u_{\mu,\boldsymbol \varphi}(0) \right>  \left<  u_{\mu,\boldsymbol \varphi}(0) \right| $, where the $E_{\mu, \boldsymbol\varphi}$ are referred to as the quasienergies, and the $\tau$-periodic states $\left| u_{\mu,\boldsymbol \varphi}(t) \right> = \hat { \mathcal U}_{\text{kick},\boldsymbol \varphi}(t) \left| u_{\mu,\boldsymbol \varphi}(0) \right> $ are denoted as Floquet states.

Next, we investigate this by using the language of Liouvillians. The imaginary parts of the eigenvalues of the effective Liouvillian in Eq.~\eqref{eq:effectiveLiouvillianEigenvalues} are associated with  the quasienergy gaps $E_{\mu_1,\boldsymbol \varphi }-E_{\mu_2,\boldsymbol \varphi } $, while the real parts vanish as the system is not dissipative. When expanding the initial state in the Floquet-state basis $ \left| u_{\mu, \overline {\boldsymbol \varphi}}\right>$
\begin{equation}
\left| \Psi (0)\right> = \sum_\mu c_\mu \left| u_{\mu, \overline {\boldsymbol \varphi}} \right> \otimes \left|  A(0) \right>,
\label{eq:genericInitialState}
\end{equation}
and using this to evaluate Eq.~\eqref{eq:dynamcalMomentGeneratingFkt}, we  find the dynamical moment-generating function at stroboscopic times $t_n=n\tau$
\begin{eqnarray}
M_{ \text{dy},\boldsymbol \chi}(t_n )   &\rightarrow&  \sum_\mu  \left|c_\mu \right|^2  e^{i \left(E_{\mu, \overline {\boldsymbol \varphi} - \frac{\boldsymbol \chi  }{2} } - E_{\mu ,\overline {\boldsymbol \varphi} + \frac{\boldsymbol \chi }{2}} \right) t_n }.
\label{eq:momentGenFkt_closedSystem}
\end{eqnarray}
In contrast to the asymptotic limit of the dissipative system, that is determined by only one eigenvalue of the effective Liouvillian in Eq.~\eqref{eq:cumulantGenertingFunctionLongTimes}, the long-time dynamics of the closed system depends on all quasienergies.  Deriving with respect to the counting field $\chi_k$, we find for the mean photon-number change $\Delta \overline n_k(t_n)   = \overline n_k(t_n)  - \overline n_k(0)  $:
\begin{equation}
\Delta \overline n_k(t_n)  =   \sum_\mu  \left|c_\mu \right|^2 \left.\frac{d E_{\mu ,\boldsymbol\varphi}}{d\varphi_k}\right|_{\boldsymbol\varphi =\overline {\boldsymbol \varphi}}  t_n ,
\label{eq:meanPhotonNumberExpansion}
\end{equation}
which is a weighted average of the quasienergies derived with respect to the driving phase. Similar to  the dissipative case in Eq.~\eqref{eq:def:cumulantsLongTime}, the first cumulant increases linearly with time. Intriguingly, when the matter system is prepared in a specific Floquet state $c_{\mu'} =\delta_{\mu,\mu'}$, the second dynamical cumulant vanishes exactly in the long-time limit $\left. \Delta \sigma_k^2 \right|_{\mu} = 0$, i.e., the photonic variance remains constant as a function of time. Together with Eq.~\eqref{eq:meanPhotonNumberExpansion}, this effect leads to a light-matter entanglement in the Floquet basis
\begin{equation}
\left| \Psi (t)\right> = \sum_\mu c_\mu e^{-iE_{\mu,\overline{\boldsymbol \varphi}}t }\left| u_{\mu, \overline {\boldsymbol \varphi}}(t) \right> \otimes \left| A_\mu (t)\right>,
\label{eq:lightMatterEntanglement}
\end{equation}
where $\left| A_\mu (t)\right>$ denotes the photonic state, which is conditioned on the Floquet initial state. \textcolor{\markColorThree}{This effect is depicted for the semiclassical Jaynes-Cummings model in Fig.~\ref{figOverview}(d)}.

 This dynamics gives rise to a super-linear scaling behavior of higher-order  cumulants $\kappa_{\text{dy},l>1}^{(k)}(t_n)$. For instance, when initializing the system in a superposition of two Floquet states with $\left| c_1\right|^2 = \left| c_2\right|^2 =0.5 $, the variance of  photon mode $k$ (i.e., the second cumulant) increases as
\begin{equation}
\Delta \sigma_k^2 (t_n)  =   \left( \frac{d E_{2 ,\overline {\boldsymbol \varphi}}}{d\varphi_k}  -  \frac{d E_{1 ,\overline {\boldsymbol \varphi} }}{d\varphi_k} \right)^2 \; t_n^2. 
\label{eq:varianceBalancedSuperposition}
\end{equation}
Similarly, one can show that higher-order cumulants can diverge with $\kappa^{(k)}_l \propto  (t_n)^{l}$.
The generation of light-matter entanglement in closed quantum systems causing the super-linear increase of the cumulants is illustrated in Fig.~\ref{figOverview}(c). 

Upon comparing Eqs.~\eqref{eq:cumulantGenertingFunctionLongTimes} for open quantum systems and ~\eqref{eq:momentGenFkt_closedSystem} for closed quantum systems, we conjecture that the single eigenvalue $\lambda_{0;\boldsymbol \chi,\boldsymbol \xi}$  in Eq.~\eqref{eq:cumulantGenertingFunctionLongTimes} comprises the weighted information of all eigenvalues $E_{\mu,\boldsymbol \varphi}$ of the closed quantum systems in Eq.~\eqref{eq:momentGenFkt_closedSystem}. The weights are thereby determined by the stationary state, and depend thus on the nature of the dissipation. These considerations are valid in the weak dissipation limit [$\gamma_j\ll E_{\mu, \boldsymbol \varphi} $ for all $\gamma_j$ in Eq.~\eqref{sec:quantumMasterEquation}], while strong dissipation ($\gamma_j\gg E_{\mu, \boldsymbol \varphi} $) will have a significant impact on this Floquet-state interpretation. In order to simplify the notation, in the remaining of the paper we will denote the stroboscopic time $t_n$ by using $t$.

{
	
\color{\markColorTwo}	
\subsection{Error discussion}

 To discuss the error of the PRFT in open quantum systems, we consider a separable coherent photonic initial state in Eq.~\eqref{eq:coherentIntialState} such that $a_{\boldsymbol n }  =  \prod_{k=1}^{N_D} a_{n_k}$. The variance of the initial photonic probability distribution is denoted by $\sigma^2$. The coefficients $a_{n_k}$ shall be Gaussian, i.e., $a_{n_k}  = \exp \left[ i\overline \varphi_k n_k - (n_k-\overline n_k)^2/(4\sigma^2) \right]/[\sigma^{1/2} (2\pi)^{1/4}]$ with mean value $\overline n_k$. The semiclassical limit is thereby defined by $\sigma\rightarrow \infty$~\cite{Engelhardt2024c}. The error analysis of the PRFT in open quantum system is carried out in Appendix~\ref{eq:errorAnalysis}. Namely, the exact cumulants $\kappa_{l}^{(\text{exact},k)}$ are related to the cumulants predicted by the PRFT $ \kappa_{l}^{(k)}$ via
 \begin{equation}
\kappa_{l}^{(\text{exact},k)} (t)  =  \kappa_{l}^{(k)}(t) +  f_{l}^{(k)}  t^{l}/\sigma^2
\label{eq:error}
\end{equation}
for long times $t$. Thereby, the error coefficients  $f_{l}^{(k)} $ are approximately constant in time, i.e., $f_{l}^{(k)} =\mathcal O (t^0)$. It follows from this relation that for a  constant time, the error vanishes quadratically with $\sigma$, which diverges in the semiclassical limit $\sigma\rightarrow \infty$. 

Exactly as in the closed system, one can show that the error terms $\propto f_{l}^{(k)} $ have a vanishing impact on the probability distribution for times of the order $t\propto \sigma$~\cite{Engelhardt2024c}. For the same reason, actually all cumulants  with $l\geq 2$ have a vanishing influence on the probability distribution in the semiclassical limit $\sigma\rightarrow \infty$ for  $t\propto \sigma$ because they scale only linearly in time $\kappa_{l}^{(k)}(t) = \kappa_{l,1}^{(k)} t   $  with constants $\kappa_{l,1}^{(k)} $ according to Eq.~\eqref{eq:def:cumulantsLongTime}~\cite{Engelhardt2024c}. Consequently  the probability distribution is solely determined by the first cumulant $\kappa_{1}^{(k)}(t)$ for $t\propto \sigma$. Nevertheless, it is worthwhile to investigate the higher-order cumulants in the long-time limit: 

(i) While the semiclassical cumulants $ \kappa_{l}^{(k)}(t)$ scale only linearly in open quantum systems, the coefficients $\kappa_{l,1}^{(k)} $ can be very large and even diverge in the small dissipation regime.  In particular, the light-matter entanglement effect introduced in Sec.~\ref{sec:photonResolvedFloquetTheory} will reveal itself in a diverging  $\kappa_{2,1}^{(k)}$. We will investigate  this situation in Sec.~\ref{sec:jcModel}.

(ii) The analysis of the  cumulants $\kappa_{l}^{(k)}(t)   $  will show that  the standard FCS, which is an established method, is consistent with the PRFT for arbitrary long times on the semiclassical level. Thus, we establish a consistent framework unifying both  methods, which advances the basic understanding of counting-statistics approaches in  quantum optics. 

(iii) The analysis of the $\kappa_{l\geq 2}^{(k)}(t) $  will  be the basis for more sophisticated investigations beyond the semiclassical limit with finite $\sigma<\infty$: The derivations in this paper are carried out in the Sambe space, for which the photonic creation and anhiliation operators are replaced by their translationally invariant counterparts, e.g,. $\hat a_k^\dagger \rightarrow \sum_n \alpha_k \left| n+1\right> \left< n \right|$ [see Eq.~\eqref{eq:def:sambeSpace}]. For coherent photonic states, this approximation is valid as long as $\overline n(t)  - \overline n(0)  \ll \sqrt{\overline n(0)} = \sigma $. Given the linear scaling of $\overline n(t)$, we find that the PRFT  breaks down at time scales of $t\gg \sqrt{\overline n} =\sigma $. On these time scales, a generalized PRFT, which takes the photon number dependence of  the creation (and annihilation) operators $ \left< n+1 \right| \hat a^\dagger\left|n \right> =\sqrt{n}$ into account, will be required.
The superlinearly scaling error in Eq.~\eqref{eq:error} is a consequence of the translational invariance of the approximate creation operators in Sambe space, which results in a ballistic expansion of the photonic wave function in the photon number space. For longer time scales $t\gg \sigma$,  the breakdown of the translational invariance will lead to a diffusive-like expansion, which will result in  a linear error scaling instead of a superlinear one. Thus, we anticipate that the semiclassical cumulants $\kappa_{l\geq 2}^{(k)}(t) $ will provide a dominant contribution for very long times $t\propto \sigma^2$ under these circumstances.

}

\section{Jaynes-Cummings model}

\label{sec:jcModel}

To illustrate the generic framework, we  consider an open Jaynes-Cummings model.   Here, we investigate the signatures of the light-matter entanglement predicted by the PRFT for closed quantum systems as explained in Sec.~\ref{sec:photonResolvedFloquetTheory}  in the presence of dissipation.

\subsection{Hamiltonian}

\label{sec:Hamiltonian}

\begin{figure}
	\includegraphics[width=\linewidth]{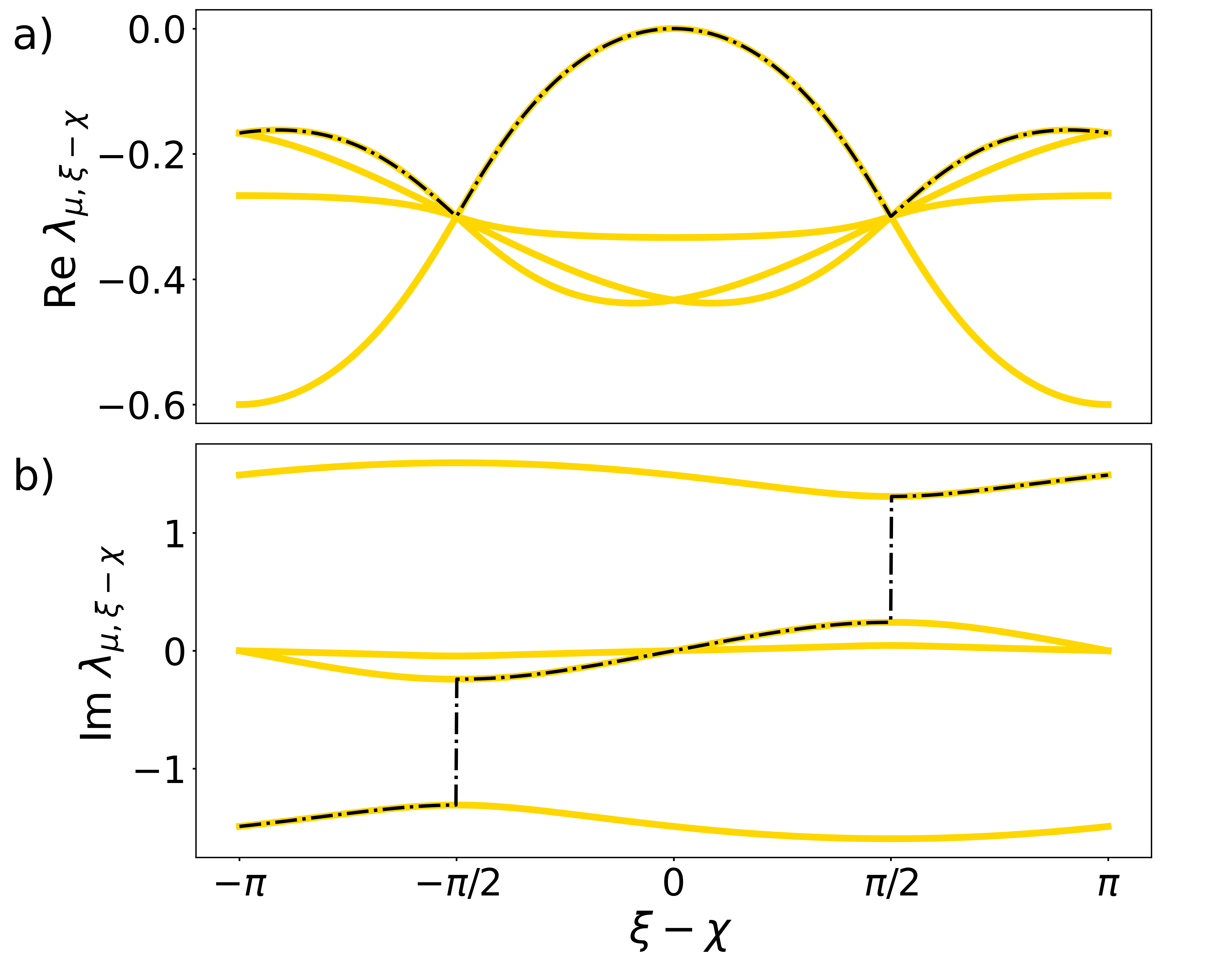}
	\caption{\textcolor{\markColorTwo}{ Real and imaginary parts of the eigenvalues of the Liouvillian in Eq.~\eqref{eq:effectLiouvillianJCmodel} are depicted in panels (a) and (b), respectively. Parameters are $\Omega_2 =\Omega_1$, $\epsilon_{\Delta}=0.5\Omega_1$, $\gamma = 0.6\Omega_1$, and we have set $\chi_1=\chi_2$.}  }
	\label{figEigenvalueAnalysi}
\end{figure}

The system is sketched in Fig.~\ref{figOverview}(a) and consists of a two-level system describing atoms, molecules or other quantum systems, that is driven by $N_{\text{D}} =2 $ coherent fields  with frequencies $\omega_k$ for $k=1,2$. Moreover, the two-level system is coupled to a continuum of photon modes $k=3,\dots N_{\text{B}}+2$ giving rise to dissipation.   The Hamiltonian reads 
\begin{eqnarray}
\hat H_{\text{JC}}  = \frac{ \epsilon }{2} \hat \sigma_{\text{z}}   +    \sum_{k=1}^{2+ N_\text{B} }  g_k  \left( \hat \sigma_{-}  \hat a_k^\dagger   +\hat \sigma_{+}  \hat a_k \right)+ \sum_{k=1}^{2+ N_\text{B}} \omega_k \hat a_k^\dagger  \hat a_k  ,	\nonumber \\
\label{eq:hamiltonian:JaynesCummings}
\end{eqnarray}
where the Pauli operators $\hat \sigma_{\text{z}} , \hat\sigma_+ ,\hat \sigma_-$ quantize the two-level system with level splitting  $\epsilon$. This model can be also realized with superconducting qubits by considering  two microwave cavities both coupled to a nonlinear transmon ancilla~\cite{Zhang2019}. In this implementation, the $N_{\text{B}}$ modes can be realized by using a multimode transmission line~\cite{Gu2017}.
The coherent photon modes  $k=1,2$ have equal frequencies $\omega_1 =\omega_2 =\omega$.  The semiclassical Hamiltonian is given as
\begin{equation}
\hat{\mathcal H}_{\boldsymbol   \varphi } (t) =  \frac{ \epsilon}{2}\hat \sigma_{\text{z}}  +  \frac {\Omega_{\text{x}, \boldsymbol  \varphi}(t)}{2} \hat \sigma_{\text{x}}   + \frac {\Omega_{\text{y}, \boldsymbol  \varphi }(t)}{2}  \hat \sigma_{\text{y}}   ,
\label{eq:semiclassicalJaymesCummingsModel}
\end{equation}
where the effective Rabi frequencies read
\begin{eqnarray}
\Omega_{\text{x}, \boldsymbol  \varphi}(t) &=& \sum_{k=1,2} \Omega_k \cos(\omega t-\varphi_k ), \nonumber \\
\Omega_{\text{y}, \boldsymbol  \varphi}(t) &=& \sum_{k=1,2} \Omega_k  \sin(\omega t -\varphi_k).
\end{eqnarray}
  The phases of the coherent states of $\hat a_k$ are denoted by  $\varphi_k$, and $\Omega_k =2  g_k \alpha_k$ denote the corresponding Rabi frequencies.  It is easy to see that the total excitation number $\hat n_{\text{ex} } = \hat \sigma_{\text{z}}+\sum_{k=1}^{2+ N_\text{B}}  \hat a_k^\dagger  \hat a_k$ is an integral of motion in the multi-mode Jaynes-Cummings model in Eq.~\eqref{eq:hamiltonian:JaynesCummings}, which we deploy  as a sanity check to verify the consistency of the FCS and the PRFT. Models, which do not exhibit  excitation number conservation will be addressed in future work investigating the consistency of the PRFT with energy conservation.

 \begin{figure*}
	\includegraphics[width=\linewidth]{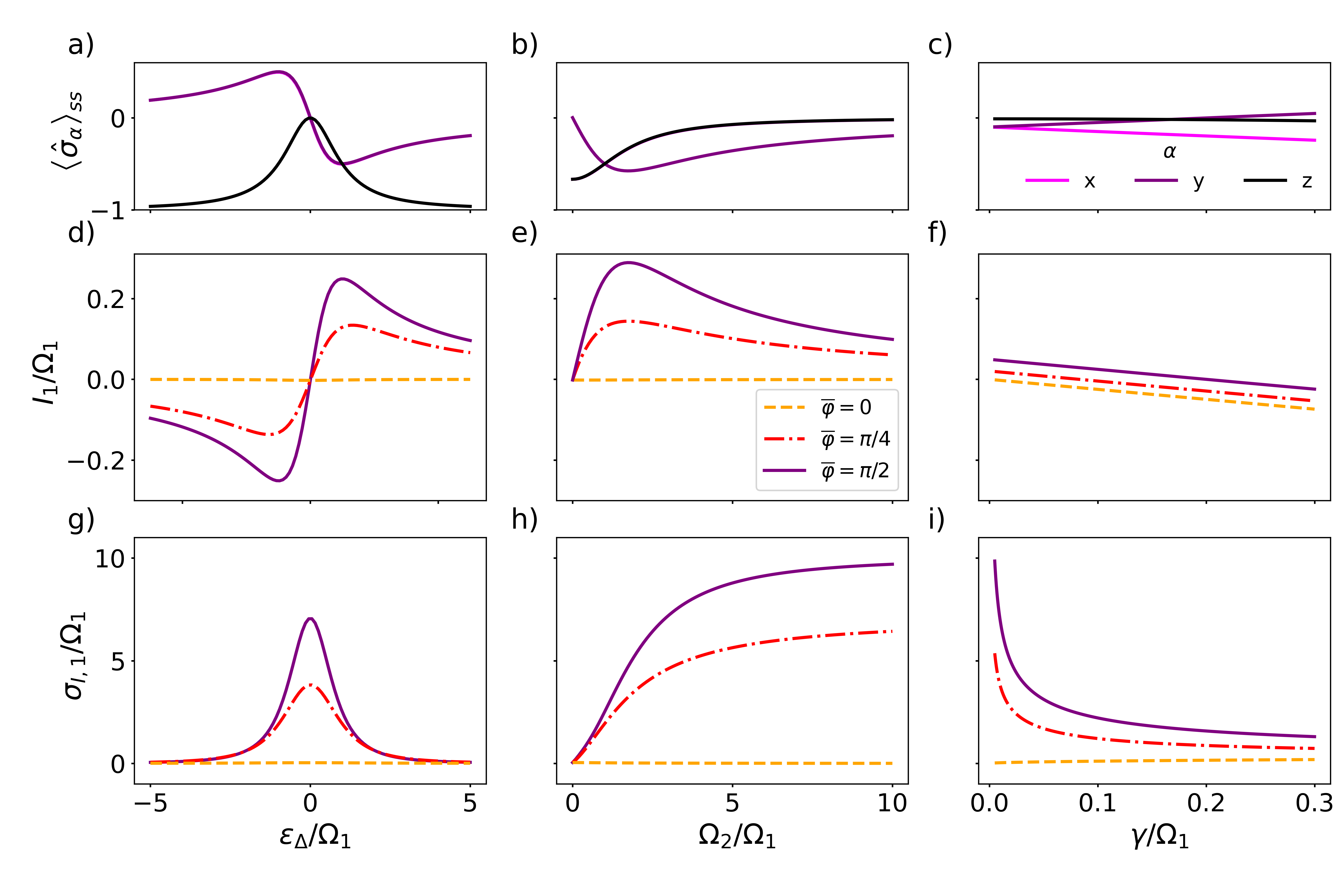}
	\caption{Analysis of the stationary state of the dissipative Jaynes-Cummings model. Expectation values of the spin-operators of the two-level system are depicted in (a), (b), and (c). The photon flux $I_1$ into mode $k=1$ is shown in (d), (e), and (f). The corresponding fluctuations $\sigma_{I,1}$  are depicted in (g), (h), and (i). (a), (d), (g) depict the observables as  a function of detuning $\epsilon_{\Delta}$ for $\Omega_2 =\Omega_1$ and $\gamma=0.002\Omega_1$, (b), (e), (h) as a function of Rabi frequency $\Omega_2$ for  $\epsilon_{\Delta}= 0.2 \Omega_1$ and $\gamma=0.002\Omega_1$, and (c), (f), (i) as a function of bath coupling $\gamma$ for $\epsilon_{\Delta}= 0.05 \Omega_1$, $\Omega_2 =\Omega_1$. The phases are $\overline \varphi_1 =0$ and $\overline \varphi_2 =\overline \varphi$ as specified in the plot legend.}
 	\label{fig:StationaryStateAnalysis}
\end{figure*}

As explained in Sec.~\ref{sec:photonResolvedFloquetTheory}, the photonic dynamics in the closed system is the basis to understand the  open quantum system in the subsequent sections. The quasienergies of the semiclassical Hamiltonian in Eq.~\eqref{eq:semiclassicalJaymesCummingsModel} are given by
$
 E_{\mu,\boldsymbol\varphi}  = \frac{(-1)^{\mu}}2\sqrt{ \epsilon_{\Delta}^2+ \Omega_1^2 + \Omega_2^2 + 2 \Omega_1\Omega_2 \cos(\varphi_2-\varphi_1) },
$
where $\mu=1,2$ and $\epsilon_{\Delta} =\epsilon -\omega$ denotes the detuning. Consequently, the Floquet-state-depended photon number change  in mode $k=1$ during time $t$ reads as
\begin{equation}
 \Delta\overline n_{1}^{ (\mu)}  = \frac {(-1)^{\mu}}{2}\frac{\Omega_1\Omega_2 \sin\varphi }{ \sqrt{ \epsilon_{\Delta}^2+ \Omega_1^2 + \Omega_2^2 + 2 \Omega_1 \Omega_2 \cos(\varphi) } }t ,
 \label{eq:photonNumberClosedJCmodel}
\end{equation}
where $\varphi = \varphi_2 -\varphi_1$ is the phase difference of the driving fields.  The dynamics in Fock space and the light-matter entanglement generation is qualitatively shown in Fig.~\ref{figOverview}(b).

\subsection{Bloch equations}

\label{sec:blochEquations}

Transforming the Hamiltonian in Eq.~\eqref{eq:hamiltonian:JaynesCummings} into a frame rotating with $\hat U_{\text{rot}} =\exp(-\omega\hat\sigma_{\text{z}} t/2)$, we can derive a quantum master equation for the  generalized density matrix of the system
 \begin{eqnarray}
 \frac{d}{dt}   \rho_{\boldsymbol  \chi,\boldsymbol \xi}  &=&  -i \left[ \hat{\mathcal H}_{ \overline {\boldsymbol \varphi} + \frac{\boldsymbol \chi  }{2}  }  \rho_{\boldsymbol  \chi,\boldsymbol \xi  }  -   \rho_{\boldsymbol  \chi,\boldsymbol \xi} \hat{\mathcal H}_{ \overline {\boldsymbol \varphi} - \frac{\boldsymbol \chi  }{2}} (t)   \right] \nonumber \\
&&+  \gamma D_{\boldsymbol \xi } \left[ \hat \sigma_- \right]\rho_{\boldsymbol  \chi,\boldsymbol \xi} , 
 \label{sec:quantumMasterEquationJC}
 \end{eqnarray}
  with the semiclassical Hamiltonian 
 \begin{equation}
 \hat {\mathcal H}_{\boldsymbol \varphi}   = \frac{ \epsilon_\Delta}{2}\hat \sigma_{\text{z}}  +  \frac {\Omega_{\text{x}, \boldsymbol  \varphi}(0)}{2} \hat \sigma_{\text{x}}   + \frac {\Omega_{\text{y}, \boldsymbol  \varphi}(0)}{2}  \hat \sigma_{\text{y}}   \nonumber ,
 \label{eq:hamiltonian:semiclassicalJC}
 \end{equation}
  and the dissipator
\begin{equation}
\mathcal D_{\xi } \left[\hat \sigma_- \right]\rho  = e^{-i\xi } \hat  \sigma_-  \rho \hat  \sigma_+   -\frac{1}{2} \hat  \sigma_+   \hat \sigma_- \rho   - \frac{1}{2}  \rho \hat  \sigma_+   \hat \sigma_-.
\label{eq:dissipatorJCmodel}
\end{equation}
\textcolor{\markColorOne}{Thereby, we have set the counting fields of the incoherent modes equal ($\xi_k =\xi$) such that $\xi$ counts the total number of all  incoherently emitted photons according to Eq.~\eqref{eq:def:momentGeneratingFkt_exact}.
}

As the master equation in Eq.~\eqref{sec:quantumMasterEquationJC}  is time-independent in the rotating frame, the effective Liouvillian is given by (see Appendix \ref{app:dissipative_JcModel:EOM} for details)
\begin{equation}
\mathcal A_{ \xi,\boldsymbol \chi} = \left( 
\begin{array}{cccc} 
 \gamma_{\xi,-}   & ic_{\boldsymbol  \chi,-} &  is_{\boldsymbol  \chi,-}  &   \gamma_{\xi,-} \\
ic_{\boldsymbol  \chi,-} &   - \frac{\gamma}{2} & - \epsilon_{\Delta}  & s_{\boldsymbol  \chi,+} \\
is_{\boldsymbol  \chi,-} & \epsilon_{\Delta}  & - \frac{\gamma}{2} & -c_{\boldsymbol  \chi,+} \\
- \gamma_{\xi,+}   & -s_{\boldsymbol  \chi,+} & c_{\boldsymbol  \chi,+} &- \gamma_{\xi,+} 
\end{array}
\right),
\label{eq:effectLiouvillianJCmodel}
\end{equation}
where the counting-field dependent coefficients read
\begin{eqnarray}
c_{\boldsymbol  \chi,\pm} &=&  \frac{1}{2} \left[ \Omega_{\text{x},\overline {\boldsymbol \varphi} - \frac{\boldsymbol \chi  }{2}}(0) \pm  \Omega_{\text{x}, \overline {\boldsymbol \varphi} + \frac{\boldsymbol \chi  }{2}}(0)\right] , \nonumber \\
s_{\boldsymbol  \chi,\pm} &=& \frac{1}{2} \left[  \Omega_{\text{y}, \overline {\boldsymbol \varphi} - \frac{\boldsymbol \chi  }{2}}(0) \pm  \Omega_{\text{y}, \overline {\boldsymbol \varphi} + \frac{\boldsymbol \chi  }{2}}(0) \right] , \nonumber \\
\gamma_{\xi,\pm}   &=&  \frac{\gamma}{2} \left( e^{-i\xi } \pm 1\right), \nonumber
\end{eqnarray}
with $\gamma$ parameterizing the coupling to the incoherent photon modes.
The Liouvillian in Eq.~\eqref{eq:effectLiouvillianJCmodel} describes the dynamics for the vector of density matrix elements $(\rho_{\text{0}},\rho_{\text{x}},\rho_{\text{y}},\rho_{\text{z}}   )$, where $\rho_\alpha = \text{tr} \left[ \rho_{ \boldsymbol \chi,\xi}  \hat \sigma_\alpha\right]$.

 At this point, it is  not obvious that both counting concepts, namely the FCS and the PRFT,  are compatible.  To demonstrate the consistency,
 we show that the total number of excitations $\hat n_{\text{ex}  }$ is conserved within the formalism. We distribute the total excitation number as  $\hat n_{\text{ex}  } =\hat \sigma_{\text{z}} +  \hat n_{\text{D}} +\hat n_{\text{B}  } $, where $\hat n_{\text{D}} =\hat n_{1} + \hat n_{2}$ denotes the total number of photons in the driving modes, and $ \hat n_{\text{B}  }$ denotes the number of photons in the incoherent  modes.

\textcolor{\markColorOne}{According to the definition of the moment-generating function in Eq.~\eqref{eq:def:momentGeneratingFkt_exact}, we obtain the statistics of $\hat n_{\text{D}}$ by setting the counting fields $\chi_1 = \chi_2 =\chi$ in the effective Liouvillian of Eq.~\eqref{eq:effectLiouvillianJCmodel}}.  The  statistics of $\hat n_{\text{B}}$ is associated to the counting field $\xi$. In this case, the characteristic polynomial of the effective Liouvillian in Eq.~\eqref{eq:effectLiouvillianJCmodel}  becomes
\begin{eqnarray}
\mathcal P_{\chi,\xi} (z)
&=& z^4 +  2\gamma z^3 + \left( \epsilon_{\Delta}^2 + \Omega_{\overline {\boldsymbol \varphi}}^2 + \frac{5}{4} \gamma^2  \right) z^2 \nonumber\\
&+& \gamma\left[  \epsilon_{\Delta}^2  +  \Omega_{\overline {\boldsymbol \varphi}}^2  \left( 1-\frac{1}{2} e^{ i\left( \xi  -\chi \right) } \right)  +\frac{\gamma^2}{4}  \right]  z \nonumber\\
&+& \frac{1}{4} \Omega_{\overline {\boldsymbol \varphi}}^2\gamma^2 \left[1- e^{ i\left( \xi  -\chi\right) }  \right],
\label{eq:simplifiedCharPolJCmodel}
\end{eqnarray}
where $\Omega_{\overline {\boldsymbol \varphi}}^2 = \Omega_{1}^2  +\Omega_{2}^2  + 2\Omega_{1}\Omega_{2} \cos (\overline \varphi_1-\overline \varphi_2)   $. Interestingly, the polynomial depends only on the difference of the counting fields $\xi$ and $\chi$, such that all eigenvalues have a functional dependence
$ \lambda_{\mu;\xi,\chi} =  \tilde{\lambda}_{\mu}(\xi -  \chi) $. From Eq.~\eqref{eq:simplifiedCharPolJCmodel} and Eq.~\eqref{eq:def:cumulantsLongTime} one can thus conclude that the cumulants of $\hat n_{\text{D}}$, $\kappa_{l}^{(\text{D})}$, and the cumulants of  $\hat n_{\text{B}}$,  $\kappa_{l}^{(\text{B})}$, fulfill the condition $\kappa_{l}^{(\text{D})} = (-1)^{l}\kappa_{l}^{(\text{B})}$, and are thus statistically consistent in the long-time limit. \textcolor{\markColorTwo}{ While a numerical benchmarking of the PRFT in open quantum systems is infeasible because of the large number of basis states of the coherent modes,  the relation $ \lambda_{\mu;\chi , \xi} =  \tilde{\lambda}_{\mu}(\xi -  \chi) $  serves as an analytical proof of the validity of the PRFT, since the FCS is an established theoretical framework.  }  Minor deviations related to the occupation of the two-level system $\hat \sigma_{\text{z}}$, are described by $\mathcal B_{\boldsymbol \chi, \boldsymbol \xi} (t )$ [introduced in Eq.~\eqref{eq:generalFloquetSolution}] rather than the eigenvalue $ \lambda_{\mu=0;\xi,\chi} $, which is responsible for  the large-scale photon flux.

\textcolor{\markColorTwo}{For illustration, we depict the eigenvalues of the Liouvillian in Fig.~\ref{figEigenvalueAnalysi}. We observe that the real parts of all eigenvalues are negative for $\chi\neq 0$, as required by the FCS and the PRFT (see Appendix~\ref{sec:upperBoundOfTrace}). As $\mathcal P_{ \chi ,\xi} (z) =\left[ \mathcal  P_{-\chi ,-\xi} (z)\right]^* $, we also find that the real (imaginary) parts  of the eigenvalues, are symmetric (antisymmetric) with respect to inversion at the origin.  At $\left| \xi-\chi\right|=\pi/2$ the real parts intersect, such that the eigenvalue with the largest real part $\lambda_{0,\chi, \xi }$ is not a continuous function of the counting field. Yet, we note that the asymptotic behavior is determined for $\left| \xi-\chi\right|\approx 0$, for which the eigenvalue with the largest real part is uniquely determined.   }

 \subsection{Photon flux and fluctuations}

Here, we investigate the photon flux into driving mode $k=1$ in the stationary state. This setup can be interpreted as a nonlinear pump-probe experiment, where  mode $k=2$ is the pump field and $k=1$ is the probe field. 
Following Sec.~\ref{sec:longTimeLimit} and Appendix~\ref{app:dissipative_JcModel}, we can analytically determine an expression for the photon flux.  The photon flux $I_1$  defined in Eq.~\eqref{eq:def:photonFluxAndVar} reads as
\begin{eqnarray}
 I_1 = \frac{\Omega_{1} \left[ \epsilon_{\Delta} \Omega_{2} \sin{\left(\overline \varphi \right)} -  \frac{\gamma}{2} \Omega_{1} - \frac{ \gamma}{2} \Omega_{2} \cos{\left(\overline \varphi \right)}\right]}{2 \epsilon_{\Delta}^{2} +  \Omega_{1}^{2}  +  \Omega_{2}^{2}  +  2\Omega_{1} \Omega_{2} \cos\left(\overline \varphi \right)  + \frac{ \gamma^{2}}{2}  } ,\nonumber\\
\label{eq:photonFlux}
\end{eqnarray}
where $\overline \varphi = \overline \varphi_2-\overline \varphi_1$.
 We emphasize that the photon flux in Eq.~\eqref{eq:photonFlux} is  non-perturbative in the driving amplitudes $\Omega_1$ and $\Omega_2$. This result can also be  obtained using the semiclassical formula in Eq.~\eqref{eq:semiclassicalCurrent}, which explicitly reads   $I_{1} = i\Omega_1\left< \hat \sigma_+e^{i\overline \varphi_1} -\hat \sigma_-e^{-i\overline \varphi_1}\right>_{\text{ss}}/2 $ for the Jaynes-Cummings model. Thereby, the expectation value is calculated in the stationary state of the two-level system at stroposcopic times $t=n\tau$, i.e., the eigenstate of $\mathcal A_{\boldsymbol \chi= \boldsymbol 0,\boldsymbol \xi= 0}$ with eigenvalue $ \lambda_{ \mu=0;  \boldsymbol \chi= \boldsymbol 0, \xi=0}=0 $.
 
 The mean photon flux in Eq.~\eqref{eq:photonFlux} has  two contributions: (i) a contribution $\propto \epsilon_{\Delta}  \sin\left(\overline \varphi \right)$ that describes the coherent pumping of photons between the modes $k=1$ and $k=2$, and can be thus interpreted as a  Josephson flux since it driven by the phase difference of the two photonic modes, akin to the Josephson effect in superconducting junctions~\cite{Golubov2004}; (ii) a dissipative flux $\propto \gamma\left[ \Omega_{1} - \Omega_{2} \cos{\left(\overline \varphi \right)}\right] $ resulting from the absorption of photons by the matter system. Noteworthy, for vanishing $\Omega_2$ and small $\Omega_1$, the flux becomes $I_1 = - \Omega_1^2\gamma/4 /(\epsilon_\Delta^2 + \gamma^2/4)$, which thus correctly resembles  the absorption in  first-order perturbation theory in $\Omega_1$ known from linear spectroscopy. Moreover, if $ \Omega_{1} = -\Omega_{2} \cos{\left(\overline \varphi \right)} $, the dissipative flux vanishes as both driving fields destructively interfere and thus decouple from the two-level system. In this case, the remaining Josephson flux is exclusively carried by virtual excitations.

In a similar fashion, we can  derive an expression for the  photon flux variance $\sigma_{I,1}^2$  defined  in Eq.~\eqref{eq:def:photonFluxAndVar}. The full expression is very complicated and is given in Eq.~\eqref{eq:fluxNoise:twolevelsystem:exact}. For a weak dissipation $\gamma$, the expression simplifies  and reads as 
\begin{eqnarray}
\sigma_{I,1}^2 = \frac{  \Omega_{1}^{2} \Omega_{2}^{2} \left[ \Omega_{1}^{2}+  \Omega_{2}^{2} + 2 \Omega_{1} \Omega_{2} \cos{\left(\overline \varphi \right)} \right]^{2} \sin^{2}{\left(\overline \varphi \right)}}{\gamma \left[2 \epsilon_{\Delta}^{2} +  \Omega_{1}^{2} +  \Omega_{2}^{2} +2  \Omega_{1} \Omega_{2} \cos{\left(\overline \varphi \right)}\right]^{3}}\nonumber ,\\
\label{eq:photonFluctuations}
\end{eqnarray}
which is nonperturbative in all parameters except of $\gamma$.
We emphasize that  standard methods of  nonlinear spectroscopy are incapable to make non-perturbative predictions about the fluctuations. 

The photon flux and its fluctuations are analyzed in Fig.~\ref{fig:StationaryStateAnalysis} as a function of $\epsilon_{\Delta}$, $\Omega_2$ and $\gamma$.  Each panel displays the results for three phase differences $\overline \varphi= \overline \varphi_2 - \overline \varphi_1=  0,\pi/4, \pi/2$ for comparison, in such that $\overline \varphi_1 =0$ and $\overline \varphi_2 =\overline \varphi$. In the first row of Fig.~\ref{fig:StationaryStateAnalysis}, we also depict the expectation values of the spin operators $\left<\hat \sigma_{\alpha}\right>_{\text{ss}}$ in the stationary state of the system. As explained above, the mean photon flux for $\varphi_1=0$ can be expressed as $I_1 =-\Omega_1 \left<\hat \sigma_{\text{y}}\right>_{\text{ss}}/2$. This correspondence  can be clearly seen when comparing Fig.~\ref{fig:StationaryStateAnalysis}(a)-(c) with Fig.~\ref{fig:StationaryStateAnalysis}(d)-(f).

\textit{Detuning.} In Fig.~\ref{fig:StationaryStateAnalysis}(d) and Fig.~\ref{fig:StationaryStateAnalysis}(g), we depict the first two cumulants as a function of detuning $\epsilon_{\Delta}$  for a weak dissipation $\gamma$ and equal driving amplitudes $\Omega_1 = \Omega_2$. 
According to Eq.~\eqref{eq:photonFlux} for $\gamma\rightarrow 0$, the current vanishes for $\epsilon_{\Delta} =0$, while $I_1<0$ ($I_1>0$) for $\epsilon_{\Delta}<0$ ($\epsilon_{\Delta}>0$).  
For a vanishing detuning $\epsilon_{\Delta}=0$, the fluctuations significantly increase. The effect  can be explained  by considering the photon flux in the closed quantum system summarized in Sec.~\ref{sec:Hamiltonian}: We recall that the flux is positive or negative depending on the Floquet state $\mu$. As shown in Appendix~\ref{sec:stationaryStateJCmodel}, for $\epsilon_{\Delta} =0$ the probabilities to be in either Floquet state $\mu=1,2$ are equal, such that the average flux vanishes $I_1=0$. Yet, the uncertainty to be in either Floquet state  is reflected in the enhanced photon flux fluctuations. A more quantitative explanation of this effect will be given in Sec.~\ref{sec:divergingFluctuations}.

 For a finite detuning,  the photon flux depends on the sign of $\epsilon_{\Delta}$ that determines in which Floquet state the system is  with higher probability. With increasing detuning, the spin system is not  resonantly excited such that the flux exhibits a turnover and vanishes with $\propto \epsilon_{\Delta}^{-1}$ for large detuning.  Likewise,  the fluctuations decrease with increasing detuning.

\textit{Pump amplitude.} The dependence of the pump amplitude $\Omega_2$ is depicted in Fig.~\ref{fig:StationaryStateAnalysis}(e). For small $\Omega_2$, the flux $I_1$ increases linearly, showing that more photons get transferred from mode $k=2$  (pump)   to mode $k=1$  (probe). Intriguingly, the flux does not increase monotonically but exhibits a turnover as a function of $\Omega_2$. 

The turnover in Fig.~\ref{fig:StationaryStateAnalysis}(e) can be understood intuitively by resorting to the relation $I_1 =-\Omega_1 \left<\hat \sigma_{\text{y}} \right>_{\text{ss}}$.  For small $\Omega_2$,  $\left<\hat \sigma_{\text{y}} \right>_{\text{ss}}$ increases because  the $\Omega_2 \hat \sigma_{\text{y}}$ term in the Hamiltonian  polarizes the stationary state in $\hat \sigma_{\text{y}}$ direction. For very large $\Omega_2$, where the closed system would be diagonal in the $\hat \sigma_{\text{y}}$ basis, the dissipation term induces random switching between the two  eigenstates of $\hat \sigma_{\text{y}}$ (i.e., the Floquet states), such that  $\left<\hat \sigma_{\text{y}} \right>_{\text{ss}}\rightarrow 0$ for increasing $\Omega_2$. However, because of the random switching between the two Floquet states, the fluctuations $\sigma_{I,1}$ increase monotonically with $\Omega_2$ as can be seen in Fig.~\ref{fig:StationaryStateAnalysis}(h).

 \begin{figure*}
 	\includegraphics[width=\linewidth]{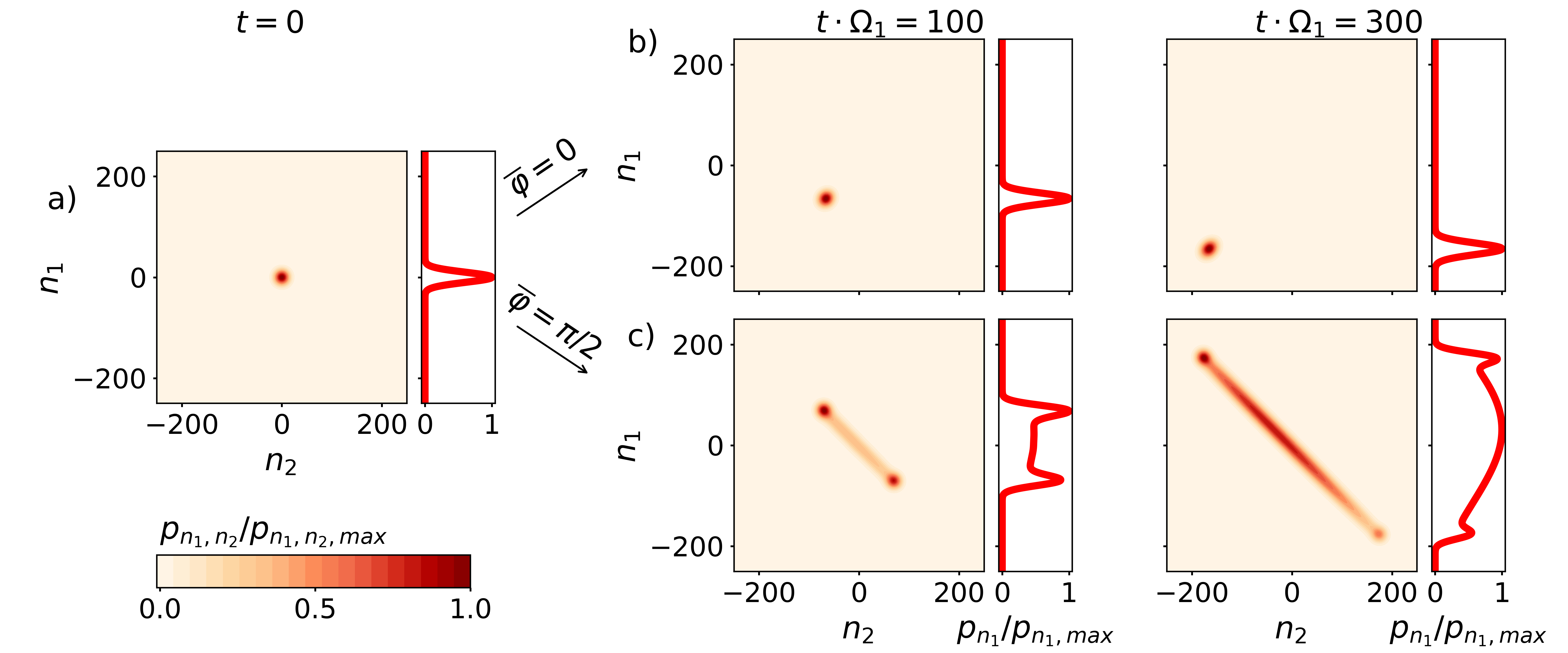}
 	\caption{Photonic probability distribution of modes $k=1$ and $k=2$  for three times. (a) depicts the initial distribution which is a Gaussian with variance $\sigma_k^2 =100$. (b) shows the time-evolved distributions for a phase difference $\overline \varphi=0$. To enhance the effect, we have chosen a large dissipation $\gamma=2\Omega_1$. (c) shows the joint distribution for $\overline \varphi=\pi/2$ and $\gamma = 0.02 \Omega_1$. Overall parameters are $\epsilon_{\Delta} = 0.1 \Omega_1$, and $\Omega_2 =\Omega_1$.}
 	\label{fig:ProbabilityDistribution}
 \end{figure*}

\textit{Dissipation.}  Figure~\ref{fig:StationaryStateAnalysis}(c) depicts the photon flux as a function of dissipation rate $\gamma$.  We observe that the flux is positive for small $\gamma$, where photons are coherently transferred from mode $k=2$ to mode $k=1$, while only a tiny fraction is emitted into the bath.  For  a sufficiently large $\gamma$, the incoherent absorption exceeds the coherent photon pumping and the overall photon flux becomes negative.

Intriguingly, for a vanishing dissipation rate $\gamma$, the fluctuations diverge. As explained in Sec.~\ref{sec:photonResolvedFloquetTheory}, the variance can scale  with $\propto t^2$ in closed quantum systems. In contrast, the framework in open quantum systems generally predicts a scaling $\propto t$. The difference in scaling is thus reflected by a divergence of $\sigma_I^2$ when approaching the closed quantum system by taking the limit $\gamma\rightarrow 0$.
As discussed in the next section, this divergence is related to   the light-matter entanglement in closed quantum systems.

\textit{Driving phase.}  The panels of Fig.~\ref{fig:StationaryStateAnalysis} analyze the dependence of the photon flux and its fluctuations on the driving phase difference $\overline \varphi$. Both the photon flux in Eq.~\eqref{eq:photonFlux} and fluctuations in Eq.~\eqref{eq:photonFluctuations} predict a strong effect for $\overline \varphi=\pi/2$ in the weak coupling regime. For  $\overline \varphi=0$,   flux and  fluctuations are diminished as  both driving fields are essentially equivalent such that the transport between modes $k=1$ and $k=2$ cancel each other. In this case, the photon number can only change by emission into the reservoir. We also note that the transport between the modes and the reservoir vanishes completely for $\overline \varphi =\pi$, where both driving modes interfere destructively (not shown).

\subsection{Diverging fluctuations}

\label{sec:divergingFluctuations}

The diverging  fluctuations observed for small $\gamma$  in Fig.~\ref{fig:StationaryStateAnalysis}(f) can be explained based on a Floquet-state analysis.  For simplicity we consider here the resonant system $\epsilon_{\Delta} = 0$. Analysis of the effective Liouvillian in Eq.~\eqref{eq:effectLiouvillianJCmodel} reveals that the system becomes fully mixed in the long-time limit
\textcolor{\markColorOne}{
\begin{equation}
	\rho_{\text{M}}(t\rightarrow \infty) = \frac 12 \mathbbm 1 = \frac 12 \sum_\mu  \left| u_{\mu, \overline {\boldsymbol \varphi}} \right> \left< u_{\mu, \overline {\boldsymbol \varphi}}\right|,
	\label{eq:stationaryState_jcModel_resonantSystem}
\end{equation}
}
i.e.,
an incoherent mixture of both Floquet states (see Appendix~\ref{sec:stationaryStateJCmodel} for a detailed derivation). Because of the dissipation, the system state randomly switches between two Floquet states with rate $\gamma/4$ (the factor $1/4$ appears when representing the dissipator in Eq.~\eqref{eq:dissipatorJCmodel} in the Floquet basis). We thus divide the time dynamics into intervals of length $\Delta t = 4/\gamma$, i.e.,   $t_i \in \left[ i \Delta t, (i+1 ) \Delta t \right)$. In each interval, the Floquet state  can be considered as a random variable $\mu(t_i) =1,2$. According to the analysis of the  closed system, the photon mode can accumulate $\Delta n_{1}^{(\mu)} $ photons in each interval. The photon number $\Delta n_{1}^{(\mu)} $, which is explicitly given in Eq.~\eqref{eq:photonNumberClosedJCmodel} for $t=\Delta t$ , can be regarded as a binomial random variable for each interval, with mean
\begin{equation}
	\overline { \Delta n_{1}^{(\mu)}} = 0 
\end{equation}
and variance
\begin{equation}
\text{Var} \left[ \Delta n_{1}^{(\mu)} \right]  = \frac 14\frac{\Omega_1^2 \Omega_2^2 \sin^2 \overline\varphi }{ \Omega_1^2 + \Omega_2^2 + 2 \Omega_1\Omega_2 \cos(\overline \varphi)  }  \Delta t^2.
\end{equation}
This expression agrees with the photon flux fluctuation in Eq.~\eqref{eq:photonFluctuations} for $\epsilon_{\Delta}=0$ within each time-interval  $\Delta t = 4/\gamma$. Thus, for decreasing $\gamma$, the system can evolve coherently for a longer time until the Floquet state randomly switches, such that fluctuations can add up quadratically within $\Delta t$. This analysis also demonstrates that the fluctuations are determined by the fluxes in the Floquet states in the closed quantum system, which thus reflects the importance of the Floquet-state fluxes in closed quantum system discussed in the companion paper~\cite{Engelhardt2024c}.

Interestingly, the diverging fluctuations appear for parameters which feature a strong light-matter entanglement in the closed system that lead to a quadratically diverging variance~\cite{Engelhardt2024}.  We  therefore interpret the diverging photon flux fluctuations as an open quantum system precursor of the light-matter entanglement appearing in closed quantum systems.

\subsection{Probability redistribution}

In Fig.~\ref{fig:ProbabilityDistribution}, we investigate the probability redistribution for short times. The time-evolved probabilities are calculated by first  evaluating the time-dependent moment-generating function in Eq.~\eqref{eq:unifyingMomentGeneratingFkt}, and then carrying out the  Fourier transformation  in Eq.~\eqref{eq:relation_momGenFkt_probabilites}. We consider the phase differences $\overline \varphi=0$ and $\overline \varphi =\pi/2$ for a finite detuning $\epsilon_{\Delta}$. \textcolor{\markColorThree}{The dynamics for vanishing detuning can be found in Fig.~\ref{figOverview}(d).} The photons in the  driving modes are initially Poisson distributed, which approaches a Gaussian distribution for a large mean photon number. Here we consider an initial variance  $\sigma_1^2(0) = 100$ for illustration.

For $\varphi=0$ depicted in Fig.~\ref{fig:ProbabilityDistribution}(b), there is no transfer of photons between the driving modes. The photon numbers  change only because of emission into the reservoir. To enhance this effect, we have chosen a relatively large dissipation $\gamma$. For this reason, the photon distribution is translated to smaller photon numbers in both driving modes, while overall preserving the shape of the initial probability distribution.

For $\varphi=\pi/2$  depicted in Fig.~\ref{fig:ProbabilityDistribution}(c), the probability redistribution is mainly determined by the coherent photon exchange between the driving modes, i.e., the Josephson flux. At time $\Omega_1 t= 100$, we observe two peaks in the  probability distribution of mode $k=1$, shifted towards smaller and larger photon numbers, respectively. These peaks are a consequence of the light-matter entanglement and resemble the peaks observed in the isolated system shown in Fig.~\ref{figOverview}(b).  Here, one peak is higher because of the finite detuning $\epsilon_{\Delta}$. This results into a net photon flux in agreement with Eq.~\eqref{eq:photonFlux}.  Additionally, probability also accumulates between   the peaks because of the incoherent dynamics, i.e., the probabilistic transition between the Floquet states as explained in Sec.~\ref{sec:divergingFluctuations}. For larger times, the incoherentely accumulated probability distribution increases and finally exceeds the coherent peaks as seen at $\Omega_1t=300$.

 \section{ac-driven lambda system}
 
 \label{sec:acDrivenLambdaSystem}

To demonstrate its feasibility, we apply the  PRFT to an open ac-driven lambda system. Thereby, we show how to design a  driving protocol for an efficient up-conversion of photons from a low to a high frequency that features a high signal-to-noise ratio. The proposed lambda system can be also implemented  with microwave photons using metastable states of a capacitively shunted Fluxonium~\cite{Earnest2018}.

\subsection{Equations of motion}

 \begin{figure}
	\includegraphics[width=\linewidth]{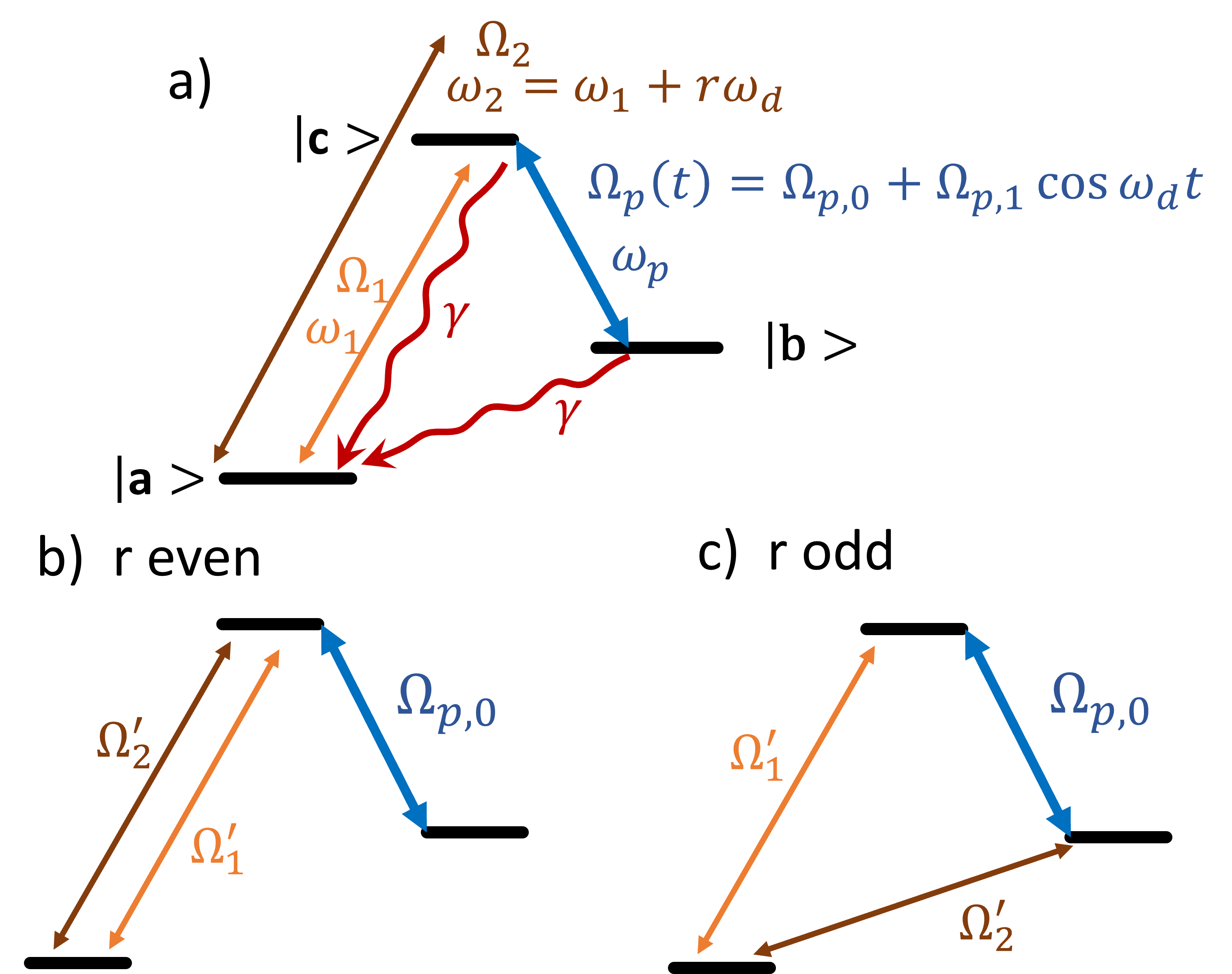}
	\caption{ (a) Sketch of the ac-driven lambda system. The levels $\left| b\right>$ and $\left| c \right>$ are connected by a pump field with frequency $\omega_{\text{p}} $ and ac-modulated Rabi frequency $\Omega_{\text{p} }(t) =\Omega_{\text{p},0} + \Omega_{\text{p},1} \cos \omega_{\text{d}}  t $. The signal field frequencies $\omega_1$ and $\omega_2$ differ by $r$ times the driving frequency $\omega_{\text{d} }$, i.e., $\omega_2  -\omega_1 = r \omega_{\text{d}}$. (b) For  $r$ even, the signal Rabi frequency is effectively renormalized. (c) For $r$ odd, the coupling structure  is changed to a triangular one, which is  dipole forbidden in atomic systems for unpumped systems.}
	\label{fig:DrivenLambdaSystem}
\end{figure}

The ac-driven lambda system considered here consists of three energy levels coupled by three different coherent fields as sketched in Fig.~\ref{fig:DrivenLambdaSystem}(a). The states $\left| a \right>$ and $\left|c \right>$ are coupled by two coherent signal fields with frequencies $\omega_1$ and $\omega_2$, respectively. For simplicity we assume that the corresponding Rabi frequencies are equal $\Omega_{1} = \Omega_{2} = \Omega_{\text{s}}$. The transition $\left|b  \right> \leftrightarrow \left| c \right>$ is driven by a pump field with frequency $\omega_{\text{p} }$. Its Rabi frequency is periodically modulated as
\begin{equation}
	\Omega_{\text{p} }(t)  =  \Omega_{\text{p},0}+  \Omega_{\text{p},1}  \cos\left( \omega_{\text{d}} t \right)
\end{equation}
with dc component $\Omega_{\text{p},0}$, driving amplitude $\Omega_{\text{p},1} $ and driving frequency $\omega_{\text{d}}$. This driving protocol can be experimentally implemented using concatenated driving~\cite{wang2021observation}. Moreover, we impose the frequency matching condition
\begin{equation}
	\omega_2 -\omega_1 = r \omega_{\text{d}}
\end{equation}
for integer $r$ to establish a resonant transition between the two signal fields.

The equation of motion of the generalized density matrix  in Eq.~\eqref{sec:quantumMasterEquation} including the counting fields  reads as
\begin{eqnarray}
\frac{d}{dt}\rho_{\boldsymbol \chi} &=& -i \hat {\mathcal H}_{\overline {\boldsymbol \varphi} +\frac{\boldsymbol \chi  }{2}}(t) \rho_{\boldsymbol \chi} + i\rho_{\boldsymbol \chi} \hat {\mathcal H}_{\overline {\boldsymbol \varphi} - \frac{\boldsymbol \chi  }{2}}(t)   \nonumber\\ 
&&+ \gamma  D\left[ \braket{a}{b} \right] \rho_{\boldsymbol \chi} +\gamma D\left[ \braket{a}{c}   \right] \rho_{\boldsymbol \chi},
\label{eq:eomLambdaSystem}
\end{eqnarray}
where we have added two dissipators describing the incoherent relaxation from the states $\left| b\right>$ and $\left| c \right>$ to state $\left| a \right>$. For simplicity, we have assumed that both processes have equal relaxation constants $\gamma$. The counting fields of the incoherent processes have been omitted. The semiclassical Hamiltonian reads
\begin{eqnarray}
\hat {\mathcal H}_{\boldsymbol \varphi}(t) &=& \sum_{\alpha=a,b,c} \epsilon_\alpha  \braket{\alpha}{\alpha} +  \frac{\Omega_{\text{p} }(t)}{2} \left( e^{ - i\omega_{\text{p}} t} \braket{c}{b} +\text{ h.c.}\right) \nonumber  \\
&+&  \left[ \left(\frac{\Omega_{1} }{2} e^{i\varphi_1 - i\omega_{1} t }  + \frac{\Omega_{2} }{2} e^{i\varphi_2 - i\omega_{2} t} \right)  \braket{c}{a}   +\text{ h.c.} \right].\nonumber  \\
\end{eqnarray}
The photonic dynamics of the system  can be numerically calculated along the lines  presented in Sec.~\ref{sec:longTimeLimit}.  The behavior in the stationary state is depicted in Figs.~\ref{fig:DrivenLambdaSystem_detuning} and~\ref{fig:DrivenLambdaSystem_drivingAmplitude}, where we observe a sensitive dependence of the photon flux, the fluctuations, and the signal-to-noise ratio as a function of the signal frequency $\omega_1$ and the driving amplitude $\Omega_{\text{p},1}$. To check the consistency of the FCS and the PRFT predictions, we have compared the number of photons absorbed  by the dissipator with the total change of photons in modes $k=1$ and $k=2$ and found a perfect agreement (not shown).

\subsection{Effective treatment}

\begin{figure*}
	\includegraphics[width=\linewidth]{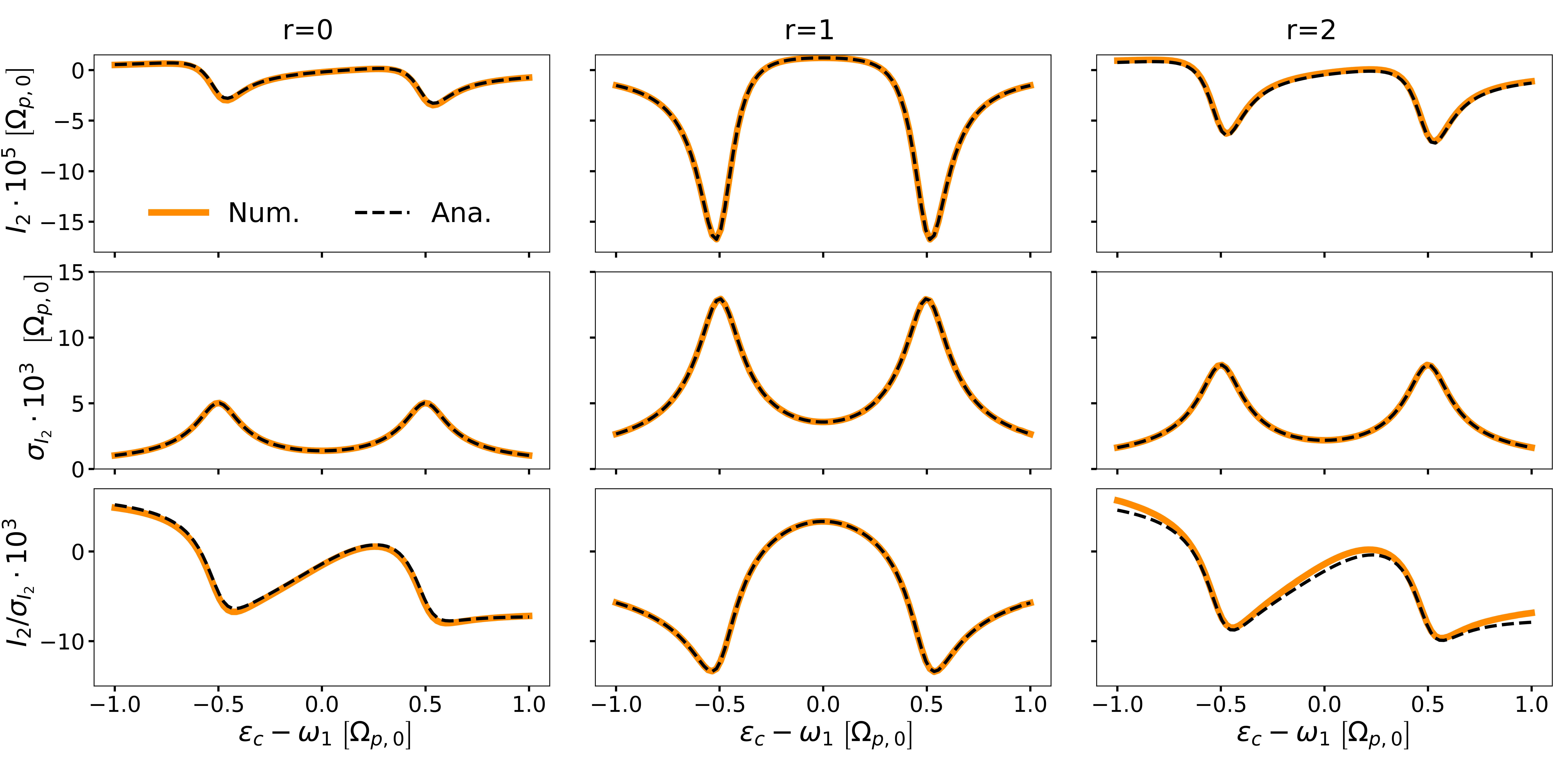}
	\caption{ Photon flux, its fluctuations and the signal-to-noise ratio into the signal field $k=2$ for different resonance orders $r =0$, $r =1$, and $r =3$. The analytical predictions using the cumulant-generating function in Eq.~\eqref{eq:cumulantGeneratingFunctionLambdaSystem} (dashed) are benchmarked against the exact numerical results (solid). Overall parameters are $\omega_{\text{d}} = 20 \Omega_{\text{p},0}$, $\Omega_{\text{p},1} = 4\omega_{\text{d}} $,  $\epsilon_c = \epsilon_b + \omega_{\text{p} } $, and $\gamma = 0.2  \Omega_{\text{p},0} $.}
	\label{fig:DrivenLambdaSystem_detuning}
\end{figure*}

To obtain a deeper understanding of the underlying physical processes, we derive an analytical expression for the stationary cumulant-generating function in the limit $\Omega_{\text{s}} \ll   \epsilon_{b,\Delta}-\epsilon_{a,\Delta}, \epsilon_{c,\Delta}-\epsilon_{a,\Delta}, \gamma \ll\omega_{\text{d}} $, where  $ \epsilon_{a,\Delta} =\epsilon_{a}$,  $ \epsilon_{b,\Delta} =\epsilon_{b}+\omega_{\text{p}}-\omega_1$, $ \epsilon_{c,\Delta} =\epsilon_{c}-\omega_1$.   Experimentally, this limit is relevant when tuning the frequencies $\omega_1,  \omega_{\text{p}}$  close to resonance of the respective transitions in the lamba system, and the Rabi frequencies of the signal fields $\Omega_{\text{s}}$  are very weak. To this end, we first transform  Eq.~\eqref{eq:eomLambdaSystem} into a rotating frame defined by
\begin{eqnarray}
U_{\text{rot}} (t)&=&  \exp\left[-i  \frac{\Omega_{\text{p},1}}{2\omega_{\text{d} }} \sin(\omega_{\text{d}} t) \left( \braket{b}{c}+\braket{c}{b}  \right)  \right] 
\label{eq:rotatingFrame}
\end{eqnarray}
and apply then a rotating-wave approximation following Refs.~\cite{Engelhardt2013,Bastidas2012}. In doing so, the dissipator in Eq.~\eqref{eq:eomLambdaSystem} remains unchanged, and  we obtain an effective  time-independent Hamiltonian
\begin{eqnarray}
\hat {\mathcal H}_{\boldsymbol \varphi} &=&  \epsilon_{a,\Delta} \braket{a}{a} +   \frac{\epsilon_{b,\Delta}+\epsilon_{c,\Delta}}{2}  \left( \braket{b}{b} +  \braket{c}{c}\right) \nonumber  \\
&+&  \mathcal J_0 \left( \frac{\Omega_{p,1} }{\omega_{\text{d} }} \right) \frac{\epsilon_{b,\Delta}-\epsilon_{c,\Delta}}{2} \left[\braket{b}{b}-\braket{c}{c} \right] \nonumber \\
&+& \frac{\Omega_{\text{p},0} }{2} \left( \braket{b}{c}   + \braket{c}{b} \right)  \nonumber \\
&+& \hat {\mathcal H}_{\text{sig},\boldsymbol \varphi}^{(r )},
\end{eqnarray}
where $\mathcal J_{r }\left( x\right)$ denotes the $r$-th order Bessel function of the first kind. Interestingly, the structure of the Hamiltonian $H_{\text{sig} }^{(r )}$ describing the transitions induced by the signal fields depends on the parity of the order  $r $. For  $r $ even, the signal-field Hamiltonian
\begin{eqnarray}
 \hat {\mathcal H}_{\text{sig},\boldsymbol \varphi }^{(r )}&=&    \frac{ \Omega_{\text{s} }}{2} \left[ e^{i\varphi_1} \mathcal J_0 \left( \frac{\Omega_{\text{p},1}}{2\omega_{\text{d} }} \right)  \braket{c}{a}  + \text{h.c.} \right]  \nonumber \\ 
  &+&  \frac{ \Omega_{\text{s} }}{2} \left[   e^{i\varphi_2} \mathcal J_{r } \left( \frac{\Omega_{\text{p} ,1}}{2\omega_{\text{d} }}     \right) \braket{c}{a}   + \text{h.c.}\right]
\end{eqnarray}
is a renormalized  version of the original transition terms. Yet, for
$r $ odd,  the signal-field Hamiltonian
\begin{eqnarray}
\hat {\mathcal H}_{\text{sig},\boldsymbol \varphi } ^{(r )}&=&   \frac{ \Omega_{\text{s} }}{2} \left[ e^{i\varphi_1} \mathcal J_0 \left( \frac{\Omega_{\text{p},1}}{2\omega_{\text{d} }} \right)  \braket{c}{a}  + \text{h.c.}\right]  \nonumber \\ 
&+&   \frac{ \Omega_{\text{s} }}{2} \left[   e^{i\varphi_2} \mathcal J_{r } \left( \frac{\Omega_{\text{p},1}}{2\omega_{\text{d} }}     \right) \braket{b}{a}   + \text{h.c.}\right]
\end{eqnarray}
exhibits an additional transition between levels $a\leftrightarrow b$, establishing a triangular structure as sketched in Fig.~\ref{fig:DrivenLambdaSystem}(c).  Intriguingly, as such a triangular structure is dipole-forbidden in atomic systems, the ac-driving  protocol provides a way to circumvent this selection rule:
Selection rules appear due to spatial symmetries of the probed system, e.g., isotropy for atoms.  The periodic driving glorifies these spatial symmetries to time-spatial symmetries, which is reflected in the modification of optical selection rules~\cite{Engelhardt2021} as observed in the lambda system in Fig.~\ref{fig:DrivenLambdaSystem}.

As shown in detail in Appendix~\ref{sec:lambdaSystemPerturbationTheory}, the counting-field dependent eigenvalue $\lambda_{0;\boldsymbol \chi}  $ determining the asymptotic moment-generating function in Eq.~\eqref{eq:cumulantGenertingFunctionLongTimes} can be evaluated using second-order perturbation theory for non-Hermitian systems. For the case $\epsilon_c -\epsilon_b =\omega_{\text{p}}$, we obtain
\begin{eqnarray}
\lambda_{0;\boldsymbol  \chi} 
&=& \sum_{\alpha=b,c}  \frac{ \Omega_{\alpha,\overline {\boldsymbol \varphi} +\frac{\boldsymbol \chi  }{2}} }{4}\left[-  \Omega_{\alpha,\overline {\boldsymbol \varphi} +\frac{\boldsymbol \chi  }{2}}^*   +   \Omega_{\alpha,\overline {\boldsymbol \varphi} -\frac{\boldsymbol \chi  }{2}}^*   \right]  \frac{1}{i \tilde \epsilon_\alpha  - \gamma } \nonumber \\
&+& \sum_{\alpha=b,c} \frac{ \Omega_{\alpha,\overline {\boldsymbol \varphi} -\frac{\boldsymbol \chi  }{2}}^* }{4} \left[ -   \Omega_{\alpha,\overline {\boldsymbol \varphi}-\frac{\boldsymbol \chi  }{2}} + \Omega_{\alpha,\overline {\boldsymbol \varphi} +\frac{\boldsymbol \chi  }{2} }  \right]   \frac{1}{-i \tilde \epsilon_\alpha  - \gamma } \nonumber \\
&+& \mathcal O\left( \Omega_{\text{s}}^3\right)
\label{eq:cumulantGeneratingFunctionLambdaSystem},
\end{eqnarray}
where
\begin{eqnarray}
\Omega_{b,\boldsymbol \varphi} &=&  \frac{\Omega_{\text{s} } } {\sqrt{2}} \left[  e^{i\varphi_1} \mathcal J_0 \left( \frac{\Omega_{\text{p},1}}{2\omega_{\text{d} }} \right) +   e^{i\varphi_2+ir\pi} \mathcal J_{r } \left( \frac{\Omega_{\text{p},1}}{2\omega_{\text{d} }} \right)   \right]\nonumber , \\  
\Omega_{c,\boldsymbol \varphi}  &=& \frac{\Omega_{\text{s} }} {\sqrt{2}} \left[  e^{i\varphi_1} \mathcal J_0 \left( \frac{\Omega_{\text{p} ,1}} {2\omega_{\text{d} }} \right) +   e^{i\varphi_2} \mathcal J_{r } \left( \frac{\Omega_{p,1}}{2\omega_{\text{d} }} \right)   \right] . \nonumber \\
\label{eq:effectiveRabifrequencies}
\end{eqnarray}
Moreover, $\tilde \epsilon_b  = \epsilon_b  - \omega_{1}   -\Omega_{\text{p} ,0} $ and $\tilde \epsilon_c  =  \epsilon_c  - \omega_{1} + \Omega_{\text{p},0}$ are two of the quasienergies of the closed system for $\Omega_{\text{s}}=0$. In Fig.~\ref{fig:DrivenLambdaSystem_detuning} we benchmark the analytical calculation for three orders $r =0,1,2$ and find a very good agreement.

\subsection{Discussion}

 \begin{figure}
	\includegraphics[width=\linewidth]{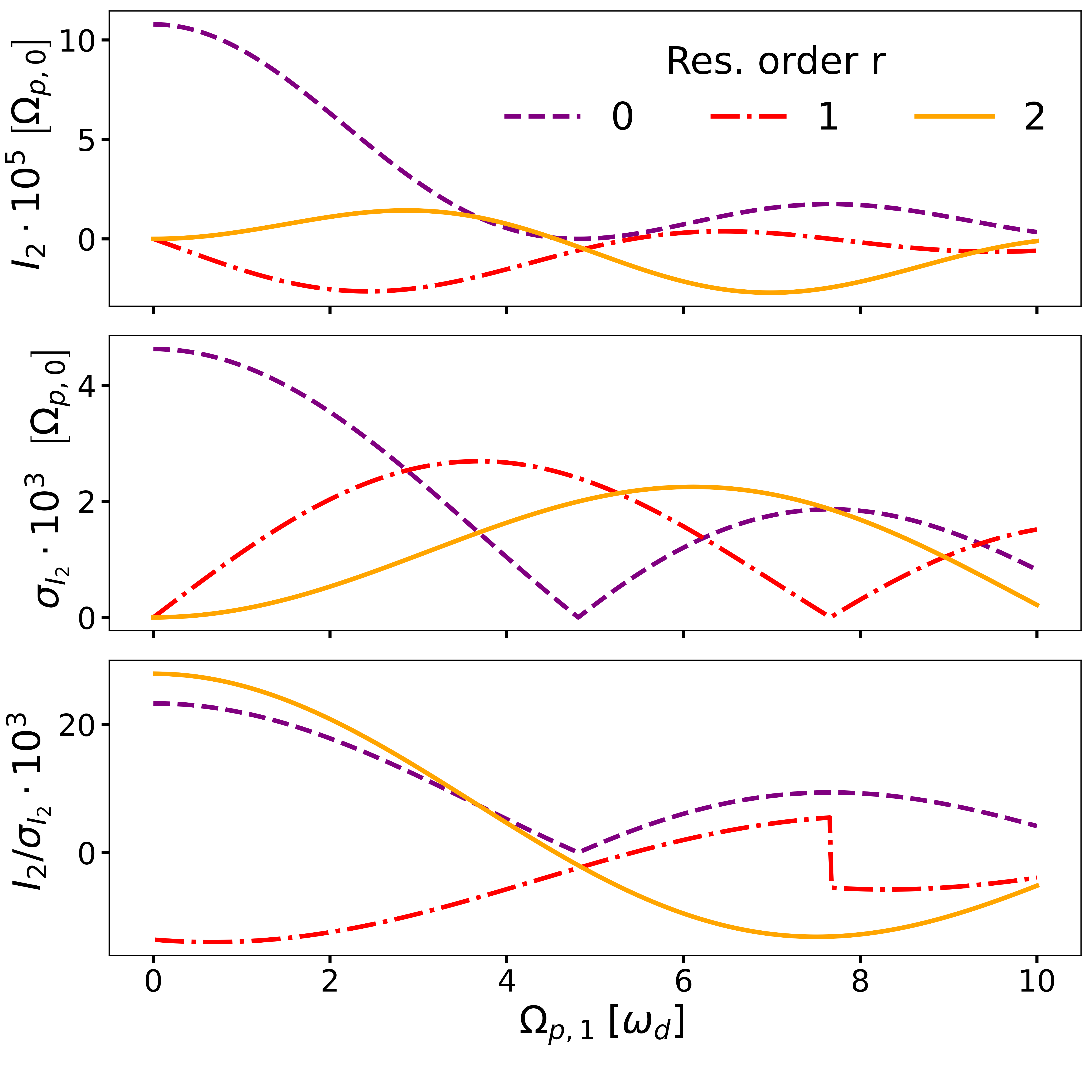}
	\caption{ Analysis of the photon flux from signal field $k=1$ into signal field $k=2$ as a function of driving amplitude for the three resonance orders $r=0,1,2$. Overall parameters are the same as in Fig.~\ref{fig:DrivenLambdaSystem_detuning}. }
	\label{fig:DrivenLambdaSystem_drivingAmplitude}
\end{figure}
 
Here we discuss the  lambda system for a possible application as a frequency  up-conversion device, which converts photons from mode $k=1$ to mode $k=2$. To this end, we seek for combinations of the detuning  $\omega_{\Delta} =\epsilon_c - \omega_{1}$ and driving amplitude $\Omega_{\text{p},1}$ with  a significant photon flux $I_2$, while exhibiting a high signal-to-noise ratio $ I_2/\sigma_{I,2}$.

\textit{Detuning}. Figure \ref{fig:DrivenLambdaSystem_detuning} investigates the photon transport for a driving amplitude $\Omega_{\text{p},1} = 4\omega_{\text{d} } $. In all orders $r $, the photon flux shows  two resonances located at the energies of the Rabi splitting  $\omega_{\Delta} = \pm\Omega_{\text{p},0}/2$. Close to the two resonance peaks, the flux is negative, as net absorption due to the dissipation dominates over the coherent energy flux. Accordingly,  the noise is significantly enhanced. Away from the resonances, the photon flux can be positive, meaning photons are being pumped from  field $k=1$ to field $k=2$ via the assistance of the pump field.

The signal-to-noise ratio is particularly high for a large negative detuning  $\omega_{\Delta}\ll -\Omega_{\text{p},0}$ for the resonances $r =0$ and $r =2$, and around $\omega_{\Delta}\approx 0$ for the resonance order $r =1$. In the former cases, the detuning suppresses the absorption of photons, while in the latter case, the absorption is partly  suppressed by a destructive interference known as electromagnetically-induced transparency~\cite{Fleischhauer2002}.
 
\textit{Driving amplitude.} In Fig.~\ref{fig:DrivenLambdaSystem_drivingAmplitude}, we investigate the photon transport for a negative detuning $\omega_{\Delta}= -2 \Omega_{\text{p},0} $. Qualitatively similar observations can be found for $\omega_{\Delta}= 0 $ (not shown). The photon flux $I_2$ is always positive for $r =0$ because of the chosen signal phases $\varphi_1 - \varphi_2 = \pi/2 $, resembling the photon transport in the two-level system of Sec.~\ref{sec:jcModel}. Contrary, the photon flux can be either positive or negative for the resonance orders $r =1,2$. Because of the Bessel functions, the photon flux shows overall an oscillatory dependence on the pump field amplitude. A similar Bessel function dependence can be observed for the fluctuations $\sigma_{I,2}$.

At particular values of the driving amplitude $\Omega_{\text{p},1}$ in Fig.~\ref{fig:DrivenLambdaSystem_drivingAmplitude}, the Bessel function has a root. According to Eqs.~\eqref{eq:cumulantGeneratingFunctionLambdaSystem} and ~\eqref{eq:effectiveRabifrequencies}, this leads to a vanishing dependence of the eigenvalue $\lambda_{0;\boldsymbol  \chi} $  on either counting field $\chi_1$ or $\chi_2$. As a consequence, the  photon flux vanishes completely at these values in Fig.~\ref{fig:DrivenLambdaSystem_drivingAmplitude}, meaning that the mean photon flux and all  higher cumulants are zero.  Thus, the transport into signal mode $k=2$ is completely suppressed and the system becomes transparent for the frequency $\omega_2$. This effect is coined coherent destruction of tunneling, and has been identified in various contexts, such as electronic transport~\cite{FernandezFernandez2023}, many-body dynamics in interacting systems~\cite{Gong2009,Bastidas2012,Engelhardt2013}, and topological band structures~\cite{Benito2014}. In the present context, the coherent-destruction of tunneling suppresses the  Josephson flux between two photonic modes due to the control of the pump field. 

From Fig.~\ref{fig:DrivenLambdaSystem_drivingAmplitude} we can determine a driving amplitude for which the signal-to-noise ratio is very high, i.e., where the lambda system can be deployed as a frequency upconversion device. The figure suggests to use the $r =2$ resonance around $\Omega_{\text{p},1} \approx  2\omega_{\text{d} }$ or the $r =1 $ resonance around $\Omega_{\text{p},1} \approx 7 \omega_d$. Both parameter combinations feature a significant photon flux $I_2$ while the fluctuations are relatively small. We note that the jump in the $r =1$ signal-to-noise ratio appears as $I_2$ changes from positive to negative at the same point where the fluctuations hit a zero.

\section{Conclusions and outlook}

\label{sec:conclusions}

In this paper, we have developed a  framework unifying the standard FCS and the recently developed PRFT. To this end, we have defined a  moment-generating function that reduces to the FCS and the PRFT in the respective limits. Importantly, the unified framework is non-perturbative in all parameters and paves thus the way to a non-perturbative spectroscopic framework, that  allows the development of metrological methods which can characterize the probability distribution of stimulated processes. 

The framework has been applied to the multimode Jaynes-Cummings model, where two modes coherently drive a two-level system, that is additionally coupled to a continuum of incoherent photon modes.  We have found that the unified cumulant-generating function consistently describes the photon statistics of coherent and incoherent emission. Moreover, we have found that the light-matter entanglement predicted by the PRFT in the closed quantum systems leads to diverging transport fluctuations in the dissipative system. These fluctuations should be observable in  spectroscopic experiments, which we will discuss in a future work. 

As a more sophisticated application, we have investigated an ac-driven lambda system for its application as a frequency-conversion device. In doing so, we have found that the ac-driving can create a triangular coupling  structure, which is actually electrically-dipole forbidden in atoms. Using our unified framework, we could identify parameters that feature both a significant photon conversion and a high signal-to-noise ratio.

The calculations in this paper could be carried out analytically.  To this end, we  utilized the method of the truncated characteristic polynomial. In the weak driving regime, we have generalized the common stationary perturbation theory known for Hermitian systems, to non-Hermitian ones. As an alternative method, one can reduce the dimension of the Liouvillian equation by means of adiabatic elimination. Depending on the system, either method will be more suitable. For more sophisticated models, efficient numerical evaluation  can be easily implemented. Besides,  the polynomial perturbation theory can be applied even for larger systems  if the system parameters are on different orders of magnitude~\cite{Engelhardt2022}.

We have derived a quantum master equation in second-order perturbation theory of the  system-bath interaction. The application of more advanced methods  such as the hierarchy equations of motion or the polaron transformation is likewise possible and can lead to new insights into fundamental concepts of light-matter interactions. Using the modified FCS in the hierarchy equations of motion in Ref.~\cite{cerrillo2016nonequilibrium}, it is also possible to investigate the consistency of energy conservation within the PRFT for open quantum systems, which do not feature an excitation number conservation as the Jaynes-Cummings model considered here. The PRFT will also give rise to the development of new spectroscopic  and metrological methods, which help to  improve  detection protocols for  electric fields~\cite{Jing2020,Meyer2020} and  dark matter ~\cite{Bloch2020,Bloch2022,Engelhardt2024a}. Moreover, the unified framework introduced here can find application to heat engines and quantum batteries.

In the present work, the PRFT has been formally investigated in the context of cavity and circuit quantum electrodynamics,  while the application to spectroscopic setups, where a propagating laser field drives a quantum system, will be considered in future work. The flexibility of the PRFT will thereby allow to spectroscopically describe complex many-body systems of interacting quantum emitters, such as arrays of Rydberg atoms, which can exhibit a pronounced photonic non-linearity~\cite{Bekenstein2020,Rusconi2021,MorenoCardoner2021}, and circuit QED systems~\cite{Zhang2022a}. Being a semiclassical method,  the PRFT is  compatible with sophisticated design and control strategies in Floquet phase space~\cite{Guo2023}, or deploying  exceptional points~\cite{Simonson2022}. \textcolor{\markColorThree}{ The non-perturbative character of the PRFT allows also to investigate spectroscopic signatures of phase transitions in light-matter interacting systems~\cite{Hwang2015,Cai2021}.}  Other possible spectroscopic applications are higher-harmonic generation~\cite{Yuan2022} and spectroscopy with non-classical light~\cite{Dorfman2021,Zhang2022a}.

\section*{Acknowledgments}

G.E. acknowledges the  support by the Guangdong Provincial Key Laboratory (Grant No.2019B121203002)
J.Y.L. acknowledges the support from the National Natural Science Foundation of China (Grant Nos. 11774311).
V.M.B. wish to thank NTT Research for their financial and technical support.
G.P. acknowledges the Spanish Ministry of Economy and Competitiveness for financial support through the grant: PID2020-117787GBI00 and  
support from CSIC Interdisciplinary Thematic Platform on Quantum  
Technologies (PTI-QTEP+). The authors thank Ming Li for helpful comments.

\appendix

{ \color{\markColorTwo}

\section{Derivation of the PRFT in open quantum systems}

In this Appendix, we provide the mathematical derivations of the PRFT in open quantum systems, which unifies it with the standard FCS.

\subsection{Definition of the moment-generating function}
\label{app:defMomentGenFct}

Here, we derive the formal expression for the moment-generating function in Eq.~\eqref{eq:def:momentGeneratingFkt_exact}. We start by expressing the definition of the conditional probabilities in Eq.~\eqref{eq:conditionalProbabilities} as an integral over the counting fields
\begin{eqnarray}
p_{\boldsymbol n   ,  \boldsymbol \nu +\Delta \boldsymbol \nu  \mid \boldsymbol  \nu }(t) &=& \text{tr} \left[ \rho_{\text{tot},\boldsymbol \nu}(t) \sum_{\boldsymbol n_1 , \boldsymbol \nu_1} \delta_{\boldsymbol n,\boldsymbol n_1} \hat P_{\boldsymbol n_1} \delta_{\boldsymbol \nu +\Delta \boldsymbol \nu,\boldsymbol \nu_1} \hat P_{\boldsymbol \nu_1}    \right] \nonumber  \\
&=& \text{tr} \left[ \rho_{\text{tot},\boldsymbol \nu}(t) \phantom{ \int_{-\pi}^{\pi}} \right.  \nonumber \\
&\times&  
\int_{-\pi}^{\pi} \frac{d\boldsymbol \chi}{(2\pi)^{N_{\text{D}} }} e^{i\boldsymbol \chi \cdot \boldsymbol n}   \sum_{\boldsymbol n_1} e^{-i\boldsymbol \chi \cdot  \boldsymbol n_1}  \hat P_{\boldsymbol n_1}
 \nonumber \\
&\times& \left.  \int_{-\pi}^{\pi} \frac{d\boldsymbol \xi}{(2\pi)^{N_{\text{B}} }} e^{i\boldsymbol \xi \cdot (\boldsymbol \nu +\Delta\boldsymbol \nu  )}   \sum_{\boldsymbol \nu_1} e^{-i\boldsymbol \xi \cdot  \boldsymbol \nu_1}  \hat P_{\boldsymbol \nu_1 }  \right] \nonumber
 \\
&=& \int_{-\pi}^{\pi} \frac{d\boldsymbol \chi}{(2\pi)^{N_{\text{D}} }}\int_{-\pi}^{\pi}\frac{d\boldsymbol \xi}{(2\pi)^{N_{\text{B}} }}   e^{i\boldsymbol \chi\cdot {\boldsymbol n} }  e^{i\boldsymbol \xi \cdot (\boldsymbol \nu +\Delta\boldsymbol \nu  )} \nonumber \\
 &\times &  \text{tr} \left[   e^{ -i \boldsymbol \chi\cdot \hat {\boldsymbol n}_{\text{D}} -i\boldsymbol \xi \cdot \hat {\boldsymbol n}_{\text{B}}     }  \rho_{\text{tot},\boldsymbol \nu}(t)   \right].
\end{eqnarray}
Next, we use that the conditional initial density matrix in Eq.~\eqref{eq:conditionalDensityMatrix} is diagonal in the basis of Fock states of the incoherent modes, such that we can express the time-evolved conditional density matrix as
\begin{multline}
\rho_{\text{tot},\boldsymbol \nu}(t)  =    \hat U(t)  \rho_{\text{tot},\boldsymbol \nu}(0)\hat U^\dagger(t)  \\
 = \hat U(t)  \rho_{\text{M} }(0) \otimes \left| A_0 \right> \left< A_0  \right|    \otimes \left|\boldsymbol \nu \right> \left< \boldsymbol \nu \right|U^\dagger(t)   \\
= \hat U(t)  \rho_{\text{M} }(0) \otimes \left| A_0 \right> \left< A_0  \right|    \otimes \left|\boldsymbol \nu \right> \left< \boldsymbol \nu \right|e^{-i\boldsymbol  \xi \cdot \boldsymbol \nu}    e^{i\boldsymbol \xi \cdot  \hat {\boldsymbol n}_{\text{B}} } U^\dagger(t)     \\
=   e^{-i\boldsymbol  \xi \cdot \boldsymbol \nu}    \hat U(t)   \rho_{\text{tot},\boldsymbol \nu}(0)  e^{i\boldsymbol \xi \cdot  \hat {\boldsymbol n}_{\text{B}} } U^\dagger(t) .
\end{multline}
Consequently, the conditional probability distribution reads 
\begin{multline}
p_{\boldsymbol n, \boldsymbol \nu + \Delta \boldsymbol \nu  \mid \boldsymbol  \nu }(t)
=   \int_{-\pi}^{\pi} \frac{d\boldsymbol \chi}{(2\pi)^{N_{ \text{D} }}} \int_{-\pi}^{\pi}\frac{d\boldsymbol \xi}{(2\pi)^{N_{\text{B}} }}   e^{i\boldsymbol \chi \cdot \boldsymbol n} e^{i\boldsymbol \xi\cdot \Delta \boldsymbol \nu  }    \\
\times \text{tr} \left[ e^{ -i \boldsymbol \chi\cdot \hat {\boldsymbol n}_{ \text{D} } -i\boldsymbol \xi \cdot \hat {\boldsymbol n}_{\text{B}}    } \hat U (t)\rho_{\text{tot},\boldsymbol \nu}(0) e^{i\boldsymbol \xi \cdot \hat {\boldsymbol n}_{\text{B}} }  \hat U^\dagger(t)   \right].
\end{multline}
Using this expression, we can evaluate the averaged probabilities in Eq.~\eqref{eq:def:jointProbabilityDistribution}
\begin{eqnarray}
\overline p_{\boldsymbol  n  , \Delta \boldsymbol  \nu   }& &\equiv \sum_{\boldsymbol  \nu} p_{\boldsymbol  n , \boldsymbol  \nu +\Delta \boldsymbol  \nu \mid \boldsymbol  \nu  }(t) p_{\boldsymbol  \nu }(0) \nonumber  \\
&&= \int_{-\pi}^{\pi} \frac{d\boldsymbol \chi}{(2\pi)^{N_{ \text{D} } }}   \int_{-\pi}^{\pi}\frac{d\boldsymbol \xi}{(2\pi)^{N_{\text{B}} }} e^{i\boldsymbol \chi \cdot \boldsymbol n}   e^{i\boldsymbol \xi\cdot \Delta \boldsymbol \nu   }   \nonumber \\
&&\times \text{tr} \left[ e^{-i \boldsymbol \chi\cdot \hat {\boldsymbol n}_{ \text{D} } -i\boldsymbol \xi \cdot \hat {\boldsymbol n}_{\text{B}}     } \hat U (t)\rho_{\text{tot}}(0) e^{i\boldsymbol \xi \cdot \hat {\boldsymbol n}_{\text{B}} }  \hat U^\dagger(t)   \right] \nonumber \\
=& &\int_{-\pi}^{\pi} \frac{d\boldsymbol \chi}{(2\pi)^{N_{ \text{D} } }}  \int_{-\pi}^{\pi}\frac{d\boldsymbol \xi}{(2\pi)^{N_{\text{B}} }}  e^{  i\boldsymbol \chi \cdot \boldsymbol n+ i\boldsymbol \xi\cdot \Delta \boldsymbol  \nu }    M_{ \boldsymbol \chi,\boldsymbol \xi }(t), \nonumber \\
\end{eqnarray}
where  we have identified the moment-generating function in Eq.~\eqref{eq:def:momentGeneratingFkt_exact} in the third equality.

\subsection{Exact moment-generating function in Sambe space}

\label{sec:exact_momGenFct}

Here, we derive an exact expression for the moment-generating function in Eq.~\eqref{eq:def:momentGeneratingFkt_exact} in the Sambe space, which is convenient to analyze. The Sambe space is similar to the Fock space, but by replacing the photonic operators of the coherent modes as 
\begin{subequations}
	\label{eq:def:sambeSpace}
	\begin{eqnarray}
	\hat a_k^\dagger \hat a_k &\rightarrow& \sum_{n_k=-\infty}^{\infty}  n_k \left| n_k\right> \left< n_k\right|,  \\
	\hat a_k^\dagger   &\rightarrow&    \sum_{n_k=-\infty}^{\infty} \alpha_k \left| n_k+1\right> \left< n_k\right| . 
	\end{eqnarray} 
\end{subequations}
This establishes a translational invariance in the photon-number space, which we utilize in the following derivations. The derivation of the PRFT in open quantum system follows essentially the same lines as the derivation in closed quantum system in Ref.~\cite{Engelhardt2024c} besides   some modifications which we explain here.

Transforming  the system into an interaction picture specified by the unitary operator
\begin{equation}
\hat U^{(\text{free})}(t)  =e^ { -i \sum_{k =1}^{N_{ \text{D} }} \omega_k  \hat a_k^\dagger \hat a_k   t },
\end{equation}
 the exact time-evolution operator corresponding to the Hamiltonian in Eq.~\eqref{eq:hamiltonian:quantum} can be expressed as
\begin{equation}
\hat U(t)  =  \hat U^{(\text{free})}(t) \hat U^{(\text{int})}(t).
\label{eq:timeEvolutionOperator_quantum}
\end{equation}
Thereby, we have introduce the time-evolution operator in the interaction picture 
\begin{equation}
\hat U^{(\text{int})}(t)  = \hat{\mathcal T} e^{-i \int_{0}^{t} \hat H^{(\text{int})} (t') dt' } ,
\end{equation}
where the Hamiltonian in the interaction picture reads as
\begin{eqnarray}
\hat H^{(\text{int})} (t)  &=& \hat  H_{0} \nonumber\\
&+&    \sum_{k=1}^{N_{{ \text{D} }} }  g  \hat V_k   \left(\alpha_k e^{i\omega_k t}  \sum_{n_k}  \left| n_k+1\right> \left< n_k\right|   +\text{h.c.} \right)	\nonumber\\
\label{eq:hamiltonian:interaction}
\end{eqnarray}
with
\begin{equation}
	\hat  H_{0} =\hat  H_{\text{M}} + \sum_{k=N_{ \text{D} }+1}^{ N_{ \text{D} } + N_{\text{B} } }\omega_k \hat a_k^\dagger  \hat a_k  +  \sum_{k=N_{ \text{D} }+1}^{ N_{ \text{D} } + N_{\text{B} } }  g  \hat V_k \left(  \hat a_k^\dagger   + \hat a_k \right)
\end{equation}
describing the matter system and incoherent photonic modes.

Using Eq.~\eqref{eq:timeEvolutionOperator_quantum}, we find
\begin{eqnarray}
\left[ \hat U(t) \right]_{\boldsymbol  n,\boldsymbol  n^\prime}  &\equiv&  \left< \boldsymbol n \right| \hat U(t) \left|\boldsymbol n^\prime   \right> \nonumber \\
&=&   e^{-i\boldsymbol \omega \cdot \boldsymbol n t } \left< \boldsymbol n \right|\hat  U^{(\text{int} )}(t)\left|\boldsymbol n^\prime  \right> \nonumber  \\
&\equiv&   e^{-i\boldsymbol \omega \cdot \boldsymbol n t } \hat U^{(\text{int})}_{\boldsymbol n - \boldsymbol  n^\prime }(t),
\label{eq:translationalinvarianceRel}
\end{eqnarray}
where $\boldsymbol \omega  = (\omega_1 ,\dots,\omega_{N_{ \text{D} }}) $ is a vector of frequencies.   In  the third equality we utilized the translational invariance of the interaction Hamiltonian in Eq.~\eqref{eq:hamiltonian:interaction} in the photon-number space.

 Evaluating the moment-generating function in Eq.~\eqref{eq:def:momentGeneratingFkt_exact} using Eq.~\eqref{eq:translationalinvarianceRel} and the initial state in Eq.~\eqref{eq:coherentIntialState}, we find
\begin{widetext}
	\begin{eqnarray}
	M_{\boldsymbol\chi,\boldsymbol \xi}(t)   &\equiv& \sum_{\boldsymbol n, \boldsymbol n_1, \boldsymbol  n_2 }  \text{tr}\left\lbrace      \left[ \hat U^\dagger(t) \right]_{\boldsymbol n-\boldsymbol n_{1}, \boldsymbol n }  e^{-i\boldsymbol \xi \cdot \hat {\boldsymbol n}_{\text{B}}    -i \boldsymbol \chi\cdot \hat {\boldsymbol n}_{ \text{D} }  }  \left[ \hat U(t) \right]_{\boldsymbol n,\boldsymbol n-\boldsymbol n_{2}} \rho_{\text{MB}}(0) e^{i\boldsymbol \xi \cdot \hat {\boldsymbol n}_{\text{B}} }   \right\rbrace  a_{ \boldsymbol n- \boldsymbol n_{1}}^{*}  a_{ \boldsymbol n-\boldsymbol n_{2}} \nonumber  \\
	&=&  \sum_{\boldsymbol n, \boldsymbol n_1, \boldsymbol  n_2 }   \text{tr}\left[    \hat  U^{(\text{int} )\dagger } _{\boldsymbol n_1}(t)  e^{-i\boldsymbol \xi \cdot \hat {\boldsymbol n}_{\text{B}}    -i \boldsymbol \chi\cdot \hat {\boldsymbol n}_{ \text{D} }  }   \hat   U_{\boldsymbol  n_2}^{(\text{int})} (t)    \rho_{\text{MB}}(0) e^{i\boldsymbol \xi \cdot \hat {\boldsymbol n}_{B} }   \right]    a_{ \boldsymbol n-\boldsymbol n_{1}}^{*}  a_{ \boldsymbol n- \boldsymbol n_{2}} \nonumber  \\
	&=&\frac{1}{(2\pi)^{3N_{ \text{D} }} }\iiiint d\boldsymbol \varphi_1 d\boldsymbol\varphi_2 d\boldsymbol\varphi_3d\boldsymbol\varphi_4   \sum_{\boldsymbol n,\boldsymbol  n_1,  \boldsymbol n_2 }  \text{tr}  \left[    \hat  U^{\boldsymbol \dagger}_{\boldsymbol \varphi_1} (t)  e^{-i\boldsymbol \xi \cdot \hat {\boldsymbol n}_{\text{B}}   }  \hat   U_{\boldsymbol \varphi_2} (t)    \rho_{\text{MB}}(0) e^{i\boldsymbol \xi \cdot \hat {\boldsymbol n}_{\text{B}} }    \right]  \nonumber \\
	&& \qquad \qquad \times a_{\boldsymbol \varphi_3}^{*}  a_{\boldsymbol \varphi_4}  e^{i\boldsymbol  n \cdot(\boldsymbol \varphi_4-\boldsymbol \chi-\boldsymbol\varphi_3) + i \boldsymbol n_1\cdot  (\boldsymbol \varphi_3-\boldsymbol \varphi_1)  - i \boldsymbol n_2\cdot (\boldsymbol\varphi_4-\boldsymbol \varphi_2)  }    ,
	\label{eq:tra1:momGenFunction}
	\end{eqnarray}
\end{widetext}
where $ \rho_{\text{MB}}(0) =\rho_{\text{M}}(0)\otimes \rho_{\text{B} }(0) $ denotes the reduced density matrix of the matter and bath systems.  Moreover, we have introduced the Fourier transformations
\begin{eqnarray}
\hat   U_{\boldsymbol n}^{(\text{int})} (t) &=&   \frac{1}{(2\pi)^{N_{ \text{D} }} } \int_{-\pi}^{\pi} d\boldsymbol \varphi\;\hat   U_ {\boldsymbol \varphi } (t ) e^{i   \boldsymbol n \cdot   \boldsymbol \varphi } , \nonumber \\
a_{ \boldsymbol  n} &=&  \frac{1}{(2\pi)^{N_{ \text{D} }/2}}\int_{-\pi}^{\pi} d\boldsymbol \varphi\;  a_{\boldsymbol \varphi} e^{i  \boldsymbol  n \cdot \boldsymbol  \varphi } .
\label{eq:timeEvolutionOp_fourierTrans}
\end{eqnarray}
The summations over  $\boldsymbol n,\boldsymbol n_1,\boldsymbol n_2$ give rise to delta functions, which we can use to evaluate three of the integrals over $\boldsymbol \varphi_1,\dots , \boldsymbol \varphi_4$. In doing so,  the moment-generating function becomes
\begin{multline}
M_{\boldsymbol \chi,\boldsymbol \xi}(t) 
=\int_{-\pi}^{\pi}      a_{\boldsymbol \varphi -\frac{\boldsymbol \chi}{2} }^{*}  a_{ \boldsymbol \varphi+\frac{\boldsymbol \chi}{2}  } \\   
 \times\text{tr}  \left[   \hat  U^{\boldsymbol \dagger}_{\boldsymbol \varphi - \frac{\boldsymbol \chi}{2}  } (t)  e^{-i\boldsymbol \xi \cdot \hat {\boldsymbol n}_{\text{B}}   }  \hat   U_{\boldsymbol \varphi + \frac{\boldsymbol \chi}{2}  } (t)    \rho_{\text{MB}}(0) e^{i\boldsymbol \xi \cdot \hat {\boldsymbol n}_{\text{B}} }  \right]  d\boldsymbol \varphi  . 
\end{multline}
Exactly as for closed quantum systems (Ref.~\cite{Engelhardt2024c}, Appendix A), we can replace the $\boldsymbol \varphi $-depended time-evolution operator by its semiclassical counterpart 
\begin{equation}
\hat   U_{\boldsymbol \varphi } (t)   = \hat  {\mathcal U} _{\boldsymbol \varphi} (t),
\label{eq:equivalenceRelationQuantumClassical}
\end{equation}
where
\begin{eqnarray}
\hat  {\mathcal U} _{\boldsymbol \varphi} (t) &=& \hat{\mathcal T} e^{-i \int_{0}^{t} \hat {\mathcal H}_{\text{MB},\boldsymbol \varphi} (t') dt' } ,\nonumber \\
\hat {\mathcal H}_{\text{MB},\boldsymbol \varphi} (t )&=& \hat  H_{0}  +
    \sum_{k=1}^{N_{\text{D}} }  g_k  \hat V_k   \left(\alpha_k e^{i\omega_k t -\varphi_k}    +\text{h.c.} \right).
\end{eqnarray}
We note that this replacement is  an exact identity. In doing so, the moment-generating function becomes
\begin{multline}
M_{\boldsymbol \chi,\boldsymbol \xi}(t) 
=\int_{-\pi}^{\pi}     a_{\boldsymbol \varphi -\frac{\boldsymbol \chi}{2} }^{*}  a_{ \boldsymbol \varphi+\frac{\boldsymbol \chi}{2}  }\\
 \times \text{tr}  \left [    \hat   {\mathcal U}^{\boldsymbol \dagger}_{ - \frac{\boldsymbol \xi}{2} ,  \boldsymbol \varphi - \frac{\boldsymbol \chi}{2}  } (t)   \hat    {\mathcal U}_{\frac{\boldsymbol \xi}{2} ,  \boldsymbol \varphi + \frac{\boldsymbol \chi}{2}  } (t)    \rho_{\text{MB}}(0)    \right ]  d\boldsymbol \varphi ,
\label{eq:momentGenFct_exact}
\end{multline}
where we have introduced the counting-field dependent time-evolution operator
\begin{eqnarray}
\hat    {\mathcal U}_{\boldsymbol \varphi,  \boldsymbol \xi   } (t) &=&  e^{-i\boldsymbol \xi \cdot \hat {\boldsymbol n}_{\text{B}}   }   \hat  {\mathcal U} _{\boldsymbol \varphi} (t) e^{i\boldsymbol \xi \cdot \hat {\boldsymbol n}_{\text{B}}   }   \nonumber \\
 &=& \hat{\mathcal T} e^{-i \int_{0}^{t} \hat {\mathcal H}_{\text{MB},\boldsymbol \varphi ,\boldsymbol \xi } (t') dt' }
 \label{eq:countingFieldDependentTimeEvOperator}
\end{eqnarray}
in terms of the counting-field dependent Hamiltonian
\begin{equation}
	\hat {\mathcal H}_{\text{MB},\boldsymbol \varphi ,\boldsymbol \xi }   =e^{-i\boldsymbol \xi \cdot \hat {\boldsymbol n}_{\text{B}}   }   \hat {\mathcal H}_{\text{MB},\boldsymbol \varphi  } e^{i\boldsymbol \xi \cdot \hat {\boldsymbol n}_{\text{B}}   }  .
	\label{eq:countingFieldDependentHamiltonian}
\end{equation}
We emphasize that Eq.~\eqref{eq:momentGenFct_exact}  is an exact expression of the moment-generating function in Sambe space.

\subsection{Semiclassical limit}
\label{sec:semiclassicalLimit}

To investigate the semiclassical limit of Eq.~\eqref{eq:momentGenFct_exact}, we assume a Gaussian form for the photonic expansion coefficients in Eq.~\eqref{eq:coherentIntialState}
\begin{equation}
a_{ \boldsymbol  n } = \frac{1}{(2\pi )^{\frac{N_{\text{D}} }{4} } \sqrt{\text{det} \boldsymbol \Sigma } } e^{-\frac{1}{4}   \left( \boldsymbol n- \overline {\boldsymbol n} \right)   \boldsymbol \Sigma^{-2}  \left( \boldsymbol n- \overline {\boldsymbol n} \right)   }  e^{i \overline {\boldsymbol \varphi}  \cdot \boldsymbol  n},
\label{eq:coefficientsParameterization}
\end{equation}
where $\overline {\boldsymbol n}  = ( \overline n_1 ,\dots, \overline n_{N_\text{D}})  $  is a vector of mean-photon numbers of the photonic modes $k$ and  $\overline {\boldsymbol \varphi}  = ( \overline \varphi_1 ,\dots, \overline \varphi_{N_\text{D}} ) $ are the corresponding mean phases. The matrix $\boldsymbol \Sigma$ denotes the covariance matrix of the photonic probability distribution.

 The semiclassical limit  is defined as follows: First we introduce a scaled covariance matrix $ \tilde {\boldsymbol \Sigma}=\text{const.}$ by $\boldsymbol \Sigma =\sigma \tilde {\boldsymbol \Sigma}$ with scalar $\sigma$. The semiclassical limit is then controlled by the diverging $\sigma \rightarrow \infty$. 
 
 The Fourier transform of the expansions coefficients  defined in Eq.~\eqref{eq:coefficientsParameterization} is given by
\begin{eqnarray}
a_{ \boldsymbol  \varphi} 
&\rightarrow & \frac{\sqrt{ \text{det}\sqrt{2} \boldsymbol \Sigma }  }{(2\pi )^{\frac{N_{\text{D}}}{4} }  }    e^{-\left(  \boldsymbol \varphi  -\boldsymbol {\overline \varphi}     \right)    \boldsymbol \Sigma^2   \left(   \boldsymbol \varphi -\boldsymbol {\overline \varphi}\right)  }e^{i\left(  \boldsymbol  \varphi     -  \boldsymbol {\overline \varphi} \right)  \cdot  \boldsymbol {\overline n}}. \label{eq:expansionCoefficientGausFourier}
\end{eqnarray}
Inserting into Eq.~\eqref{eq:momentGenFct_exact},   the moment-generating function exactly factories
\begin{equation}
M_{\boldsymbol  \chi,\boldsymbol \xi}(t) =M_{\text{dy} ,\boldsymbol \chi,\boldsymbol \xi}^{(\text{exact})}(t)  M_{\boldsymbol  \chi,\boldsymbol \xi}(0). 
\label{eq:momentGenFkt_factorized}
\end{equation}
Thereby, the second term is the moment-generating function at time $t=0$, which for a  Gaussian state reads
\begin{equation}
M_{\boldsymbol  \chi,\boldsymbol \xi}(0) =  e^{-i \overline{ \boldsymbol  n } \cdot \boldsymbol \chi  -  \frac{1}{2}\boldsymbol \chi \boldsymbol \Sigma^2 \boldsymbol \chi  }.
\label{eq:initalMomentGenaratingFunction}
\end{equation}
The first term in Eq.~\eqref{eq:momentGenFkt_factorized} incorporates the light-matter interaction and is given by
\begin{multline}
M_{\text{dy} ,\boldsymbol \chi,\boldsymbol \xi }^{(\text{exact})}(t) 
=\int   \frac{ e^{-\left(  \boldsymbol  \varphi     -  \boldsymbol {\overline \varphi}   \right)    \boldsymbol \Sigma^2   \left(   \boldsymbol  \varphi     -  \boldsymbol {\overline \varphi}  \right)  }}{\mathcal N}   \\
 \times  \text{tr}  \left [   \hat   {\mathcal U}^{\boldsymbol \dagger}_{   \boldsymbol \varphi - \frac{\boldsymbol \chi}{2},- \frac{\boldsymbol \xi}{2}   } (t)   \hat    {\mathcal U}_{\boldsymbol \varphi + \frac{\boldsymbol \chi}{2} ,\frac{ \boldsymbol \xi}{2}  } (t)    \rho_{\text{MB}}(0)    \right  ]    d\boldsymbol \varphi
    ,   
\label{eq:dynCumulantGenFct}
\end{multline}
which  is the  dynamical moment-generating function. Thereby, $\mathcal N$ is the normalization of the Gaussian probability distribution. For $\sigma \rightarrow \infty$, the Gaussian function approaches a multi-dimensional delta function, such that the moment-generating function in the semiclassical limit becomes
\begin{equation}
M_{\text{dy},\boldsymbol \chi,\boldsymbol \xi,}( t)  =   \text{tr}  \left[   \rho_{\text{MB},\boldsymbol \chi,\boldsymbol \xi}(t)     \right],  
\label{eq:dynCumulantGenFct_PRFT}
\end{equation}
where we have defined the generalized density matrix of the matter and bath subsystems as
\begin{eqnarray}
  \rho_{\text{MB},\boldsymbol \chi,\boldsymbol \xi}(t) =  \hat    {\mathcal U}_{  \overline{\boldsymbol \varphi} + \frac{\boldsymbol \chi}{2} ,\frac{\boldsymbol \xi}{2} } (t)    \rho_{\text{MB}}(0)\hat   {\mathcal U}^{\boldsymbol \dagger}_{   \overline {\boldsymbol \varphi} - \frac{\boldsymbol \chi}{2} ,- \frac{\boldsymbol \xi}{2}  } (t)  .
\end{eqnarray}
Thus, by simulating the dynamics of $   \rho_{\text{MB},\boldsymbol \chi,\boldsymbol \xi}(t)$, we obtain the full statistical information of the photonic modes.

Because of the definition of the time-evolution operator in Eq.~\eqref{eq:countingFieldDependentTimeEvOperator}, it is not hard to see that the generalized reduced density matrix fulfills
\begin{eqnarray}
	\frac{d}{dt}\rho_{\text{MB},\boldsymbol \chi,\boldsymbol \xi} &=& -i \hat {\mathcal H}_{\text{MB},\overline {\boldsymbol \varphi} + \frac{\boldsymbol \chi}{2} , \frac{\boldsymbol \xi}{2}}(t) \rho_{\text{MB},\boldsymbol \chi,\boldsymbol \xi} \nonumber \\
	&+&i  \rho_{\text{MB},\boldsymbol \chi,\boldsymbol \xi}  \hat {\mathcal H}_{\text{MB},\overline {\boldsymbol \varphi} - \frac{\boldsymbol \chi}{2} ,- \frac{\boldsymbol \xi}{2} } (t).
	\label{eq:vanNeumanEquation}
\end{eqnarray}
In general, it will be impossible to solve this differential equation because of the macroscopic number of photonic modes in the bath, such that the Hilbert space is infinite dimensional. However, we can use that 
\begin{eqnarray}
M_{\text{dy},\boldsymbol \chi,\boldsymbol \xi,}( t) &=& \text{tr} \left\lbrace  \text{tr}_{\text{B}}    \left [\rho_{\text{MB},\boldsymbol \chi,\boldsymbol \xi}(t)   \right]  \right\rbrace  \nonumber \\
&=& \text{tr} \left[ \rho_{\boldsymbol \chi,\boldsymbol \xi}(t)   \right] ,
\end{eqnarray}
where $\text{tr}_{\text{B}}  \left[ \bullet \right]$  denotes the partial trace over the bath degrees of freedom.   In the second line, we have introduced the generalized reduced density matrix of the  matter system as
\begin{equation}
	 \rho_{\boldsymbol \chi,\boldsymbol \xi}(t)  \equiv \text{tr}_{\text{B}}    \left [\rho_{\text{MB},\boldsymbol \chi,\boldsymbol \xi}(t)   \right].
	 \label{eq:definition:reducedDensityMarix}
\end{equation}
As $\rho_{\boldsymbol \chi,\boldsymbol \xi}(t)$ has a small dimensionality, its time-evolution can be conveniently described by methods of open quantum systems. The specific equation of motion for $\rho_{\boldsymbol \chi,\boldsymbol \xi}(t)$ depends thereby on the system properties and the applied method. In Appendix ~\ref{app:dissipative_JcModel}, we will showcase how to construct and analyze a master equation  in second-order perturbation theory for a multi-mode Jaynes-Cummings model.
}

{\color{\markColorTwo}
	
\subsection{Error analysis}

\label{eq:errorAnalysis}

The error analysis of the semiclassical approximation follows  the same steps as the one in the closed system performed in the companion paper~\cite{Engelhardt2024c}. 
For simplicity, we perform the error analysis for a single coherent photonic mode with expansion coefficients $a_{n}  = \exp \left[ i\overline \varphi n - (n-\overline n)^2/(4\sigma^2) \right]/[\sigma^{1/2} (2\pi)^{1/4}]$. Moreover, without loss of generality, we set $\overline \varphi= 0$ and $\overline n =0$. In doing, the moment-generating function in Eq.~\eqref{eq:dynCumulantGenFct} in the long-time limit reads
\begin{equation}
M_{\text{dy} , \chi}^{(\text{exact})}(t\rightarrow \infty) 
= \frac{\sigma}{\sqrt{\pi} }\int     e^{\lambda_{\varphi , \chi} t }  e^{-\sigma^2 \varphi^2 } d  \varphi
,   
\label{eq:dynCumulantGenFctExact}
\end{equation}
where $\lambda_{\varphi , \chi}$  is the same as $ \lambda_{0;\boldsymbol \xi,\boldsymbol \chi} $ introduced in Eq.~\eqref{eq:cumulantGenertingFunctionLongTimes}, but we have explicitly emphasized the microscopic phase dependency before taking the semiclassical limit. The dynamical cumulants are defined as the derivatives of the dynamical moment-generating function, i.e., 
\begin{eqnarray}
m_{\text{dy} ,l}^{(\text{exact})}  (t)&\equiv &  \left. \frac{d^l}{d\chi^l} M_{\text{dy} , \chi}^{(\text{exact})}(t) \right|_{\chi=0}\nonumber \\
&=& \frac{\sigma}{\sqrt{\pi} } \int  d\varphi     \left( \lambda_{\varphi ,0}  ^\prime  t  \right)^l  
e^{-    \sigma^2   \varphi ^{ 2}  }+\mathcal O\left( t^{l-1}\right) .\nonumber \\
\label{eq:kth-moment_floquetExpansion} 
\end{eqnarray}
In the second line, we have evaluated the leading order terms in time, where $\lambda_{\varphi ,0}  ^\prime =d \lambda_{\varphi ,\chi =0}/(-id\chi) $. Expanding $\lambda_{\varphi ,0}  ^\prime $ around $\varphi=0$, i.e., $\lambda_{\varphi ,0}  ^\prime =  \lambda_{\varphi = 0 ,0}  ^\prime + (\frac{d}{d\varphi}\lambda_{\varphi=0 ,0}  ^\prime)\varphi + \frac{1}{2} (\frac{d^2}{d\varphi^2}\lambda_{\varphi=0 ,0}  ^\prime) \varphi^2 +\dots$, and evaluating the integral in Eq.~\eqref{eq:kth-moment_floquetExpansion}, we obtain
  \begin{multline}
m_{\text{dy} ,l}^{(\text{exact})}  (t)
=   \left( \lambda_{0 ,0}  ^\prime  t  \right)^l  \nonumber   \\
+  \left[ \left( \lambda_{0 ,0}  ^\prime   \right)^{l-1}    \frac{d^2\lambda_{0 ,0}  ^\prime}{d^2\varphi} +\left( \lambda_{0,0}  ^\prime   \right)^{l-2}   \left(  \frac{d\lambda_{0 ,0}  ^\prime}{d\varphi} \right)^2   \right] \frac{t^l}{\sigma^2} \nonumber  \\
+  \mathcal O\left( t^{l-1},\frac{t^l}{\sigma^4}\right) .
\label{eq:kth-moment_floquetExpansioner} 
\end{multline}
The first term is the semiclassical  contribution predicted by the PRFT, while the second term is the leading-order error term due to the semiclassical approximation. 

As explained in the companion paper (Ref.~\cite{Engelhardt2024c}, Appendix B), the dynamical moments and the cumulants share the same error scaling properties. For this reason, we can conclude that the exact cumulants are related to the PRFT cumulants  via
\begin{eqnarray}
\kappa_l^{(\text{exact})} (t)= \kappa_l (t)  + \mathcal O \left( \frac{t^l}{\sigma^2}   \right) .
\label{eq:errorScaling}
\end{eqnarray}
We see that the exact cumulants feature a super linear scaling because of the error terms $\propto t^l$. This is in contrast to the common behavior of the cumulants in the standard FCS, which  typically scale linearly with time due to a probabilistic dynamics. 

The error terms $\propto t^l$ are a consequence of the translational invariance of the operators in Sambe space in Eq.~\eqref{eq:def:sambeSpace}. For this reason, the Hamiltonian in Eq.~\eqref{eq:hamiltonian:interaction} is block diagonal in the basis 
\begin{eqnarray}
\left|\boldsymbol  \varphi \right>  &=&\frac{1  }{(2\pi)^{N_{ \text{D} }/2}}  \sum_{\boldsymbol n} e^{i\boldsymbol n \cdot \boldsymbol \varphi}\left| \boldsymbol n\right> ,
\end{eqnarray}
with the continuous multidimensinal quantum number  $\boldsymbol \varphi \in \left[ -\pi,\pi \right)^{\otimes N_{ \text{D} }} $.
Due to this block-diagonal structure, the Hamiltonian in Sambe space does not exhibit ergodicity because different $\left|\boldsymbol  \varphi \right> $ do not couple to each other.
Thus, the moment-generating function of each phase $\boldsymbol  \varphi $ in the integrand in Eq.~\eqref{eq:kth-moment_floquetExpansion}  evolves independently, i.e., non-ergodic, in time. This simple superposition of the moment-generating function finally gives rise to the superlinear growth of the error terms, which can be thus interpreted as a ballistic dynamics of the photonic wave function in the photon-number basis.

Consequently, we consider the superlinear error terms as an artifact of the representation of the photonic operators in Sambe space in Eq.~\eqref{eq:def:sambeSpace}. When taking the photon-number dependence of the creation and annihilation operators into account, the translational invariance is broken, and we restore ergodicity into the dynamics. This will result into a diffusive-like dynamics in the photon-number space.  Based on this discussion, we conjecture that the error terms will only grow linearly even for long time.}

{ \color{\markColorTwo}
\subsection{Mathematical properties}
	\label{sec:upperBoundOfTrace}
	
	In the following, we analyze some basic mathematical properties of the generalized density matrix and the resulting cumulants.
	First, we prove  the following property:
	\begin{equation}
	\left| M_{\boldsymbol  \chi,\boldsymbol \xi } (t) \right| = \left| M_{\text{dy}, \boldsymbol  \chi,\boldsymbol \xi } (t) \right|  \,\left| M_{\boldsymbol  \chi,\boldsymbol \xi} (0) \right|   \leq 1 ,
	\label{eq:traceNormBound}
	\end{equation}
	where we have used the exact factorization of the moment-generating function for a Gaussian photonic state in Eq.~\eqref{eq:momentGenFkt_factorized}. From Eq.~\eqref{eq:initalMomentGenaratingFunction} it is easy to see that $ \left| M_{\boldsymbol  \chi,\boldsymbol \xi } (0) \right|   \leq 1 $.
	
	To bound the dynamical moment-generating function, we describe the time evolution of the density matrix by the semiclassical time evolution operator in Eq.~\eqref{eq:countingFieldDependentTimeEvOperator}, where we have not traced out the incoherent photonic modes yet.  In doing so, we can represent the  moment-generating function as:
	\begin{multline}
\left| M_{\text{dy},\boldsymbol  \chi,\boldsymbol \xi } (t) \right| 
    = \left| \text{tr}\left[ \rho_{\text{MB},\boldsymbol  \chi,\boldsymbol \xi } (t) \right] \right| \\
	=\left|  \text{tr}\left[ \hat {\mathcal U}_{ \overline {\boldsymbol \varphi}  + \frac{\boldsymbol \chi}{2} ,\frac{\boldsymbol \xi}{2} } (t)  \rho_{\text{MB}}(0)  \hat  {\mathcal U}^{\dagger}_{ \overline {\boldsymbol \varphi}   - \frac{\boldsymbol \chi}{2},-\frac{\boldsymbol \xi}{2} } (t) \right] \right| , \nonumber \\
	= \left|  \sum_{\mu} p_\mu \text{tr}\left[ \hat {\mathcal U}_{ \overline {\boldsymbol \varphi}   + \frac{\boldsymbol \chi}{2} ,\frac{\boldsymbol \xi}{2} } (t)  \left|\mu \right>\left<\mu\right|  \hat  {\mathcal U}^{\dagger}_{ \overline {\boldsymbol \varphi}  - \frac{\boldsymbol \chi}{2} ,-\frac{\boldsymbol \xi}{2}  } (t)   \right] \right| \nonumber \\
	=  \left|  \sum_{\mu} p_\mu \text{tr}\left[  \left|\mu_{\overline {\boldsymbol \varphi}   + \frac{\boldsymbol \chi}{2},\frac{\boldsymbol \xi}{2}  } (t) \right>\left<\mu_{\overline {\boldsymbol \varphi}  - \frac{\boldsymbol \chi}{2},-\frac{\boldsymbol \xi}{2} } (t)  \right|  \right] \right|. \nonumber 
	\end{multline}
	In the third line, we have represented the initial density matrix of the system and the incoherent photonic modes in its diagonal basis, i.e., $ \rho(0) = \sum_{\mu} p_\mu \left|\mu \right>\left<\mu\right|  $.  In the third line, we have introduced the short-hand notation $\left|\mu_{\boldsymbol  \chi,\boldsymbol \xi } (t) \right>  =  \hat {\mathcal U}_{\boldsymbol  \chi,\boldsymbol \xi } (t) \left|\mu \right>  $. Applying the triangle inequality, we find
	\begin{eqnarray}
	\left| \text{tr}\left[ \rho_{\text{MB},\boldsymbol  \chi,\boldsymbol \xi } (t) \right] \right| &\leq &
	\sum_{\mu} p_\mu  \left| \text{tr}\left[  \left|\mu_{\overline {\boldsymbol \varphi}   + \frac{\boldsymbol \chi}{2} , \frac{\boldsymbol \xi}{2} } (t)   \right>\left<\mu_{\overline {\boldsymbol \varphi}  - \frac{\boldsymbol \chi}{2} ,-\frac{\boldsymbol \xi}{2}} (t) \right|  \right] \right| \nonumber \\
	&=&  \sum_{\mu} p_\mu  \left| \left<\mu_{\overline {\boldsymbol \varphi}  - \frac{\boldsymbol \chi}{2},-\frac{\boldsymbol \xi}{2} } (t) \mid \mu_{\overline {\boldsymbol \varphi}   + \frac{\boldsymbol \chi}{2},\frac{\boldsymbol \xi}{2}  (t)  } \right>   \right| \nonumber \\
	&\leq &  \sum_{\mu} p_\mu   =1,
	\end{eqnarray}
	where we have used that the inner product of the two states is bounded by $1$, as each is time evolved with a different unitary operator.

	Because of the definitions of the moment-generating function in Eq.~\eqref{eq:def:momentGeneratingFkt_exact} and the cumulant-generating function, it is easy to see that
	\begin{equation}
	K_{\boldsymbol  \chi,\boldsymbol \xi }  (t)   =  K_{-\boldsymbol  \chi,-\boldsymbol \xi } ^* (t) .
	\label{sec:symmetryRelationMomentGeneratingFunction}
	\end{equation}	
	By straightforward evaluation of the cumulants defined in Eq.~\eqref{eq:def:cumulantsGeneral} using Eq.~\eqref{sec:symmetryRelationMomentGeneratingFunction}, one immediately finds that all cumulants are real valued.
	 Moreover, because of Eq.~\eqref{eq:traceNormBound}, we verify that 
	\begin{eqnarray}
	\text{Re}\,  K_{\boldsymbol  \chi,\boldsymbol \xi }  (t)   \leq 0
	\label{eq:lambdaReal_bound}
	\end{eqnarray}
	for all ${\boldsymbol  \chi,\boldsymbol \xi } $. Consequently,  since $K_{\boldsymbol  \chi=0, \boldsymbol \xi=0 } (t)=0 $, the  second derivatives fulfills
		\begin{eqnarray}
      \kappa_2^{(k)}  (t)  \geq 0 , \label{eq:realEV_secondDer}
		\end{eqnarray}
	meaning, that the variance must be always positive.
}

\section{Dissipative Jaynes-Cummings model}
\label{app:dissipative_JcModel}

{\color{\markColorOne}
	
	As explained in Appendix~\ref{sec:semiclassicalLimit}, the moment-generating function is given by the trace of the reduced density matrix of the matter system in Eq.~\eqref{eq:definition:reducedDensityMarix}. Starting from the von-Neumann equation~\eqref{eq:vanNeumanEquation}, we show in this appendix how to construct and analyze a master equation in second-order perturbation theory of the system bath interaction in a multi-mode Jaynes-Cummings model.
	
	\subsection{Derivation of the master equation}

	For the Jaynes-Cummings model in Eq.~\eqref{eq:hamiltonian:JaynesCummings}, the corresponding semiclassical Hamiltonian  including the counting-fields as defined in  Eq.~\eqref{eq:countingFieldDependentHamiltonian} and appearing in von-Neumann equation~\eqref{eq:vanNeumanEquation}  is given by
	\begin{multline}
	\hat {\mathcal H}_{\text{MB}, \boldsymbol \varphi, \boldsymbol \xi} (t)= \frac{\epsilon}{2}\hat  \sigma_z  \\
	+  \sum_{k=1}^{2} g\left( \hat \sigma_-   \alpha_k e^{i \omega t- i\varphi_k  }   + \hat \sigma_+  \alpha_k e^{-i \omega t +i \varphi_k }  \right)  \\
	+ \sum_{k=3}^{N_{B}} \omega_k \hat a_k^\dagger  \hat a_k  +  \sum_{k=3}^{N_{\text{B}}} g_k  \left( \hat \sigma_{-}  \hat a_k^\dagger e^{- i \xi_k }   + \hat \sigma_{+}  \hat a_k  e^{ i \xi_k } \right)	 \\
	\end{multline}
	First, we transform the von-Neumann equation~\eqref{eq:vanNeumanEquation}  into a rotating frame defined by
	\begin{equation}
	\hat 	U_{\text{rot}  } = e^{-i\frac{\omega}{2} \hat \sigma_z t  },
	\label{eq:rotatingFrameDefinition}
	\end{equation}
	in which the semiclassical  Hamiltonian becomes
	\begin{eqnarray}
	\hat {\mathcal H}_{\text{MB}, \boldsymbol \varphi, \boldsymbol \xi} (t)&=&  \hat {\mathcal H}_{\text{MB}, \boldsymbol \varphi}^{(0)}   + \hat V_{\boldsymbol \xi} (t),
	\end{eqnarray}
	where
	\begin{eqnarray}
	\hat {\mathcal H}_{\text{MB}, \boldsymbol \varphi}^{(0)} &=& \hat {\mathcal H}_{ \boldsymbol \varphi}  +  \sum_{k=3}^{N_{\text{B}}+2} \omega_k \hat a_k^\dagger  \hat a_k \nonumber ,\\
	\hat {\mathcal H}_{ \boldsymbol \varphi} &=& \frac{ \epsilon_\Delta}{2}\hat \sigma_z +  \sum_{k=1}^{2} g_k\alpha_k \left( \hat \sigma_-    e^{-i\varphi_k }   + \hat \sigma_+   e^{ i\varphi_k  } \right) \nonumber ,\\ 
	\hat V_{\boldsymbol \xi} (t) &=&   \sum_{k=3}^{N_{\text{B}}+2} g_k  \left( \hat \sigma_{-}  \hat a_k^\dagger e^{- i \xi_k -i \omega t}   +  \hat \sigma_{+}  \hat a_k  e^{ i \xi_k +i \omega t} \right).	\nonumber \\
	\end{eqnarray}
	Thereby, we have introduced the detuning $\epsilon_{\Delta} = \epsilon -\omega $ which appears in the Hamiltonian of the isolated matter system $\hat {\mathcal H}_{ \boldsymbol \varphi}$. We note that the transformation into the rotating frame in Eq.~\eqref{eq:rotatingFrameDefinition} does not change the moment-generating function in Eq.~\eqref{eq:dynCumulantGenFct_PRFT} as the trace is invariant under unitary transformations, such that we do not have to carry-out the inverse transformation at the end of these derivations.

	Next, we carry out a transformation into an interaction picture defined by the unitary transformation
	\begin{eqnarray}
	\hat {\mathcal U}_{\boldsymbol \varphi}^{(0)}(t)  &=&  e^{-i \hat  {\mathcal H}_{\text{MB}, \boldsymbol \varphi}^{(0)} t  } .
	\end{eqnarray}
	Futhermore, we introduce the eigenstates  of the matter Hamiltonian as
	\begin{equation}
	\hat {\mathcal H}_{\boldsymbol \varphi} \left| u_{\mu,\boldsymbol \varphi}\right>  = E_{\mu,\boldsymbol \varphi} \left| u_{\mu,\boldsymbol \varphi} \right>,  
	\end{equation}
	for later purpose. More precisely, the transformation into the interaction picture is defined by the following relations
	\begin{eqnarray}
	\tilde \rho_{\text{MB},\boldsymbol  \chi,\boldsymbol \xi  }(t) &=& \hat {\mathcal U}_{\overline {\boldsymbol \varphi} + \frac{\boldsymbol \chi  }{2} }^{(0)\dagger}(t) \rho_{\text{MB},\boldsymbol  \chi,\boldsymbol \xi }(t)  \hat {\mathcal U}_{\overline {\boldsymbol \varphi} - \frac{\boldsymbol \chi  }{2} }^{(0)}(t),\nonumber \\
	\tilde A_{\boldsymbol \varphi} (t) &=& \hat {\mathcal U}_{\boldsymbol  \varphi  }^{(0)\dagger}(t) \hat A(t)  \hat {\mathcal U}_{\boldsymbol  \varphi }^{(0)}(t).
	\end{eqnarray}
	In particular, we define
	\begin{eqnarray}
	\tilde V_{\boldsymbol \varphi, \boldsymbol \xi} (t) &=& \hat {\mathcal U}_{\boldsymbol\varphi}^{(0)\dagger}(t) \hat V_{\boldsymbol \xi}(t) \hat   {\mathcal U}_{\boldsymbol\varphi}^{(0)}(t) .
	\end{eqnarray}
	In this interaction picture, the  von-Neumann equation~\eqref{eq:vanNeumanEquation} becomes
	\begin{eqnarray}
	i \frac{d}{dt}\tilde \rho_{\text{MB},\boldsymbol  \chi,\boldsymbol \xi  } (t)&=&  \tilde V_{\overline {\boldsymbol \varphi}+ \frac{\boldsymbol \chi  }{2}, \frac{\boldsymbol \xi}{2} }(t) \tilde \rho_{\text{MB},\boldsymbol  \chi,\boldsymbol \xi  }(t) \nonumber  \\
	&-&\tilde\rho_{\text{MB},\boldsymbol  \chi,\boldsymbol \xi }(t)  \tilde V_{\overline {\boldsymbol \varphi}- \frac{\boldsymbol \chi  }{2},-\frac{\boldsymbol \xi}{2}}(t).
	\label{eq:vonNeumanEq-intPicture}
	\end{eqnarray}
	Formal integration yields
	\begin{multline}
		\tilde \rho_{\text{MB},\boldsymbol  \chi,\boldsymbol \xi}(t) = \tilde \rho_{\text{MB},\boldsymbol  \chi,\boldsymbol \xi }(0)\\
		-i \int_{0}^{t}\left[  \tilde V_{\overline {\boldsymbol \varphi} + \frac{\boldsymbol \chi  }{2},\frac{\boldsymbol \xi}{2} }(t') \tilde \rho_{\text{MB},\boldsymbol  \chi,\boldsymbol \xi}(t') \right.\\\left.
		- \tilde \rho_{\text{MB},\boldsymbol  \chi,\boldsymbol \xi}(t')  \tilde V_{\overline {\boldsymbol \varphi}- \frac{\boldsymbol \chi  }{2},-\frac{\boldsymbol \xi}{2}}(t') \right] dt' ,  
	\end{multline}
	which we insert again into Eq.~\eqref{eq:vonNeumanEq-intPicture}:
	\begin{eqnarray}
	\frac{d}{dt}\tilde \rho_{\text{MB},\boldsymbol  \chi,\boldsymbol \xi }(t)& = & - i\tilde V_{\overline {\boldsymbol \varphi} + \frac{\boldsymbol \chi  }{2},\frac{\boldsymbol \xi}{2} }(t) \tilde \rho_{\text{MB},\boldsymbol  \chi,\boldsymbol \xi }(0)\nonumber \\
	&+& i \tilde \rho_{\text{MB},\boldsymbol  \chi,\boldsymbol \xi }(0)  \tilde V_{\overline {\boldsymbol \varphi}- \frac{\boldsymbol \chi  }{2},-\frac{\boldsymbol \xi}{2}}(t) \nonumber \\
	&-& \int_{0}^{t} \left[ \tilde V_{\overline {\boldsymbol \varphi} + \frac{\boldsymbol \chi  }{2},\frac{\boldsymbol \xi}{2} }(t)  \tilde V_{\frac{\boldsymbol \xi}{2},\overline {\boldsymbol \varphi} + \frac{\boldsymbol \chi  }{2}}(t')  \tilde \rho_{\text{MB},\boldsymbol  \chi,\boldsymbol \xi }(t')\nonumber \right. \\
	&-&  \tilde V_{\overline {\boldsymbol \varphi} + \frac{\boldsymbol \chi  }{2},\frac{\boldsymbol \xi}{2} }(t)    \tilde \rho_{\text{MB},\boldsymbol  \chi,\boldsymbol \xi }(t') \tilde V_{\overline {\boldsymbol \varphi}- \frac{\boldsymbol \chi  }{2},-\frac{\boldsymbol \xi}{2}}(t') \nonumber \\
	&-&  \tilde V_{\overline {\boldsymbol \varphi} + \frac{\boldsymbol \chi  }{2},\frac{\boldsymbol \xi}{2} }(t')  \tilde \rho_{\text{MB},\boldsymbol  \chi,\boldsymbol \xi } (t') \tilde V_{\overline {\boldsymbol \varphi}- \frac{\boldsymbol \chi  }{2},-\frac{\boldsymbol \xi}{2}}(t)\nonumber \\
	&+&  \left. \tilde \rho_{\text{MB},\boldsymbol  \chi,\boldsymbol \xi } (t') \tilde V_{\overline {\boldsymbol \varphi}- \frac{\boldsymbol \chi  }{2},-\frac{\boldsymbol \xi}{2}}(t')  \tilde V_{\overline {\boldsymbol \varphi}- \frac{\boldsymbol \chi  }{2},-\frac{\boldsymbol \xi}{2}}(t)\right]dt' \nonumber .\\
	\label{eq:secondOrderLiouvielleEquation}
	\end{eqnarray}
	To make process, we apply the Born approximation, which assumes that the density matrix of the matter system and the bath shall remain separable at all times such that $\tilde \rho_{\text{MB},\boldsymbol  \chi,\boldsymbol \xi }(t)   =   \tilde \rho_{\boldsymbol  \chi,\boldsymbol \xi} (t) \otimes \rho_{\text{B}}(0)$, where $\tilde \rho_{\boldsymbol  \chi,\boldsymbol \xi}(t)\equiv \text{tr}_{\text{B}}\left[\tilde \rho_{\text{MB},\boldsymbol  \chi,\boldsymbol \xi }(t)  \right]$  denotes the reduced density matrix of the matter system. This approximation is typically valid in the weak system-bath coupling regime.
	
	It is our aim to calculate the dynamics of $ \rho_{\boldsymbol  \chi,\boldsymbol \xi}(t)$. To this end, we now evaluate the partial trace over the bath states of Eq.~\eqref{eq:secondOrderLiouvielleEquation}. When evaluating the partial trace, the terms in the first line vanish as $\text{tr}_{\text{B}} \left[\tilde V_{\overline {\boldsymbol \varphi} + \frac{\boldsymbol \chi  }{2},\frac{\boldsymbol \xi}{2} }(t)  \rho_{\text{B}}(0)\right] =\text{tr}_{\text{B}} \left[ \rho_{\text{B}}(0)  \tilde V_{\overline {\boldsymbol \varphi}- \frac{\boldsymbol \chi  }{2},-\frac{\boldsymbol \xi}{2}}(t)  \right]= 0$, since the initial density matrix in Eq.~\eqref{eq:thermalInitalState} is diagonal in the photon-number basis. For the other terms, we can evaluate the time integration.
	For instance, the second term  in the integral explicitly reads
	\begin{eqnarray}
	\hat T &=&  \int_{0}^{t}  \text{tr}_{\text{B}} \left[  \tilde V_{\overline {\boldsymbol \varphi} + \frac{\boldsymbol \chi  }{2},\frac{\boldsymbol \xi}{2} }(t)  \tilde \rho_{\text{MB},\boldsymbol  \chi,\boldsymbol \xi}(t')  \tilde V_{\overline {\boldsymbol \varphi}- \frac{\boldsymbol \chi  }{2},-\frac{\boldsymbol \xi}{2}}(t')  \right] \nonumber \\
	&=&  \int_{0}^{t} dt'  \sum_k  g_k^2 \text{tr}_{\text{B}} \left[\hat a_k  \hat a_k^\dagger \tilde \rho_{\text{B}}(0) \right] e^{i\omega_k(t-t^\prime)  } e^{-i\omega (t-t')} e^{-i\xi_k}\nonumber \\
	&\times& \tilde \sigma_{-,\overline {\boldsymbol \varphi} + \frac{\boldsymbol \chi  }{2} }(t) \rho_{\boldsymbol  \chi,\boldsymbol \xi}(t') \tilde  \sigma_{+,\overline {\boldsymbol \varphi} - \frac{\boldsymbol \chi  }{2}}(t')   +\dots ,
	\end{eqnarray}
	where the dots represent three similar terms.  As the bath is initially in a thermal state, we can evaluate $ \text{tr}_{\text{B}} \left[\hat a_k^\dagger  \hat a_k \tilde \rho_{\text{B}}(0) \right] = n_{\text{B}} (\omega_k)$, where $  n_{\text{B}} (\omega)$ denotes the common Bose distribution function. From now on, we focus on counting the total photon number change in the incoherent modes, which can be calculated by setting $\xi_k =\xi$, i.e., we do not distinquish the photon number of the specific modes.  Moreover we define the bath correlation  function $\Gamma(t) = \sum_{k=3}^{N_{\text{B}}+2 }g_k^2 \left[ n_{\text{B}} (\omega_k)+1\right] e^{i\omega_kt} $. In doing so,  we obtain
	\begin{eqnarray}
	\hat T &=& \int_{0}^{t} dt'  \sum_k \Gamma(t-t')  e^{i\omega(t'-t)} e^{-i\xi}\nonumber \\
	&\times& \tilde \sigma_{-,\overline {\boldsymbol \varphi} + \frac{\boldsymbol \chi  }{2} }(t) \tilde \rho_{\boldsymbol  \chi,\boldsymbol \xi}(t') \tilde  \sigma_{+,\overline {\boldsymbol \varphi} - \frac{\boldsymbol \chi  }{2}}(t')   +\dots \nonumber, \\
	&=& \int_{0}^{t} dt'   \Gamma(t-t')  e^{i\omega (t'-t)} e^{-i\xi}\nonumber \\
	&\times& \tilde \sigma_{-,\overline {\boldsymbol \varphi} + \frac{\boldsymbol \chi  }{2} }(t) \tilde \rho_{\boldsymbol  \chi,\boldsymbol \xi}(t) \tilde  \sigma_{+,\overline {\boldsymbol \varphi} - \frac{\boldsymbol \chi  }{2}}(t')   +\dots \nonumber, \\
	&\approx& \int_{0}^{\infty} dt'   \Gamma(t')  e^{-i\omega t'} e^{-i\xi}\nonumber \\
	&\times& \tilde \sigma_{-,\overline {\boldsymbol \varphi} + \frac{\boldsymbol \chi  }{2} }(t) \tilde \rho_{\boldsymbol  \chi,\boldsymbol \xi}(t) \tilde  \sigma_{+,\overline {\boldsymbol \varphi} - \frac{\boldsymbol \chi  }{2}}(t-t')   +\dots \nonumber. \\
	\end{eqnarray}
	In the second equality, we have carried out the Markov approximation $\tilde \rho_{\boldsymbol  \chi,\boldsymbol \xi}(t') \rightarrow \tilde \rho_{\boldsymbol  \chi,\boldsymbol \xi}(t)$. Using now the expansion of the system operators in the eigenbasis, i.e.,
	\begin{equation}
	\tilde  \sigma_{\pm,\boldsymbol \varphi }  (t) \equiv \sum_{\mu_1,\mu_2} s_{\pm,\mu_1,\mu_2}^{(\boldsymbol \varphi)}\left| u_{\mu_1,\boldsymbol \varphi}\right> \left< u_{\mu_2,\boldsymbol \varphi} \right| e^{i(E_{\mu_1,\boldsymbol \varphi} -E_{\mu_2,\boldsymbol \varphi}    )t} ,
	\end{equation}
	where the $s_{\pm,\mu_1,\mu_2}^{(\boldsymbol \varphi)} $ denote the expansion coefficients in the basis $\left| u_{\mu,\boldsymbol \varphi}\right>  $,  we obtain
	\begin{eqnarray}
	\hat T &=&  \int_{0}^{\infty} dt'  \Gamma(t') e^{-i\omega t'}   e^{-i\xi} \nonumber  \\
	&\times&  \sum_{\mu_1,\mu_2}   s_{-,\mu_1,\mu_2}^{(\overline {\boldsymbol \varphi} + \frac{\boldsymbol \chi  }{2})} \left| u_{\mu_1,\overline {\boldsymbol \varphi} + \frac{\boldsymbol \chi  }{2}} \right> \left< u_{\mu_2,\overline {\boldsymbol \varphi} + \frac{\boldsymbol \chi  }{2}} \right| \nonumber \\
	&\times&  e^{i(E_{\mu_1,\overline {\boldsymbol \varphi} + \frac{\boldsymbol \chi  }{2}}  -E_{\mu_2,\overline {\boldsymbol \varphi} + \frac{\boldsymbol \chi  }{2}}    )t}  \nonumber  \\
	&\times& \tilde \rho_{\boldsymbol  \chi,\boldsymbol \xi}(t)  \nonumber \\
	&\times& \sum_{\mu_3,\mu_4}    s_{+,\mu_3,\mu_4}^{(\overline {\boldsymbol \varphi}- \frac{\boldsymbol \chi  }{2})}\left| u_{\mu_3,\overline {\boldsymbol \varphi} - \frac{\boldsymbol \chi  }{2}}\right> \left< u_{\mu_4,\overline {\boldsymbol \varphi}- \frac{\boldsymbol \chi  }{2}}\right| \nonumber \\
	 &\times& e^{i(E_{\mu_3,\overline {\boldsymbol \varphi}- \frac{\boldsymbol \chi  }{2}}  -E_{\mu_4,\overline {\boldsymbol \varphi} - \frac{\boldsymbol \chi  }{2}}    )(t-t')} \nonumber \\
	&+&\dots \ \nonumber \\ 
	&=&  \gamma \left(\omega+ E_{\mu_3,\overline {\boldsymbol \varphi}+ \frac{\boldsymbol \chi  }{2}}  -E_{\mu_4,\overline {\boldsymbol \varphi}+ \frac{\boldsymbol \chi  }{2}}  \right)  e^{-i\xi}  \nonumber \\
	&\times& \sum_{\mu_1,\mu_2}   s_{-,\mu_1,\mu_2}^{(\overline {\boldsymbol \varphi} + \frac{\boldsymbol \chi  }{2})}\left| u_{\mu_1,\overline {\boldsymbol \varphi} + \frac{\boldsymbol \chi  }{2}}\right> \left< u_{\mu_2,\overline {\boldsymbol \varphi} + \frac{\boldsymbol \chi  }{2}}\right|  \nonumber \\
	&\times& e^{i(E_{\mu_1,\overline {\boldsymbol \varphi} + \frac{\boldsymbol \chi  }{2}}  -E_{\mu_2,\overline {\boldsymbol \varphi} + \frac{\boldsymbol \chi  }{2}}    )t}  \nonumber  \\
	&\times& \tilde \rho_{\boldsymbol  \chi,\boldsymbol \xi}(t) \nonumber \\
	&\times& \sum_{\mu_3,\mu_4}  \hat  s_{+,\mu_3,\mu_4}^{(\overline {\boldsymbol \varphi} - \frac{\boldsymbol \chi  }{2})}\left| u_{\mu_3,\overline {\boldsymbol \varphi} - \frac{\boldsymbol \chi  }{2}}\right> \left< u_{\mu_4,\overline {\boldsymbol \varphi}- \frac{\boldsymbol \chi  }{2}}\right|  \nonumber \\
	 	&\times& e^{i( E_{\mu_3,\overline {\boldsymbol \varphi} - \frac{\boldsymbol \chi  }{2}}  -E_{\mu_4,\overline {\boldsymbol \varphi} - \frac{\boldsymbol \chi  }{2}}    )t }\nonumber \\
	&&+\dots 
	\end{eqnarray}
	In the second equality, we have introduced  the Fourier transform of the bath correlation function  $\gamma(\omega)= \int dt \Gamma(t) e^{i\omega t} $. 
	To further simplify this expression, we assume that
	$ \gamma \left(\omega+ E_{\mu_3,\boldsymbol \varphi}  -E_{\mu_4,\boldsymbol \varphi}  \right)   \approx   \gamma \left(\omega \right)   $
	which is valid in the broadband regime. Transforming this term back into the lab frame, it becomes
	\begin{eqnarray}
	&&\hat {\mathcal U}_{\overline{\boldsymbol\varphi} + \frac{\boldsymbol \chi  }{2} }^{(0)}  (t) \hat T \hat {\mathcal U}_{\overline{\boldsymbol\varphi}- \frac{\boldsymbol \chi  }{2} }^{(0)^\dagger}(t) \\ 
	&\approx& 
	\gamma \left(\omega \right)   e^{-i\xi} \nonumber\\
	&\times&\sum_{\mu_1,\mu_2}  \hat  s_{-,\mu_1,\mu_2}^{(\overline{\boldsymbol\varphi}+ \frac{\boldsymbol \chi  }{2} )}\left| u_{\mu_1,\overline{\boldsymbol\varphi} + \frac{\boldsymbol \chi  }{2}}\right> \left< u_{\mu_2,\overline{\boldsymbol\varphi} + \frac{\boldsymbol \chi  }{2}}\right|   \nonumber\\ 
	&\times& \rho_{\boldsymbol  \chi,\boldsymbol \xi}(t)   \sum_{\mu_3,\mu_4}   \hat  s_{+,\mu_3,\mu_4}^{(\overline{\boldsymbol\varphi} - \frac{\boldsymbol \chi  }{2})}\left| u_{\mu_3,\overline{\boldsymbol\varphi} - \frac{\boldsymbol \chi  }{2}}\right> \left< u_{\mu_4,\overline{\boldsymbol\varphi} - \frac{\boldsymbol \chi  }{2}}\right|   +\dots  \nonumber \\ 
	&=& \gamma\left(\omega \right)    e^{-i\xi}  \hat  \sigma_{-}       \rho_{\boldsymbol  \chi,\boldsymbol \xi}(t)  \hat  \sigma_{+} +\dots \nonumber
	\end{eqnarray}
	Carrying out similar calculations for the other terms in Eq.~\eqref{eq:secondOrderLiouvielleEquation}, assuming zero temperature [i.e., $ n_{\text{B}} \left(\omega \right)=0$], setting $\gamma(\omega)=\gamma$, and putting everything together, we finally obtain the generalized Bloch equation in Eq.~\eqref{sec:quantumMasterEquationJC}.

}

\subsection{Construction of the equations of motion}

\label{app:dissipative_JcModel:EOM}

Here we provide a step-by-step construction  of the Liouvillian in Eq.~\eqref{eq:effectLiouvillianJCmodel}. To this end, we define the Pauli-operator expectation value as $\rho_\alpha  \equiv \text{tr} \left[ \rho_{\boldsymbol \chi,\boldsymbol \xi} \hat \sigma_{\alpha} \right]$ with $\alpha \in\lbrace0, \text{x},\text{y} ,\text{z}\rbrace$, and use the relation $	\hat \sigma_{\text{x} } \hat \sigma_{\text{y} } = i \hat \sigma_{\text{z} } $. Thereby, $\text{tr}\left[\bullet \right]$ denotes the trace over the spin degrees of freedom. We note that $\rho_{0}=1$ for $\chi=0$ and $\xi=0$, which encodes the trace conversation of the density matrix.

We start to evaluate the terms in the $\hat \sigma_-$ dissipator in Eq.~\eqref{eq:dissipatorJCmodel}, finding
\begin{widetext}
	\begin{eqnarray}
	\text{tr} \left[ 	\mathcal D_{\xi } \left[\hat \sigma_- \right]\rho_{\boldsymbol \chi,\boldsymbol \xi} \right]  &=&  e^{i\xi } \text{tr} \left[  \hat \sigma_- \rho_{\boldsymbol \chi,\boldsymbol \xi}  \hat \sigma_+  \right] -\frac{1}{2} \text{tr} \left[\hat \sigma_+  \hat \sigma_- \rho_{\boldsymbol \chi,\boldsymbol \xi} \right] -\frac{1}{2}\text{tr} \left[ \rho_{\boldsymbol \chi,\boldsymbol \xi}  \hat \sigma_+  \hat \sigma_-  \right] \nonumber  \\
	&= &  e^{i\xi }  \text{tr} \left[ \hat \sigma_+  \hat \sigma_-  \rho_{\boldsymbol \chi,\boldsymbol \xi} \right]  - \frac{1}{4} \text{tr} \left[ \left( 1 + \hat \sigma_{\text{z}} \right) \rho_{\boldsymbol \chi,\boldsymbol \xi} \right] - \frac{1}{4} \text{tr} \left[  \left( 1 + \hat \sigma_{\text{z}} \right) \rho_{\boldsymbol \chi,\boldsymbol \xi} \right] \nonumber  \\
	&= & \frac{1}{2} \left( e^{i\xi } -1\right) \left( \rho_0 + \rho_{\text{z}} \right) , \nonumber  \\
	\text{tr} \left[ 	\mathcal D_{\xi } \left[\hat \sigma_- \right]\rho_{\boldsymbol \chi,\boldsymbol \xi} \hat \sigma_{\text{x}} \right] &=&  e^{i\xi }  \text{tr} \left[ \hat \sigma_- \rho_{\boldsymbol \chi,\boldsymbol \xi} \hat \sigma_+  \hat \sigma_{\text{x}} \right]   -\frac{1}{2}\text{tr} \left[ \hat \sigma_+  \hat \sigma_- \rho_{\boldsymbol \chi,\boldsymbol \xi} \hat \sigma_{\text{x}}\right] -\frac{1}{2}\text{tr} \left[ \rho_{\boldsymbol \chi,\boldsymbol \xi}  \hat \sigma_+  \hat \sigma_-  \hat \sigma_{\text{x}}\right]  \nonumber  \\
	&=&  e^{i\xi }\frac{1}{4} \text{tr} \left[ \left( \hat \sigma_{\text{x}}  +i\hat \sigma_{\text{y}} \right) \hat \sigma_{\text{x}}  \left( \hat \sigma_{\text{x}}  -  i\hat \sigma_{\text{y}} \right) \rho_{\boldsymbol \chi,\boldsymbol \xi} \right] - \frac{1}{4} \text{tr} \left[ \hat \sigma_{\text{x}} \left( 1 + \hat \sigma_{\text{z}} \right)\rho_{\boldsymbol \chi,\boldsymbol \xi}\right] - \frac{1}{4}\text{tr} \left[  \left( 1 + \hat \sigma_{\text{z}} \right) \hat \sigma_{\text{x}} \rho_{\boldsymbol \chi,\boldsymbol \xi} \right] \nonumber  \\
	&= & e^{i\xi }\frac{1}{4} \left(\text{tr} \left[  \hat \sigma_{\text{x}}\rho_{\boldsymbol \chi,\boldsymbol \xi} \right]  - \text{tr} \left[ \hat \sigma_{\text{x}}\rho_{\boldsymbol \chi,\boldsymbol \xi}  \right] \right)  -\frac{1}{2} \text{tr} \left[ \hat \sigma_{\text{x}}\rho_{\boldsymbol \chi,\boldsymbol \xi} \right]   = -\frac{1}{2}\rho_{\text{x}}  ,\nonumber  \\
	\text{tr} \left[ 	\mathcal D_{\xi } \left[\hat \sigma_- \right]\rho_{\boldsymbol \chi,\boldsymbol \xi} \hat \sigma_{\text{y}} \right]  &=&  e^{i\xi } \text{tr} \left[  \hat \sigma_- \rho_{\boldsymbol \chi,\boldsymbol \xi} \hat \sigma_+ \hat \sigma_{\text{y}} \right]  -\frac{1}{2}\left[ \hat \sigma_+  \hat \sigma_- \rho_{\boldsymbol \chi,\boldsymbol \xi} \hat \sigma_{\text{y}}\right] -\frac{1}{2} \text{tr} \left[ \rho_{\boldsymbol \chi,\boldsymbol \xi} \hat \sigma_+    \hat \sigma_-  \hat \sigma_{\text{y}}\right]  \nonumber  \\
	&= &  e^{i\xi } \frac{1}{4} \text{tr} \left[ \left( \hat \sigma_{\text{x}}  +  i\hat \sigma_{\text{y}} \right) \hat \sigma_{\text{y}}  \left( \hat \sigma_{\text{x}}  -  i\hat \sigma_{\text{y}} \right)\rho_{\boldsymbol \chi,\boldsymbol \xi} \right]  -\frac{1}{4} \text{tr} \left[ \hat \sigma_{\text{y}} \left( 1 + \hat \sigma_{\text{z}} \right) \rho_{\boldsymbol \chi,\boldsymbol \xi} \right] -  \frac{1}{4} \text{tr} \left[ \left( 1 + \hat \sigma_{\text{z}} \right) \hat \sigma_{\text{y}}\rho_{\boldsymbol \chi,\boldsymbol \xi}  \right] \nonumber  \\
	&= & e^{i\xi }  \frac{1}{4} \text{tr} \left[\left( -\hat \sigma_{\text{y}}  + \hat \sigma_{\text{y}}  \right)\rho_{\boldsymbol \chi,\boldsymbol \xi}  \right] -  \frac{1}{2} \text{tr} \left[ \hat \sigma_{\text{y}}\rho_{\boldsymbol \chi,\boldsymbol \xi} \right]   = -\frac{1}{2}\rho_{\text{y}}  ,\nonumber \\
	\text{tr} \left[  \mathcal D_{\xi } \left[\hat \sigma_- \right]\rho_{\boldsymbol \chi,\boldsymbol \xi} \hat \sigma_{\text{z}}  \right]  &=&  e^{-i\xi }  \text{tr} \left[  \hat \sigma_- \rho_{\boldsymbol \chi,\boldsymbol \xi} \hat \sigma_+ \hat \sigma_{\text{z}} \right]  -\frac{1}{2}\text{tr} \left[   \hat \sigma_+  \hat \sigma_- \rho_{\boldsymbol \chi,\boldsymbol \xi} \hat \sigma_{\text{z}}\right] -\frac{1}{2} \text{tr} \left[ \rho_{\boldsymbol \chi,\boldsymbol \xi} \hat \sigma_+    \hat \sigma_-  \hat \sigma_{\text{z}} \right]\nonumber   \\
	&= &  e^{-i\xi } \frac{1}{4} \text{tr} \left[ \left( \hat \sigma_{\text{x}}  +i\hat \sigma_{\text{y}} \right) \hat \sigma_{\text{z}}  \left( \hat \sigma_{\text{x}}  -  i\hat \sigma_{\text{y}} \right) \rho_{\boldsymbol \chi,\boldsymbol \xi} \right] - \frac{1}{4} \text{tr} \left[\hat \sigma_{\text{z}} \left( 1 + \hat \sigma_{\text{z}} \right) \rho_{\boldsymbol \chi,\boldsymbol \xi} \right]- \frac{1}{4} \text{tr} \left[ \left( 1 + \hat \sigma_{\text{z}} \right) \hat \sigma_{\text{z}}  \rho_{\boldsymbol \chi,\boldsymbol \xi} \right] \nonumber  \\
	&= & e^{-i\xi }\frac{1}{4} \text{tr} \left[ \left( -2 \hat \sigma_{\text{z}}  -  2  \right)\rho_{\boldsymbol \chi,\boldsymbol \xi}\right]  - \frac{1}{2} \text{tr} \left[ \left( \hat \sigma_{\text{z}}  + 1 \right)\rho_{\boldsymbol \chi,\boldsymbol \xi} \right]  =-\frac{1}{2} \left(e^{i\xi } + 1 \right) \left( \rho_{\text{z}}  + \rho_0  \right) .
	\end{eqnarray}
	For a notation reason, we define 
	\begin{eqnarray}
	\epsilon_{\Delta} &=& \epsilon  -\omega \nonumber, \\
	\Omega_{\text{x}}( \boldsymbol \varphi) &=& \Omega_1 \cos(\varphi_1) + \Omega_2 \cos(\varphi_2), \nonumber   \\
	\Omega_{\text{y}}( \boldsymbol \varphi) &=&- \Omega_1 \sin(\varphi_1) - \Omega_2 \sin(\varphi_2) ,
	\end{eqnarray}
	and evaluate the Hermitian part in the Bloch equation determined by the semiclassical Hamiltonian Eq.~\eqref{eq:semiclassicalJaymesCummingsModel}. Putting everything together, we find
	\begin{eqnarray}
	\frac{d}{dt}   \rho_0  &=& -i \frac{ \epsilon_{\Delta} }{2} \rho_{\text{z}} -i \frac{ \Omega_{\text{x}}\left( \overline {\boldsymbol \varphi}  + \frac{\boldsymbol \chi  }{2} \right)}{2}  \rho_{\text{x}}- i \frac{ \Omega_{\text{y}}\left(\overline {\boldsymbol \varphi}  + \frac{\boldsymbol \chi  }{2} \right)}{2}  \rho_{\text{y}} +i \frac{  \epsilon_{\Delta} }{2} \rho_{\text{z}} +i\frac{\Omega_{\text{x}}\left( \overline {\boldsymbol \varphi} - \frac{\boldsymbol \chi  }{2} \right)}{2}   \rho_{\text{x}}   +i\frac{\Omega_{\text{y}}\left(\overline {\boldsymbol \varphi}  - \frac{\boldsymbol \chi  }{2}  \right)}{2}  \rho_{\text{y}}\nonumber  \\
	&+& \frac{1}{2}\gamma\left( e^{-i\xi } -1\right) \left( \rho_0 + \rho_{\text{z}}\right)   \nonumber ,\\
	\frac{d}{dt}   \rho_{\text{x}}  &=&-   \epsilon_{\Delta}  \rho_{\text{y}} -i \frac{ \Omega_{\text{x}}\left(\overline{\boldsymbol\varphi}  + \frac{\boldsymbol \chi  }{2} \right)}{2}  \rho_0+ \frac{ \Omega_{\text{y}}\left(\boldsymbol \chi\right)}{2}  \rho_{\text{z}}  +i \frac{ \Omega_{\text{x}}\left( \overline{\boldsymbol\varphi} - \frac{\boldsymbol \chi  }{2}  \right)}{2}   \rho_0+ \frac{ \Omega_{\text{y}}\left(\overline{\boldsymbol\varphi}- \frac{\boldsymbol \chi  }{2}  \right)}{2}  \rho_{\text{z}} -\frac{1}{2} \gamma \rho_{\text{x}}  ,\nonumber \\
	\frac{d}{dt}   \rho_{\text{y}}  &=& \epsilon_{\Delta}  \rho_{\text{x}} - \frac{ \Omega_{\text{x}}\left( \overline{\boldsymbol\varphi} + \frac{\boldsymbol \chi  }{2} \right)}{2}  \rho_{\text{z}} -i\Omega_{\text{y}}\left(\overline{\boldsymbol\varphi} + \frac{\boldsymbol \chi  }{2} \right)  \rho_0 - \frac{\Omega_{\text{x}}\left( \overline{\boldsymbol\varphi} - \frac{\boldsymbol \chi  }{2} \right)}{2}   p_z+i \frac{ \Omega_{\text{y}}\left(\overline{\boldsymbol\varphi} - \frac{\boldsymbol \chi  }{2} \right)}{2}  \rho_0 -\frac{1}{2} \gamma \rho_{\text{y}}  ,\nonumber \\
	\frac{d}{dt}   \rho_{\text{z}}  &=&  \Omega_{\text{x}}\left( \overline{\boldsymbol\varphi} + \frac{\boldsymbol \chi  }{2} \right)  \rho_{\text{y}} -\frac{ \Omega_{\text{y}}\left(\overline{\boldsymbol\varphi} + \frac{\boldsymbol \chi  }{2} \right)}{2}  \rho_{\text{x}} + \Omega_{\text{x}}\left( \overline{\boldsymbol\varphi} - \frac{\boldsymbol \chi  }{2} \right)   \rho_{\text{y}} - \frac{ \Omega_{\text{y}}\left(\overline{\boldsymbol\varphi} - \frac{\boldsymbol \chi  }{2} \right)}{2}  \rho_{\text{x}}  \nonumber   \\
	&-& \frac{1}{2}\gamma \left( e^{-i\xi } +1\right)  \left( \rho_{\text{z}} + \rho_0    \right)  \nonumber ,\\
	\end{eqnarray}
	which after some simplifications reads 
	\begin{eqnarray}
	\frac{d}{dt}   \rho_0  &=& i  \frac{1}{2} \left [ \Omega_{\text{x}}\left( \overline{\boldsymbol\varphi} - \frac{\boldsymbol \chi  }{2}  \right)   - \Omega_{\text{x}}\left( \overline{\boldsymbol\varphi} + \frac{\boldsymbol \chi  }{2}  \right)   \right] \rho_{\text{x}} +  i \frac{1}{2} \left[\Omega_{\text{y}}\left(\overline{\boldsymbol\varphi} - \frac{\boldsymbol \chi  }{2}  \right)   - \Omega_{\text{y}}\left(\overline{\boldsymbol\varphi} + \frac{\boldsymbol \chi  }{2}  \right)      \right]\rho_{\text{y}}  \nonumber  \\
	&+& \gamma_{\xi,-}  \left( \rho_0 + \rho_{\text{z}}\right)  ,  \nonumber \\
	\frac{d}{dt}   \rho_{\text{x}}  &=& -  \epsilon_{\Delta} \rho_{\text{y}} + i  \frac{1}{2} \left [\Omega_{\text{x}}\left( \overline{\boldsymbol\varphi} - \frac{\boldsymbol \chi  }{2}  \right) -\Omega_{\text{x}}\left( \overline{\boldsymbol\varphi} + \frac{\boldsymbol \chi  }{2} \right)   \right]   \rho_0+  \frac{1}{2} \left[ \Omega_{\text{y}}\left(\overline{\boldsymbol\varphi}- \frac{\boldsymbol \chi  }{2}  \right)   + \Omega_{\text{y}}\left(\overline{\boldsymbol\varphi}+ \frac{\boldsymbol \chi  }{2} \right) \right]       \rho_{\text{z}}  -\frac{1}{2} \gamma \rho_{\text{x}} , \nonumber \\
	\frac{d}{dt}   \rho_{\text{y}}  &=&   \epsilon_{\Delta} \rho_{\text{x}} -   \frac{1}{2} \left [\Omega_{\text{x}}\left( \overline{\boldsymbol\varphi} - \frac{\boldsymbol \chi  }{2}  \right) + \Omega_{\text{x}}\left(\overline{\boldsymbol\varphi} + \frac{\boldsymbol \chi  }{2} \right)   \right]    \rho_{\text{z}} +i \frac{1}{2} \left[ \Omega_{\text{y}}\left(\overline{\boldsymbol\varphi}- \frac{\boldsymbol \chi  }{2} \right)   - \Omega_{\text{y}}\left(\overline{\boldsymbol\varphi} + \frac{\boldsymbol \chi  }{2} \right)    \right]     \rho_0 -\frac{1}{2} \gamma \rho_{\text{y}}  ,\nonumber \\
	\frac{d}{dt}   \rho_{\text{z}}  &=&   \frac{1}{2} \left [\Omega_{\text{x}}\left( \overline{\boldsymbol\varphi}- \frac{\boldsymbol \chi  }{2}  \right)  +\Omega_{\text{x}}\left( \overline{\boldsymbol\varphi} + \frac{\boldsymbol \chi  }{2} \right)   \right]    \rho_{\text{y}} -  \frac{1}{2} \left[ \Omega_{\text{y}}\left(\overline{\boldsymbol\varphi} - \frac{\boldsymbol \chi  }{2}  \right)   + \Omega_{\text{y}}\left(\overline{\boldsymbol\varphi} + \frac{\boldsymbol \chi  }{2}  \right) \right]    \rho_{\text{x}} \nonumber   \\
	&-&   \gamma_{\xi,+}  \left( \rho_{\text{z}}  + \rho_0   \right)  .
	\end{eqnarray}
\end{widetext}

For notational convenience, we define
\begin{eqnarray}
c_{\boldsymbol \chi,\pm}  &=&\Omega_{\text{x}}\left( \overline{\boldsymbol\varphi}- \frac{\boldsymbol \chi  }{2}  \right) \pm\Omega_{\text{x}}\left( \overline{\boldsymbol\varphi} + \frac{\boldsymbol \chi  }{2}  \right)\nonumber  , \\
s_{\boldsymbol \chi,\pm}  &=&\Omega_{\text{y}} \left( \overline{\boldsymbol\varphi} - \frac{\boldsymbol \chi  }{2} \right) \pm\Omega_{\text{y}}\left( \overline{\boldsymbol\varphi} + \frac{\boldsymbol \chi  }{2}  \right),
\nonumber \\
\gamma_{\xi,\pm}   &=&  \frac{\gamma}{2} \left( e^{-i\xi } \pm 1\right).  \nonumber
\end{eqnarray}
In doing so, the von-Neumann equation can be written as
\begin{equation}
\frac{d}{dt} 
\left(
\begin{array}{c}
\rho_0 \\
\rho_{\text{x}} \\
\rho_{\text{y}} \\
\rho_{\text{z}} \\
\end{array}
\right)
=
\mathcal A_{\boldsymbol \chi, \xi}
\left(
\begin{array}{c}
\rho_0 \\
\rho_{\text{x}} \\
\rho_{\text{y}} \\
\rho_{\text{z}} \\
\end{array}
\right),
\end{equation}
where the effective Liouvillian is given by
\begin{equation}
\mathcal A_{ \boldsymbol \chi, \xi} = \left( 
\begin{array}{cccc} 
\gamma_{\xi,-}  & ic_{\boldsymbol \chi,-} &  is_{\boldsymbol \chi,-}  &2   \gamma_{\xi,-} \\
ic_{\boldsymbol \chi,-} &  -\frac{\gamma}{2} & - \epsilon_{\Delta} & s_{\boldsymbol \chi,+} \\
is_{\boldsymbol \chi,-} & \epsilon_{\Delta}  & -\frac{\gamma}{2} & -c_{\boldsymbol \chi,+} \\
-\gamma_{\xi,+} & -s_{\boldsymbol \chi,+} & c_{\boldsymbol \chi,+} &- \gamma_{\xi,+}
\end{array}
\right).
\end{equation}

\subsection{Fluctuations}

Using the exact expression for the characteristic polynomial in Eq.~\eqref{eq:simplifiedCharPolJCmodel}, we can deploy the method  introduced in Appendix~\ref{sec:cumulantEvaluationTrunction} to calculate the  first and second cumulants of the photon flux probability distribution. As the calculation is straightforward but lengthy, we take advantage of the computer algebra software \textit{SymPy}. The resulting expression for the photon flux is given in Eq.~\eqref{eq:photonFlux}, while we find for the corresponding variance
\begin{equation}
\sigma_{I,1}^2	= \frac{	c_{\text{num}}}{c_{\text{den} }},
\label{eq:fluxNoise:twolevelsystem:exact}
\end{equation}
where the numerator reads 
\begin{eqnarray}
c_{\text{num}}  &=&   i  \frac{1}{64} \gamma \Omega_{1}^2 \left(i \Omega_{1} - \Omega_{2} \sin{\left(\overline \varphi \right)} + i \Omega_{2} \cos{\left(\overline \varphi \right)}\right) 
\nonumber \\ &&\times  \left(8 \epsilon_{\Delta} \Omega_{2} \sin{\left(\overline \varphi \right)} + 10 \gamma \Omega_{1} + 16 \gamma \Omega_{2} \cos{\left(\overline \varphi \right)}\right) \nonumber \\ &&\times
\left( \epsilon_{\Delta}^{2} + \gamma^{2} + 2 \Omega_{1}^{2} + \Omega_{1} \Omega_{2} \cos{\left(\overline \varphi \right)} + 2\Omega_{2}^{2}\right)\nonumber \\ &&\times
\left(4 \epsilon_{\Delta}^{2} + 16 \gamma^{2} + 8 \Omega_{1}^{2} + 16 \Omega_{1} \Omega_{2} \cos{\left(\overline \varphi \right)} + 8 \Omega_{2}^{2}\right)\nonumber \\ 
&+& \frac{1}{8} \Omega_{1}^2 \left(\frac{1}{2} \epsilon_{\Delta} \Omega_{2} \sin{\left(\overline \varphi \right)} + \gamma \Omega_{1} + \gamma \Omega_{2} \cos{\left(\overline \varphi \right)}\right)^{2}\nonumber \\ &&\times
\left( \epsilon_{\Delta}^{2} + 20 \gamma^{2} + 4 \Omega_{1}^{2} + 8 \Omega_{1} \Omega_{2} \cos{\left(\overline \varphi \right)} + 4 \Omega_{2}^{2}\right)\nonumber \\ &&\times
\left(4 \epsilon_{\Delta}^{2} + 16 \gamma^{2} + 8 \Omega_{1}^{2} + 16 \Omega_{1} \Omega_{2} \cos{\left(\overline \varphi \right)} + 8 \Omega_{2}^{2}\right)\nonumber \\
&-& \frac{1}{64} \Omega_1\left(8 i \epsilon_{\Delta} \gamma \Omega_{2} \cos{\left(\overline \varphi \right)} + 16 \gamma^{2} \Omega_{1} \right.   \nonumber \\ 
&&\quad\quad  \left. - 16 i \gamma^{2} \Omega_{2} \sin{\left(\overline \varphi \right)} + 8 \Omega_{1} \Omega_{2}^{2} \sin^{2}{\left(\overline \varphi \right)}\right) \nonumber \\ &&\times
\left( \epsilon_{\Delta}^{2} + \gamma^{2} + 2 \Omega_{1}^{2} +  \Omega_{1} \Omega_{2} \cos{\left(\overline \varphi \right)} + 2 \Omega_{2}^{2}\right)^{3}\nonumber, \\ 
\end{eqnarray}
and the denominator is given by
\begin{eqnarray}
c_{\text{den} } &=& \frac{ \gamma}{64} \left(\epsilon_{\Delta}^{2} + \gamma^{2} + 2 \Omega_{1}^{2} +  \Omega_{1} \Omega_{2} \cos{\left(\overline \varphi \right)} + 2 \Omega_{2}^{2}\right)^{3}  \nonumber \\ &&\times  
\left(4 \epsilon_{\Delta}^{2} + 16 \gamma^{2} + 8 \Omega_{1}^{2} + 16 \Omega_{1} \Omega_{2} \cos{\left(\overline \varphi \right)} + 8 \Omega_{2}^{2}\right).\nonumber \\
\end{eqnarray}
As this expression is very complicated, and does not allow for a clear  analysis, we expand both the denominator and the numerator in orders of $\gamma$, which we assume to be small. In doing so, we can understand signatures of the light-matter entanglement in closed quantum systems in the presence of weak noise. Keeping only the lowest non-vanishing terms in $\gamma$, we obtain Eq.~\eqref{eq:photonFluctuations}.

\subsection{Stationary state}

\label{sec:stationaryStateJCmodel}

In this Appendix, we evaluate the stationary state of the dissipative Jaynes-Cummings model, which we require in the discussion of the stationary transport in Sec.~\ref{sec:divergingFluctuations}. To this end, we have to solve the equations
\begin{eqnarray}
0 &=&     \Omega_{\text{y}}       \rho_{\text{z}}  -\frac{ \gamma}{2} \rho_{\text{x}} + \epsilon_{\Delta} \rho_{\text{y}} , \nonumber \\
0  &=&  -  \Omega_{\text{x}}    \rho_{\text{z}}  -\frac{ \gamma}{2} \rho_{\text{y}} - \epsilon_{\Delta} \rho_{\text{x}} , \nonumber \\
0   &=&    \Omega_{\text{x}}   \rho_{\text{y}} -   \Omega_{\text{y}}   \rho_{\text{x}} - \gamma  \left( \rho_{\text{z}}  +1  \right) .
\end{eqnarray}
The solution of this equation is given by
\begin{eqnarray}
\rho_{\text{x}} &=& \frac{1}{ \epsilon_{\Delta}^2 +\frac{\gamma^2}{4}} \left( \frac{\gamma}{2}  \Omega_{\text{y}}    - \epsilon_{\Delta}  \Omega_{\text{x}}   \right)    \rho_{\text{z}} \nonumber \\
\rho_{\text{y}} &=&  \frac{1}{ \epsilon_{\Delta}^2 +\frac{\gamma^2}{4}  } \left( -\frac{\gamma}{2} \Omega_{\text{x}}    - \epsilon_{\Delta}  \Omega_{\text{y}}   \right)    \rho_{\text{z}} \nonumber \\
\rho_{\text{z}} &=& - \frac{\gamma}{  \frac{\Omega_{\text{x}}  \left( \frac{\gamma}{2} \Omega_{\text{x}}    - \epsilon_{\Delta}  \Omega_{\text{y}}   \right)    + \Omega_{\text{y}} \left( \frac{\gamma}{2} \Omega_{\text{y}}    + \epsilon_{\Delta}  \Omega_{\text{x}}   \right)   }{\epsilon_{\Delta}^2 +\frac{\gamma^2}{4} } + \gamma}\nonumber \\
&=&  -\frac{\epsilon_{\Delta}^2 +\frac{\gamma^2}{4}}{  \frac{\Omega_{\text{x}} ^2}{2}+ \frac{\Omega_{\text{y}} ^2}{2}  +\epsilon_{\Delta}^2 +\frac{\gamma^2}{4} }
\end{eqnarray}
For  $\epsilon_{\Delta}\rightarrow 0 $ and  $\gamma \rightarrow 0$, we find that $\rho_{\text{x}} =\rho_{\text{y}} =\rho_{\text{z}} =0$  and thus obtain the stationary state in  Eq.~\eqref{eq:stationaryState_jcModel_resonantSystem}.

\section{Evaluation of the cumulants in the long-time limit}

\label{app:cumulantEvaluationLongTimeLimit}

In this Appendix, we introduce three different methods which allow to analytically determine information about the photonic probability distributions. The methods are applied in Secs.~\ref{sec:jcModel} and \ref{sec:acDrivenLambdaSystem} to the Jaynes-Cummings model and the ac-driven lambda system for illustration. 

\subsection{Truncation of the characteristic polynomial}

\label{sec:cumulantEvaluationTrunction}

The analytical calculation of the eigenvalues $\lambda_{\mu; \boldsymbol\chi, \boldsymbol \xi}$  in Eq.~\eqref{eq:effectiePropagatorExpansion} will be impossible for large  Liouvillians. Here we introduce a method  to non-perturbatively determine  low-order cumulants  in the long-time limit without depending on the exact eigenvalues of the Liouvillian. We restrict the explanations to  a single counting field for simplicity. 

In general, the characteristic polynomial of the effective Liovillian can be written as
\begin{equation}
\mathcal P_{\chi}(z) =  \sum_{j =0}^{N} a_{j }( \chi) z^{j }
\end{equation}
with counting-field dependent coefficients $ a_{j }( \chi)$. The roots of this polynomial are the eigenvalues of the Liouvillian. 
Importantly, $a_0(0)  = 0 $ such that the $\mu=0$ root of the polynomial is $\lambda_{0;0} =0$, reflecting the existence of a stationary state.
To determine the photonic cumulants $\kappa_l(t)$ for long times $t\rightarrow \infty$, we need to calculate the $l$-th derivatives of the root $\lambda_{0;\chi}$. To this end, one can use the truncated characteristic polynomial instead of the full one:
\begin{equation}
\mathcal P^{(R)} _{ \chi}(z) =  \sum_{j =0}^{R} a_{j }( \chi) z^{j },
\end{equation}
with $R <N$.
Denoting the corresponding roots by $ \lambda_{\mu;\chi}^{(R )} $, the $\mu=0$ derivatives fulfill 
\begin{equation}
 \frac{d^{l }}{d(-i\chi)^{l} } \lambda_{0;\chi=0}= \frac{d^{l}}{d(-i\chi)^{l} } \lambda_{0;\chi=0}^{(R)} 
\end{equation}
for $R\geq l$, which we can harness to exactly calculate low-order cumulants.
For instance, the $\mu=0$ root of the first-order truncated characteristic polynomial reads as
\begin{eqnarray}
\lambda_{0,\chi}^{(1)} =  -\frac{a_0(\chi)}{a_1( \chi)},
\end{eqnarray}
such that the first cumulant in the long-time limit is given by
\begin{eqnarray}
\kappa_1(t) =  - \frac{1}{a_1(0)}\left. \frac{da_0  (\chi)}{d(-i\chi)}  \right|_{\chi=0}  t .
\label{eq:firstCumulant}
\end{eqnarray}
Likewise, we find for the second-order truncated eigenvalue
\begin{eqnarray}
\lambda^{(2)}_{0;\chi} =  \frac{1}{2a_2}\left(-a_1 + \sqrt{a_1^2 - 4 a_0 a_2}  \right) \nonumber ,\\
\end{eqnarray}
where we have suppressed the counting field arguments for a notational reason. Deriving $\lambda^{(2)}_{0;\chi} $ with respect to $\chi$ reproduces the first cumulant in Eq.~\eqref{eq:firstCumulant}. Evaluation of the second derivative yields
\begin{eqnarray}
\kappa_2(t)
&=&     \frac{1}{a_1} \left\lbrace   a_0^{(2)}  -2ia_2 \tilde \kappa_1  \left[\frac{a_1^{(1)}}{a_2} -i\tilde \kappa_1       - \frac{a_1 a_2^{(1)} } {a_2^2}  \right] \right\rbrace t ,  \nonumber \\
\end{eqnarray}
where $a_j=a_j(0)$, and $a_j^{(l)}=d^{l}a_j/d\chi^l\mid_{\chi=0}$ for $i=0,1,2$. Moreover, $\tilde \kappa_1 =\kappa_1/t$ with $\kappa_1$ given in Eq.~\eqref{eq:firstCumulant}. The coefficients are typically lengthy analytical expression, which can be evaluated efficiently with computer algebra systems such as SymPy or Mathematica. Finally, we note that we can evaluate the coefficients $a_j^{(l)}$ by calculating derivatives of the characteristic polynomial, i.e.,
\begin{eqnarray}
a_j^{(l)}  = \frac{1}{j!}\left. \frac{d^l}{d\chi^l}\frac{d^j}{dz^j} \mathcal P_{\chi}(z)\right|_{\chi=0,z=0}
\end{eqnarray}
which is numerically very efficient even for larger systems.

\subsection{Stationary perturbation theory in non-Hermitian systems}

\label{sec:nonHermitianPerturbationTheory}

Here, we generalize the standard stationary perturbation theory from Hermitian systems to non-Hermitian systems. Given that the total  Liouvillian can be written as
\begin{equation}
\mathcal L = \mathcal L_0 + g \mathcal L_1,
\label{eq:perturbationExpantionLiovillian}
\end{equation}
where $\mathcal L_0$ is the Liouvillian of the unperturbed system, $\mathcal L_1 $ is the perturbation Liouvillian and $0\leq g \ll 1$ parameterizes the strength of the perturbation.

The goal is to approximatively determine the right and left eigenvectors and their corresponding eigenvalues of the Liouvillian, that are defined by
\begin{eqnarray}
\mathcal L  \dket{ u_\mu}    &=&  \lambda_\mu   \dket{ u_\mu}  , \nonumber \\
 \dbra{ u_\mu}  \mathcal L       &=&  \lambda_\mu \dbra{ u_\mu}  ,  
\label{eq:eigenvectorDefinition}
\end{eqnarray}
respectively.
We recall that  the eigenvectors corresponding to  eigenvalues $\lambda_\mu$ fulfill the orthogonality relation
\begin{equation}
\dbraket{u_{\mu_1}  }{ u_{\mu_2}  } = \delta_{\mu_1,\mu_2}.
\label{eq:nonHermitianOrthogonalityRelation}
\end{equation}
The normalization can be achieved by  properly rescaling the eigenvectors.

Our aim is to find an expansion of the eigenvalues and eigenvectors  in orders of $g$, i.e.,
 \begin{eqnarray}
\lambda_\mu  &=& \lambda_\mu^{(0)} + g \lambda_\mu^{(1)}  + g^2  \lambda_\mu^{(2)} +\dots    \nonumber , \\
\dket{ u_\mu}  &= &\dket{ u_\mu^{(0)}} + g \dket{ u_\mu^{(1)}}  + g^2  \dket{ u_\mu^{(2)} } +\dots     \nonumber ,\\
\dbra{ u_\mu} &= & \dbra{ u_\mu^{(0)}} + g \dbra{u_\mu^{(1)}}  + g^2  \dbra{  u_\mu^{(2)} }+\dots   , \nonumber \\
\label{eq:eigenvalue/vectorExpansion}
 \end{eqnarray}
where we define the unperturbed eigenvalues and eigenstates by
\begin{eqnarray}
\mathcal L_0  \dket{ u_\mu^{(0)} }    &=&  \lambda_\mu^{(0)} \dket{  u_\mu^{(0)} }, \nonumber \\
\dbra{ u_\mu^{(0)}} \mathcal L_0       &=&  \lambda_\mu^{(0)} \dbra{  u_\mu^{(0)}} .  
\label{eq:UnperturpbedEigenvectors}
\end{eqnarray}
We want  to express $ \lambda_\mu^{(n>0)}$ and the $\dket{ u_\mu^{(n>0)}} ,  \dbra{u_\mu^{(n>0)} } $ in terms of $ \lambda_\mu^{(0)}$ and the $\dket{ u_\mu^{(0)}} ,\dbra{ u_\mu^{(0)} } $. Inserting these expansions into  the right-eigenvector equation in Eq.~\eqref{eq:eigenvectorDefinition}, we find
\begin{eqnarray}
\left(   \mathcal L_0 + g \mathcal L_1   \right)&&\left(\dket{  u_\mu^{(0)}} + g \dket{ u_\mu^{(1)} } + g^2\dket{   u_\mu^{(2)}} +\dots    \right)\nonumber \\
&&= \left(\lambda_\mu^{(0)} + g \lambda_\mu^{(1)}  + g^2  \lambda_\mu^{(2)} +\dots    \right)\nonumber\\
&&\times\left( \dket{ u_\mu^{(0)}} + g \dket{ u_\mu^{(1)} } + g^2 \dket{  u_\mu^{(2)}} +\dots    \right)\nonumber \\
\end{eqnarray}
These equations must be fulfilled in all orders of $g$. The zeroth-order is just the unperturbed eigenvalue equation for $\mathcal L_0$ in Eq.~\eqref{eq:UnperturpbedEigenvectors}. In first order, we find
\begin{eqnarray}
\mathcal L_0  \dket{ u_\mu^{(1)} }  +  \mathcal L_1 \dket{ u_\mu^{(0)}  }   &=&  \lambda_\mu^{(1)}  \dket{ u_\mu^{(0)} }   +  \lambda_\mu^{(0)}  \dket{ u_\mu^{(1)} } .\nonumber \\
\label{eq:firstOrderEigenvalueEquation}
\end{eqnarray}
Multiplying from the left with $ \dbra{ u_\mu^{(0)} } $ and using the orthogonality relation in Eq.~\eqref{eq:nonHermitianOrthogonalityRelation}, we readily find the first order correction for the eigenvalue
\begin{equation}
\lambda_\mu^{(1)}   = \dbra{  u_\mu^{(0)} }    \mathcal L_1  \dket{  u_\mu^{(0)} } .
\label{eq:eigenvalueFirstOrderCorrection}
\end{equation}
To obtain the first-order correction for the related right eigenvector, we expand it as
\begin{equation}
\dket{ u^{(1)}_\mu}    =\sum_{\mu_1} a_{\mu,\mu_1} \dket{ u^{(0)}_{\mu'}} .
\end{equation}
Inserting into  Eq.~\eqref{eq:firstOrderEigenvalueEquation}, appropriately left-multiplying with $  \dbra{ u^{(0)}_{\mu'\neq \mu} } $, and resolving for $a_{\mu,\mu'}$, we finally find
\begin{equation}
\dket{u^{(1)}_\mu }   =\sum_{\mu_1} \frac { \dbra{  u_{\mu_1}^{(0)} }    \mathcal L_1  \dket{  u_\mu^{(0)} }   }  { \lambda_\mu^{(0)}  -  \lambda_{\mu_1}^{(0)} } \dket{  u^{(0)}_{\mu_1}} . 
\end{equation}
 The expression for the left eigenvector can be obtained accordingly.
The derivation of the second-order terms can be obtained along the same lines. The second-order eigenvalue equation reads as 
\begin{multline}
\mathcal L_0  \dket{ u_\mu^{(2)}}   +  \mathcal L_1 \dket{ u_\mu^{(1)}}     = \\ \lambda_\mu^{(2)} \dket{ u_\mu^{(0)} }    +  \lambda_\mu^{(0)}  \dket {u_\mu^{(2)}}  +  \lambda_\mu^{(1)} \dket{  u_\mu^{(1)}} .
\end{multline}
Multiplying from left with $\dbra{  u_\mu^{(0)}}    $ and resolving for $\lambda_\mu^{(2)} $, we find
\begin{equation}
\lambda_\mu^{(2)}  =\sum_{\mu_1} \frac {\dbra{  u_{\mu_1}^{(0)} }    \mathcal L_1  \dket{  u_\mu^{(0)} }  \dbra{  u_{\mu}^{(0)} }    \mathcal L_1  \dket{  u_{\mu_1}^{(0)} }    }  { \lambda_\mu^{(0)}  -  \lambda_{\mu_1}^{(0)}   }  .
\label{eq:eigenvalueSecondOrderCorrection}
\end{equation}
Thus, the formulas known from Hermitian systems are modified by appropriately replacing the left eigenvectors in the scalar products.

\subsection{Adiabatic elimination}

\label{sec:adiabaticElimination}

Here, we introduce a method to reduce the dimension of the system based on adiabatic elimination. Thereby, we again consider systems that can be separated in an unperturbed and a perturbation part as in Eq.~\eqref{eq:perturbationExpantionLiovillian}.  
For $g=0$, we assume that the system density matrix in the stationary state can be distributed in two subsystems (s) and (t), such that the corresponding probabilities fulfill 
\begin{equation}
	\sum_{j=1}^{d^{(\text{s})}}  \rho_{jj} =1,\qquad \qquad \sum_{j=d^{(\text{s})}+1}^{d }  \rho_{jj} =0,
\end{equation}
where $ d = d^{(\text{s})} +d^{(\text{t})} $ is the total dimension of the matter system. Thereby the subsystems `$\text{s}$' and `$\text{t}$' refer to the stationary and transient dynamics, respectively.
The corresponding  elements of  the density matrices are  arranged as vectors   $\rho^{(\text{s})} $ and $\rho^{(\text{t})}  $, respectively, such that  the  Liouvillian operator can be written as
\begin{equation}
\frac{d}{dt}\left(
\begin{array}{cc}
\rho^{(\text{s})}  \\
\rho^{(\text{t})} 
\end{array}
\right)
=
 \left(
\begin{array}{cc}
   \mathcal L^{(\text{s,s})}  & \mathcal L^{(\text{s,t})} \\
   \mathcal L^{(\text{t,s})}  & \mathcal L^{({t,t})}
\end{array}
\right)
\left(
\begin{array}{cc}
\rho^{(\text{\text{s}})}  \\
\rho^{(\text{t})}  
\end{array}
\right).
\end{equation}
The aim is  now to find an effective Liouvillian acting only on  $ \rho^{(\text{s})}$. To this end, we apply the adiabatic approximation to the transient subsystem, i.e.,
\begin{equation}
	\frac{d}{dt} \rho^{(\text{t})} =0,
	\label{eq:adiabaticElimination_condition}
\end{equation}
which is  valid if $g$ is small compared to the energy scales in $\mathcal L_0$. Resolving for $ \rho^{(\text{t})} $, and inserting back into the equation for $\rho^{(\text{s})}$, we readily find the effective equation of motion
\begin{eqnarray}
\frac{d}{dt}
\rho^{(\text{s})} 
&=&
\left[ \mathcal L^{(\text{s,s})}  - \mathcal L^{(\text{s,t})} \mathcal L^{(\text{t,s})-1}   \mathcal L^{(\text{\text{t,s} })}  \right]
\rho^{(\text{s})} \nonumber   \\
&=&
\left[ \mathcal L^{(\text{s,s})}  - \mathcal L^{(\text{s,t})} \mathcal L^{(\text{t,s})-1}   \mathcal L^{(\text{t,s})}  \right]_{\mathcal O\left( g^n \right)}\rho^{(\text{s})}\nonumber  \\
 &&+ \mathcal O\left( g^{n+1} \right) .
 \label{eq:adiabaticEliminationEffLiouvillian}
\end{eqnarray}
In the first line, the main challenge is to evaluate the inverse $\mathcal L^{(t,s)-1} $, which will be non-trivial in general. In the second line, the notation $\left[ \bullet \right]_{\mathcal O\left( g^n \right)} $ refers to the terms scaling as $g^m$ with $m\leq n$   in the effective Liouvillian, while terms of higher order will be ignored.
The identification of the $\left[ \bullet \right]_{\mathcal O\left( g^n \right)} $ terms requires a skill-full analysis of the equation structure. As the resulting effective Liouvillian in Eq.~\eqref{eq:adiabaticEliminationEffLiouvillian} has a reduced dimension, it can be diagonalized more easily. In many cases of interested, the effective Liouvillian will have only a single element, e.g., the ground state of the matter system.

The non-Hermitian perturbation theory in Sec.~\ref{sec:nonHermitianPerturbationTheory} and the adiabatic-elimination procedure introduced here often deliver equivalent results, as they involve  similar constrains and approximations. The non-Hermitian perturbation theory offers a clear formal procedure, however, it requires the evaluation of all left and right eigenvectors of $\mathcal L_0$, which is challenging in many cases. In contrast, the adiabatic elimination relies on a skill-full expansion of the linear equation, which has not been formalized here. An example is given in Sec.~\ref{sec:lambdaSystemAdiabaticElimination} for the  lambda system.

\section{ ac-driven lambda system}

In this Appendix, we demonstrate how to calculate the cumulant-generating function  of the effective Liovillian of the ac-driven lambda system. For illustration, we carry out the calculation using both the non-Hermitian perturbation theory introduced in Appendix~\ref{sec:nonHermitianPerturbationTheory}, and the adiabatic elimination approach introduced in Appendix~\ref{sec:adiabaticElimination}, which both deliver the same result.

\subsection{Non-Hermitian perturbation theory}
\label{sec:lambdaSystemPerturbationTheory}

In this Appendix, we apply the stationary perturbation theory for non-Hermitian systems to the ac-driven lambda system to derive an expression for the cumulant-generating function in the limit of weak signal fields. 
Here we focus on the $r$-odd case, as  the $r$-even case works along the same lines.

In the rotating frame defined by Eq.~\eqref{eq:rotatingFrame}, the effective Hamiltonian  reads as
\begin{eqnarray}
\hat {\mathcal H}_{\boldsymbol \varphi} &=& \epsilon_{a,\Delta} \braket{a}{a} +   \frac{\epsilon_{b,\Delta}+\epsilon_{c,\Delta}}{2}  \left( \braket{b}{b} +  \braket{c}{c}\right) \nonumber  \\
&+&  \mathcal J_0 \left( \frac{\Omega_{\text{p},1}}{\omega_{\text{d}}} \right) \frac{\epsilon_{b,\Delta}-\epsilon_{c,\Delta}}{2} \left[\braket{b}{b}-\braket{c}{c} \right] \nonumber \\
&+& \frac{\Omega_{\text{p},0} }{2} \left( \braket{b}{c}   + \braket{c}{b} \right) \nonumber  \\
&+&  \frac{ \Omega_{\text{s} }}{2} e^{i\varphi_1} \mathcal J_0 \left( \frac{\Omega_{p,1}}{2\omega_{d}} \right)   \braket{c}{a}\nonumber  \\
 &+&   \frac{ \Omega_{\text{s} }}{2} e^{i\varphi_2} \mathcal J_{r } \left( \frac{\Omega_{\text{p},1}}{2\omega_{\text{d} }} \right)   \braket{b}{a} ,  
\end{eqnarray}
where  $ \epsilon_{a,\Delta} =\epsilon_{a}$,  $ \epsilon_{b,\Delta} =\epsilon_{b}+\omega_{\text{p} } -\omega_1$, $ \epsilon_{c,\Delta} =\epsilon_{c}-\omega_1$.
Applying a unitary transformation to diagonalize the  subsystem of the states $\left| b\right>$ and $\left| c\right>$, the effective Hamiltonian becomes
\begin{eqnarray}
\tilde  {\mathcal H}_{\boldsymbol \varphi } &=& \tilde  {\mathcal H}_{\boldsymbol \varphi}^{(0)} + \tilde  {\mathcal H}_{\boldsymbol \varphi} ^{(1)} \nonumber, \\
\tilde  {\mathcal H}_{\boldsymbol \varphi}^{(0)}  &=& \tilde \epsilon_{a}  \braket{a}{a} +   \tilde \epsilon_{b}  \braket{b}{b}  + \tilde \epsilon_{c} \braket{c}{c}  , \nonumber  \\
%
\tilde  {\mathcal H}_{\boldsymbol \varphi} ^{(1)}  &=& \left[ \frac{\Omega_{c,\boldsymbol  \varphi}}{2} \braket{c}{a} + \text{h.c.} \right] \nonumber  \\
&+&  \left[\frac{ \Omega_{b,\boldsymbol  \varphi}}{2}  \braket{b}{a} +  \text{h.c.} \right],
\end{eqnarray}
where  the effective energies reads as
\begin{eqnarray}
\tilde \epsilon_{a}  &=&   \epsilon_{a,\Delta} , \nonumber \\
\tilde \epsilon_{b}  &=&  \frac{\epsilon_{b,\Delta} + \epsilon_{c,\Delta}}{2}\nonumber  \\
&& - \frac{1}{2}\sqrt{ \mathcal J_0 \left( \frac{\Omega_{p,1} }{\omega_{d}} \right)^2 \left(\epsilon_{b,\Delta} - \epsilon_{c,\Delta} \right)^2 +\Omega_{\text{p},0}^{2}   }, \nonumber  \\
\tilde \epsilon_{c}  &=&  \frac{\epsilon_{b,\Delta} + \epsilon_{c,\Delta}}{2}\nonumber  \\
&&+ \frac{1}{2}\sqrt{\mathcal J_0 \left( \frac{\Omega_{\text{p},1}}{\omega_{\text{d}}} \right)^2 \left(\epsilon_{b,\Delta } - \epsilon_{c,\Delta } \right)^2 +\Omega_{\text{p},0}^{2}  }. \nonumber \\
\label{eq:lambdaSystem:paramterRenormalization}
\end{eqnarray}
The effective Rabi couplings of the state $\left| a\right> $ with the states $\left| b\right> $ and $\left| c\right> $ are given by
\begin{eqnarray}
\Omega_{b,\boldsymbol \varphi} &=& -  \Omega_{\text{s} } \sin\frac{\theta}{2}  e^{i\varphi_1} \mathcal J_0 \left( \frac{\Omega_{\text{p},1}}{2\omega_{\text{d} }} \right) \nonumber \\
&&+  \Omega_{\text{s} } \cos\frac{\theta}{2}  e^{i\varphi_2} \mathcal J_{r } \left( \frac{\Omega_{\text{p},1}}{2\omega_{d}} \right) , \nonumber  \\
\Omega_{c,\boldsymbol  \varphi}  &=& \Omega_{\text{s} } \cos\frac{\theta}{2}  e^{i\varphi_1} \mathcal J_0 \left( \frac{\Omega_{\text{p},1}}{2\omega_{\text{d} }} \right) \nonumber \\
&&+  \Omega_{\text{s} } \sin\frac{\theta}{2}  e^{i\varphi_2} \mathcal J_{r } \left( \frac{\Omega_{\text{p},1}}{2\omega_{\text{d} }} \right) ,
\label{eq:lambdaSys:effCoupling}
\end{eqnarray}
respectively, where $\theta$ parameterizes the transformation of the eigenstates and is defined via
\begin{equation}
	\tan \theta =  \frac{\Omega_{\text{p},0}}{\epsilon_{c,\Delta} - \epsilon_{b,\Delta }}.
\end{equation}

The unperturbed Liouvillian is now given by
\begin{eqnarray}
\mathcal  L_0\rho &=&  -i \tilde H_{ \overline{\boldsymbol\varphi} + \frac{\boldsymbol \chi  }{2} }^{(0)}  \rho +i \rho \tilde H_{ \overline{\boldsymbol\varphi} - \frac{\boldsymbol \chi  }{2} }^{(0)}\nonumber \\
 &&+   \gamma D\left[ \braket{a}{b} \right] \rho +\gamma D\left[ \braket{a}{c}   \right] \rho,
\end{eqnarray}
where the dissipator has remained unchanged in all transformations because both dissipators  are proportional to $\gamma$.
The  eigenvalues as well as  the corresponding right and left eigenvectors of $\mathcal L_0$ are given by
\begin{widetext}
\begin{align}
	\lambda_0^{(0)}  &= 0, \qquad    &\dket{ u_0^{(0)} } &=  \braket{a}{a}, \qquad    &\dbra{ u_0^{(0)} } &=  \braket{a}{a} + \braket{b}{b} + \braket{c}{c} \nonumber , \\
	\lambda_1 ^{(0)} &= i \tilde \epsilon_b  - \frac{\gamma}{2}, \qquad    &\dket{ u_1^{(0)} } &=  \braket {a}{b}, \qquad    &\dbra{ u_1^{(0)} } &=  \braket {a}{b}   \nonumber ,  \\
	\lambda_2^{(0)}  &=-i \tilde \epsilon_b - \frac{\gamma}{2} , \qquad    &\dket{ u_2^{(0)} } &=  \braket{b}{a} , \qquad    &\dbra{ u_2^{(0)} } &=  \braket{b}{a}  \nonumber ,  \\
	\lambda_3^{(0)}  &=i \tilde \epsilon_c -\frac{\gamma}{2}, \qquad    &\dket{ u_3 ^{(0)}} &=  \braket{a}{c}, \qquad    &\dbra{ u_3^{(0)}}  &=  \braket{a}{c}\nonumber   ,\\
	\lambda_4 ^{(0)} &=-i \tilde \epsilon_c  - \frac{\gamma}{2} , \qquad    &\dket{ u_4^{(0)} } &=  \braket{c}{a}, \qquad    &\dbra{ u_4^{(0)} } &=  \braket{c}{a}  \nonumber  ,\\
	\lambda_5^{(0)}  &= - \gamma, \qquad  &\dket{ u_5^{(0)} }  &= \braket{b}{b}  - \braket{a}{a} , \qquad  &\dbra{  u_5^{(0)}}  &=  \braket{b}{b} \nonumber ,  \\
	\lambda_6 ^{(0)} &= - \gamma ,\qquad  &\dket{ u_6^{(0)} }  &= \braket{c}{c}  - \braket{a}{a} ,\qquad  &\dbra{  u_6^{(0)} } &=  \braket{c}{c}   \nonumber , \\
	\lambda_7^{(0)}  &= - i(\tilde \epsilon_b -\tilde \epsilon_c  ) - \gamma, \qquad  &\dket{ u_7^{(0)} }  &= \braket{b}{c} , \qquad  &\dbra{   u_7^{(0)} } &= \braket{b}{c}  \nonumber , \\
	\lambda_8^{(0)} &=  i(\tilde \epsilon_b -\tilde \epsilon_c  ) - \gamma ,  \qquad  &\dket{ u_8^{(0)} }  &= \braket{c}{b}  ,\qquad  &\dbra{  u_8^{(0)} } &= \braket{c}{b}    .
\end{align}
\end{widetext}
The perturbation part of the Liouvillian is accordingly
\begin{eqnarray}
\mathcal  L_1\rho &=&  -i\left[  \frac{ \Omega_{b,\overline{\boldsymbol\varphi} + \frac{\boldsymbol \chi  }{2}  } }{2}\braket{b}{a}+  \frac{ \Omega_{c,\overline{\boldsymbol\varphi} + \frac{\boldsymbol \chi  }{2} }    }{2} \braket{c}{a}  + \text{ h.c.}\right]\rho \nonumber \\
&&+i \rho\left[  \frac{  \Omega_{b,\overline{\boldsymbol\varphi} - \frac{\boldsymbol \chi  }{2} } }{2} \braket{b}{a} + \frac{ \Omega_{c,\overline{\boldsymbol\varphi}- \frac{\boldsymbol \chi  }{2}  } }{2}   \braket{c}{a}  +\text{ h.c.} \right]  , \nonumber\\
\end{eqnarray}
where $\Omega_{\text{s} } $ takes the role of the perturbation parameter $g$.

We apply the perturbation theory to the  eigenvalue $\lambda_{0}^{(0)}$, as it constitutes the stationary state in the limit $\Omega_{\text{s}} \rightarrow 0$.
It is easy to see that the first-order correction of the eigenvalue in Eq.~\eqref{eq:eigenvalueFirstOrderCorrection} vanishes
\begin{equation}
\lambda_{\mu=0}^{(1)}   =\dbra{  u_0^{(0)} }    \mathcal L_1  \dket{  u_0^{(0)} } =0,
\end{equation}
where the scalar product  is defined by
\begin{equation}
	 \dbraket{ w  }{  v } =  \text{tr} \left[w^\dagger v  \right].
\end{equation} 
To evaluate the second-order correction in Eq.~\eqref{eq:eigenvalueSecondOrderCorrection}, we have to evaluate the four scalar products appearing in the sum
\begin{eqnarray}
\lambda_{\mu=0}^{(2)}  &=& \sum_{\mu_1 =1,2,3,4} \frac {\dbra{  u_{\mu_1}^{(0)} }    \mathcal L_1  \dket{  u_\mu^{(0)} }  \dbra{  u_{\mu}^{(0)} }    \mathcal L_1  \dket{  u_{\mu_1}^{(0)} }    }  { \lambda_\mu^{(0)}  -  \lambda_{\mu_1}^{(0)}   }  , \nonumber \\
\end{eqnarray}
as all other terms vanish. The scalar products are explicitly given as
\begin{widetext}
\begin{eqnarray}
 \dbra{  u_{1}^{(0)} }\mathcal L_1 \dket{    u_0^{(0)} } \dbra{  u_{0}^{(0)}  }\mathcal L_1  \dket  {   u_{1}^{(0)}  } &=& \text{tr} \left(\braket{a}{b}      \mathcal L_1   \braket{a}{a}   \right) \text{tr} \left( \braket{a}{a} + \braket{b}{b} + \braket{c}{c}     \mathcal L_1   \braket{b}{a}  \right) \nonumber \\
&=&-\frac{i}{4}  \Omega_{b,\overline{\boldsymbol\varphi} + \frac{\boldsymbol \chi  }{2} } \left[-i  \Omega_{b,\overline{\boldsymbol\varphi} + \frac{\boldsymbol \chi  }{2} }^*   +i   \Omega_{b,\overline{\boldsymbol\varphi} - \frac{\boldsymbol \chi  }{2} }^*   + 0 \right] ,\nonumber   \\
 \dbra{  u_{2}^{(0)} }\mathcal L_1 \dket{    u_0^{(0)} } \dbra{  u_{0}^{(0)}  }\mathcal L_1  \dket  {   u_{2}^{(0)}  }  &=& \text{tr} \left(\braket{b}{a}      \mathcal L_1   \braket{a}{a}   \right) \text{tr} \left( \braket{a}{a} + \braket{b}{b} + \braket{c}{c}     \mathcal L_1   \braket{a}{b}  \right) \nonumber \\
&=& \frac{i}{4}  \Omega_{b,\overline{\boldsymbol\varphi}- \frac{\boldsymbol \chi  }{2}}^*  \left[i  \Omega_{b,\overline{\boldsymbol\varphi}- \frac{\boldsymbol \chi  }{2} }  -i \Omega_{b,\overline{\boldsymbol\varphi} + \frac{\boldsymbol \chi  }{2} } + 0 \right] \nonumber,  \\
 \dbra{  u_{3}^{(0)} }\mathcal L_1 \dket{    u_0^{(0)} } \dbra{  u_{0}^{(0)}  }\mathcal L_1  \dket  {   u_{3}^{(0)}  } &=& \text{tr} \left(\braket{a}{c}      \mathcal L_1   \braket{a}{a}   \right) \text{tr} \left( \braket{a}{a} + \braket{b}{b} + \braket{c}{c}     \mathcal L_1   \braket{c}{a}  \right) \nonumber \\
&=&-\frac{i}{4}  \Omega_{c,\overline{\boldsymbol\varphi} + \frac{\boldsymbol \chi  }{2} }\left[-i  \Omega_{c,\overline{\boldsymbol\varphi} + \frac{\boldsymbol \chi  }{2} }^*   +i   \Omega_{c,\overline{\boldsymbol\varphi} - \frac{\boldsymbol \chi  }{2} }^*  + 0 \right]  \nonumber ,  \\
 \dbra{  u_{4}^{(0)} }\mathcal L_1 \dket{    u_0^{(0)} } \dbra{  u_{0}^{(0)}  }\mathcal L_1  \dket  {   u_{4}^{(0)}  } &=& \text{tr} \left(\braket{c}{a}      \mathcal L_1   \braket{a}{a}   \right) \text{tr} \left( \braket{a}{a} + \braket{b}{b} + \braket{c}{c}     \mathcal L_1   \braket{a}{c}  \right) \nonumber ,\\
&=& \frac{i}{4}  \Omega_{c,\overline{\boldsymbol\varphi} - \frac{\boldsymbol \chi  }{2} }^*  \left[ i  \Omega_{c,\overline{\boldsymbol\varphi} - \frac{\boldsymbol \chi  }{2} }   -i \Omega_{c,\overline{\boldsymbol\varphi} + \frac{\boldsymbol \chi  }{2} } + 0 \right] .
\end{eqnarray}
\end{widetext}
Putting everything together, we obtain the expression for the eigenvalue in second-order perturbation
\begin{eqnarray}
\lambda_{0;\boldsymbol \chi}
&=& -\frac{i}{4}  \Omega_{b,\overline{\boldsymbol\varphi} + \frac{\boldsymbol \chi  }{2}} \left( -i  \Omega_{b,\overline{\boldsymbol\varphi} + \frac{\boldsymbol \chi  }{2}}^*   +i    \Omega_{b,\overline{\boldsymbol\varphi} - \frac{\boldsymbol \chi  }{2} }^*  \right)  \frac{1}{i \tilde \epsilon_b  - \frac{\gamma}{2} }\nonumber\\
&&-\frac{i}{4}  \Omega_{c,\overline{\boldsymbol\varphi} + \frac{\boldsymbol \chi  }{2} } \left (-i  \Omega_{c,\overline{\boldsymbol\varphi} + \frac{\boldsymbol \chi  }{2} }^*   +i   \Omega_{c,\overline{\boldsymbol\varphi} - \frac{\boldsymbol \chi  }{2} }^{*}    \right )  \frac{1}{i \tilde \epsilon_c  - \frac{\gamma}{2} }   \nonumber  \\
&+& \frac{ i}{4}  \Omega_{b,\overline{\boldsymbol\varphi} - \frac{\boldsymbol \chi  }{2} }^* \left (  i  \Omega_{b,\overline{\boldsymbol\varphi} - \frac{\boldsymbol \chi  }{2} }  -i  \Omega_{b,\overline{\boldsymbol\varphi} + \frac{\boldsymbol \chi  }{2} }  \right )   \frac{1}{-i \tilde \epsilon_b  - \frac{\gamma}{2} } \nonumber\\
&&+\frac{i}{4}  \Omega_{c,\overline{\boldsymbol\varphi} -\frac{\boldsymbol \chi  }{2} }^*  \left (  i   \Omega_{c,\overline{\boldsymbol\varphi}- \frac{\boldsymbol \chi  }{2} } -i  \Omega_{c,\overline{\boldsymbol\varphi} + \frac{\boldsymbol \chi  }{2} } \right )   \frac{1}{-i \tilde \epsilon_c  - \frac{\gamma}{2} }\nonumber\\
 && +\mathcal O\left(\Omega_s^3 \right)  , \nonumber\\
\label{eq:stationaryEigenvalueLambdaSystem}
\end{eqnarray}
which   can be used to construct the moment and cumulant-generating functions according to Eq.~\eqref{eq:cumulantGenertingFunctionLongTimes}.

\subsection{Adiabatic elimination}

\label{sec:lambdaSystemAdiabaticElimination}

Complementary to the non-Hermitian perturbation theory,  we apply here the adiabatic-elimination approach introduced in Appendix~\ref{sec:adiabaticElimination}  to derive the moment- and cumulant-generating functions. To start with, we construct the Liouvillian equation of motion for the  density-matrix elements  $\rho_{\alpha\beta} = \text{tr}\left( \left|\alpha \right> \left<\beta \right| \rho_{\boldsymbol \chi ,\boldsymbol \xi  }\right) $, i.e.,
\begin{eqnarray}
\frac{d}{dt}   \rho_{aa}   &=&   - \gamma  \rho_{bb}  - \gamma  \rho_{cc} \nonumber \\
&-& \frac{ i }{2} \Omega_{b,\overline{\boldsymbol\varphi} + \frac{\boldsymbol \chi  }{2}  }^{*}        \rho_{ba}  - \frac{ i} {2}\Omega_{c,\overline{\boldsymbol\varphi} + \frac{\boldsymbol \chi   }{2}  }^{*}    \rho_{ca} \nonumber \\
  &+&   \frac{i}{2}  \Omega_{b,\overline{\boldsymbol\varphi} - \frac{\boldsymbol \chi  }{2}}   \rho_{ab} +   \frac{i}{2}\Omega_{c,\overline{\boldsymbol\varphi} - \frac{\boldsymbol \chi   } {2}  }  \rho_{ac} \nonumber,  \\
\frac{d}{dt}   \rho_{ab}    &=&  -\frac{\gamma}{2} \rho_{ab}  - i \left(\tilde \epsilon_a  -\tilde \epsilon_b \right)\rho_{ab}  \nonumber \\
&-& \frac{i}{2}\Omega_{b,\overline{\boldsymbol\varphi} + \frac{\boldsymbol \chi  }{2} }^{*}   \rho_{bb} -\frac{i}{2}\Omega_{c,\overline{\boldsymbol\varphi} + \frac{\boldsymbol \chi  }{2} }^{*}   \rho_{cb} +\frac{i}{2}\Omega_{b,\overline{\boldsymbol\varphi} - \frac{\boldsymbol \chi  }{2}}^{*}   \rho_{aa}   \nonumber , \\
\frac{d}{dt}   \rho_{ac}    &=&  -\frac{\gamma}{2} \rho_{ac}  - i \left(\tilde \epsilon_a -\tilde \epsilon_c \right) \rho_{ac}    \nonumber \\
&-&\frac{i}{2}\Omega_{c,\overline{\boldsymbol\varphi} + \frac{\boldsymbol \chi  }{2}}^{*}   \rho_{cc} -\frac{i}{2}\Omega_{b,\overline{\boldsymbol\varphi} + \frac{\boldsymbol \chi  }{2}}^{*}   \rho_{bc} +\frac{i}{2}\Omega_{c,\overline{\boldsymbol\varphi} - \frac{\boldsymbol \chi  }{2} }^{*}   \rho_{aa}    \nonumber ,\\
\frac{d}{dt}   \rho_{ba}    &=&  -\frac{\gamma}{2} \rho_{ba}  - i \left(\tilde \epsilon_b -\tilde \epsilon_a \right)\rho_{ba} \nonumber \\
&-&\frac{i}{2}\Omega_{b,\overline{\boldsymbol\varphi} + \frac{\boldsymbol \chi  }{2} } \rho_{aa}  +\frac{i}{2}\Omega_{b,\overline{\boldsymbol\varphi} - \frac{\boldsymbol \chi  }{2}} \rho_{bb} +\frac{i}{2}\Omega_{c,\overline{\boldsymbol\varphi} - \frac{\boldsymbol \chi  }{2}} \rho_{bc}     \nonumber, \\
\frac{d}{dt}   \rho_{bb}    &=&  -\gamma \rho_{bb}  -\frac{i}{2}\Omega_{b,\overline{\boldsymbol\varphi} + \frac{\boldsymbol \chi  }{2}} \rho_{ab}  +\frac{i}{2}\Omega_{b,\overline{\boldsymbol\varphi} - \frac{\boldsymbol \chi  }{2}}^{*}   \rho_{ba}   \nonumber, \\
\frac{d}{dt}   \rho_{bc}    &=&   -\gamma \rho_{bc}  -i\left(\tilde \epsilon_b -\tilde \epsilon_c \right)\rho_{bc}  \nonumber ,\\
\frac{d}{dt}   \rho_{ca}    &=&   -\frac{\gamma}{2} \rho_{ca}  - \frac{i}{2} \left(\tilde \epsilon_c -\tilde \epsilon_a \right)\rho_{ca} \nonumber \\
&-& \frac{i}{2}\Omega_{c,\overline{\boldsymbol\varphi} + \frac{\boldsymbol \chi  }{2}} \rho_{aa}  +\frac{i}{2}\Omega_{c,\overline{\boldsymbol\varphi} - \frac{\boldsymbol \chi  }{2}} \rho_{cc} +\frac{i}{2}\Omega_{b,\overline{\boldsymbol\varphi} - \frac{\boldsymbol \chi  }{2}} \rho_{cb} \nonumber,\\
\frac{d}{dt}   \rho_{cb}    &=&  -\gamma \rho_{cb}  - i \left(\tilde \epsilon_c -\tilde \epsilon_b \right)\rho_{cb}  \nonumber,\\
\frac{d}{dt}   \rho_{cc}    &=&  -\gamma \rho_{cc}  -\frac{i}{2} \Omega_{c,\overline{\boldsymbol\varphi} + \frac{\boldsymbol \chi  }{2}} \rho_{ac}  +\frac{i}{2}\Omega_{c,\overline{\boldsymbol\varphi} - \frac{\boldsymbol \chi  }{2}}^{*}   \rho_{ca} ,
\label{eq:lambdaSystem_EOMs}
\end{eqnarray}
where  the $\tilde \epsilon_\alpha$ with $ \alpha \in \left\lbrace a,b,c \right\rbrace$,  $\Omega_{b,\boldsymbol  \varphi}   $ are defined in  Eq.~\eqref{eq:lambdaSystem:paramterRenormalization} and $\Omega_{c,\boldsymbol  \varphi }   $  is defined in Eq.~\eqref{eq:lambdaSys:effCoupling}. 
Inspection of these equations for vanishing $\Omega_{\text{s}}$ reveals that  in the stationary state $\rho_{aa}(t\rightarrow \infty)=1 $, while all other matrix elements vanish. For this reason, we can identify the stationary subsystem (introduced in Sec.~\ref{sec:adiabaticElimination}) as $\rho_{\text{ss}}= \rho_{aa} $, while the transient subsystem vector $\rho^{(\text{t})} $ contains all remaining matrix elements. 
We thus continue to calculate an expression for $\frac{d}{dt}   \rho_{aa}$ in second order of $\Omega_{\text{s}}$.

As the matrix elements $\rho_{ab} ,\rho_{ba} , \rho_{ac}   $ appear in the equation for $\rho_{aa}$ with coefficients proportional to  $\Omega_s$, we set the left-hand side to zero according to Eq.~\eqref{eq:adiabaticElimination_condition}, and resolve these equations such that
\begin{eqnarray}
\rho_{ab}    &=&   \frac{1}{ 2i\tilde \epsilon_{ab} } \left( -i\Omega_{b,\overline{\boldsymbol\varphi} + \frac{\boldsymbol \chi  }{2} }^{*}   \rho_{bb} -i\Omega_{c,\overline{\boldsymbol\varphi} + \frac{\boldsymbol \chi  }{2} }^{*}   \rho_{cb} +i\Omega_{b,\overline{\boldsymbol\varphi} - \frac{\boldsymbol \chi  }{2}}^{*}   \rho_{aa} \right)   ,\nonumber \\
\rho_{ba}    &=&   \frac{1}{2i\tilde \epsilon_{ba} } \left(-i\Omega_{b,\overline{\boldsymbol\varphi} + \frac{\boldsymbol \chi  }{2} } \rho_{aa}  +i\Omega_{b,\overline{\boldsymbol\varphi} - \frac{\boldsymbol \chi  }{2}} \rho_{bb} +i\Omega_{c,\overline{\boldsymbol\varphi} - \frac{\boldsymbol \chi  }{2}} \rho_{bc}       \right)   , \nonumber\\
\rho_{ac}    &=&   \frac{1}{2 i\tilde \epsilon_{ac} } \left(-i\Omega_{c,\overline{\boldsymbol\varphi}+ \frac{\boldsymbol \chi  }{2}}^{*}   \rho_{cc} -i\Omega_{b,\overline{\boldsymbol\varphi} + \frac{\boldsymbol \chi  }{2} }^{*}   \rho_{bc} +i\Omega_{c,\overline{\boldsymbol\varphi} - \frac{\boldsymbol \chi  }{2} } ^{*}  \rho_{aa}      \right)   , \nonumber\\
\rho_{ca}    &=&   \frac{1}{ 2i\tilde \epsilon_{ca} } \left( -i\Omega_{c,\overline{\boldsymbol\varphi} + \frac{\boldsymbol \chi  }{2} }   \rho_{aa}  +i\Omega_{c,\overline{\boldsymbol\varphi}- \frac{\boldsymbol \chi  }{2} } \rho_{cc} +i\Omega_{b,\overline{\boldsymbol\varphi} - \frac{\boldsymbol \chi  }{2} } \rho_{cb}  \right)  ,\nonumber \\  
\label{eq:lambdaSystem:firstOrderTerms}
\end{eqnarray}
where we have defined $\tilde \epsilon_{\alpha\beta} =\tilde \epsilon_{\alpha} -\tilde \epsilon_{\beta}  $. From these equations, we can obtain an expression for $\mathcal L^{(\text{t,t})-1}$ [introduced in Eq.~\eqref{eq:adiabaticEliminationEffLiouvillian}] in first order of $\Omega_s$.
Assuming $\rho_{aa}\gg \rho_{bb},\rho_{bc},\rho_{cb},\rho_{cc}\propto \Omega_{\text{s}}^2$, we neglect corresponding terms. This assumption can be  further justified by considering the solution of the  $\rho_{bb}$  and $\rho_{cc} $ equations
\begin{eqnarray}
\rho_{bb}    &=& \frac{1}{2\gamma  }\left(  -i\Omega_{b,\overline{\boldsymbol\varphi} + \frac{\boldsymbol \chi  }{2}} \rho_{ab}  +i\Omega_{b,\overline{\boldsymbol\varphi} - \frac{\boldsymbol \chi  }{2}}^{*}   \rho_{ba}   \right) \nonumber  \\
&=& \frac{1}{4\gamma  }\left( \Omega_{b,\overline{\boldsymbol\varphi} + \frac{\boldsymbol \chi  }{2} } \frac{ \Omega_{b,\overline{\boldsymbol\varphi} - \frac{\boldsymbol \chi  }{2}}^{*}  }{ i\tilde \epsilon_{ab} }   +\Omega_{b,\overline{\boldsymbol\varphi} - \frac{\boldsymbol \chi  }{2} }^{*}  \frac{\Omega_{b,\overline{\boldsymbol\varphi} + \frac{\boldsymbol \chi  }{2} }   }{ i\tilde \epsilon_{ba} }  \right) \rho_{aa} , \nonumber\\
\rho_{cc}    &=& \frac{1}{2\gamma  }\left(  -i\Omega_{c,\overline{\boldsymbol\varphi} + \frac{\boldsymbol \chi  }{2} } \rho_{ac}  +i\Omega_{c,\overline{\boldsymbol\varphi} - \frac{\boldsymbol \chi  }{2}} ^{*} \rho_{ca}   \right) \nonumber  \\
&=& \frac{1}{4\gamma  }\left( \Omega_{c,\overline{\boldsymbol\varphi} + \frac{\boldsymbol \chi  }{2}}\frac{ \Omega_{c,\overline{\boldsymbol\varphi} - \frac{\boldsymbol \chi  }{2}}^{*}    }{ i\tilde \epsilon_{ac} }   +\Omega_{c,\overline{\boldsymbol\varphi} - \frac{\boldsymbol \chi  }{2}} ^{*} \frac{\Omega_{c,\overline{\boldsymbol\varphi} + \frac{\boldsymbol \chi  }{2}}   }{ i\tilde \epsilon_{ca} }  \right) \rho_{aa}  ,\nonumber \\
\label{eq:diagonalElementsStationaryState}
\end{eqnarray}
which shows that these terms already scale with $\Omega_{\text{s}}^2$ and can be thus neglected in Eq.~\eqref{eq:lambdaSystem:firstOrderTerms}. The same is true for the coherence terms $\rho_{bc}$ and $\rho_{cb}$.  

Inserting the expressions in Eq.~\eqref{eq:lambdaSystem:firstOrderTerms}  and Eq.~\eqref{eq:diagonalElementsStationaryState} into the equation for $\rho_{aa}$ in Eq.~\eqref{eq:lambdaSystem_EOMs}, we find the effective Liouvillian equation
\begin{eqnarray}
\frac{d}{dt}   \rho_{aa}   
&=& \lambda_{0;\boldsymbol \chi}\rho_{aa} ,
\end{eqnarray}
with the only eigenvalue $\lambda_{0;\boldsymbol  \chi}$ given in Eq.~\eqref{eq:stationaryEigenvalueLambdaSystem}.

%

\end{document}